\documentclass[11pt]{article}

\usepackage{amsthm,amsmath,amssymb,bbm}
\usepackage{natbib}
\usepackage{multirow}
\usepackage[pdftex]{graphicx}
\usepackage{subfigure}
\usepackage{makecell}

\usepackage{array}
\usepackage{tabularx}
\usepackage{tabulary}
\usepackage{caption}
\usepackage{booktabs}
\usepackage{url}
\usepackage{algorithm}
\usepackage{algorithmic}
\usepackage{bm}
\usepackage{wrapfig}
\usepackage{lipsum}
\usepackage{mathrsfs}
\usepackage{dsfont}
\usepackage{titling}
\usepackage{smile}
\usepackage{verbatim}

\usepackage{xr}

\usepackage[usenames,dvipsnames,svgnames,table]{xcolor}
\usepackage[colorlinks,
linkcolor=red,
anchorcolor=blue,
citecolor=blue
]{hyperref}
\usepackage{pgf}

\newcommand{\sgn}{\mathop{\mathrm{sign}}}

\newcommand*\diff{\mathop{}\!\mathrm{d}}
\usepackage{smile}

\newcommand{\nn}{\nonumber}

\def\##1\#{\begin{align}#1\end{align}}
\def\$#1\${\begin{align*}#1\end{align*}}

\newcommand {\vecc}{\textnormal {vec}}
\def\T{^{{\rm T}}} %%%transpose operator
\def\sn{\sum_{i=1}^n}

\newcommand{\BB}{\mathbb{B}}

\newcommand{\wt}{\widetilde}

\newcommand{\bfsym}[1]{\ensuremath{\boldsymbol{#1}}}
\def \balpha   {\bfsym{\alpha}}       \def \bbeta    {\bfsym{\beta}}
\def \bgamma   {\bfsym{\gamma}}       \def \bdelta   {\bfsym{\delta}}

\newcommand{\Rom}[1]{\text{\uppercase\expandafter{\romannumeral #1\relax}}}

\usepackage{geometry}
\usepackage{enumitem}

\newcommand{\tp}{^{{\rm T}}}

\numberwithin{equation}{section}

\begin{document}
	
	\title{High-Dimensional Composite Quantile Regression:  \\Optimal Statistical Guarantees and Fast Algorithms}

	\author{Haeseong Moon\thanks{Department of Mathematics, University of California, San Diego, La Jolla, CA 92037, USA. E-mail:\href{mailto:h5moon@ucsd.edu}{\textsf{h5moon@ucsd.edu}}.}~~~and~~Wen-Xin Zhou\thanks{Department of Mathematics, University of California, San Diego, La Jolla, CA 92093, USA. E-mail:\href{mailto:wez243@ucsd.edu}{\textsf{wez243@ucsd.edu}}.}
	}

	\date{}
	\maketitle
	
	\vspace{-0.75in}
	
	\begin{abstract}
		The composite quantile regression (CQR) was introduced by Zou and Yuan [{\it Ann. Statist.} {\bf 36} (2008) 1108--1126] as a robust regression method for linear models with heavy-tailed errors while achieving high efficiency. Its penalized counterpart for high-dimensional sparse models was recently studied in Gu and Zou [{\it IEEE Trans. Inf. Theory} {\bf 66} (2020) 7132--7154], along with a specialized optimization algorithm based on the alternating direct method of multipliers (ADMM).  Compared to the various first-order algorithms for penalized least squares, ADMM-based algorithms are not well-adapted to large-scale problems. To overcome this computational hardness, in this paper we employ a convolution-smoothed technique to CQR, complemented with iteratively reweighted $\ell_1$-regularization.
		The smoothed composite loss function is convex, twice continuously differentiable, and locally strong convex with high probability. We propose a gradient-based algorithm for penalized smoothed CQR via a variant of the majorize-minimization principal, which gains substantial computational efficiency over ADMM.
		Theoretically, we show that the iteratively reweighted $\ell_1$-penalized smoothed CQR estimator achieves near-minimax optimal convergence rate under heavy-tailed errors without any moment constraint, and further achieves near-oracle convergence rate under a weaker minimum signal strength condition than needed in Gu and Zou (2020).
		Numerical studies demonstrate that the proposed method exhibits significant computational advantages without compromising statistical performance compared to two state-of-the-art methods that achieve robustness and high efficiency simultaneously.
		%near-oracle rate of convergence after a few iterations. We also propose a gradient-based algorithm for the smoothed composite quantile regression, which gains substantial computational efficiency over its non-smooth counterpart
		%is a robust alternative to the least squares while maintaining reasonable relative efficiency under light-tailed noises. It is known that the composite quantile regression is robust against general heavy-tailed error contamination, such as asymmetric or even having unbounded moments.  In this paper, we study the sparse composite quantile regression under ultrahigh dimensionality, where the feature dimension exceeds the sample size. 
		%Due to highly non-smooth nature of  the quantile loss function, the computational cost to calculate the estimator is heavy, especially in high dimensions. To overcome the lack of smoothness, we propose a convolution-type smoothed composite quantile regression with iteratively reweighted $\ell_1$-regularization. 
		%The resulting smoothed loss is provably locally strongly convex, and twice continuously differentiable. 
	\end{abstract}
	
	
	%%%%%%%%%%%%%%%%%%%%%%%%%%%%%%%%%%%%%%%%%%%
	%%%%%%%%%%%%%%%%%%%%%%%%%%%%%%%%%%%%%%%%%%%
	% Introduction
	%%%%%%%%%%%%%%%%%%%%%%%%%%%%%%%%%%%%%%%%%%%
	%%%%%%%%%%%%%%%%%%%%%%%%%%%%%%%%%%%%%%%%%%%
	\section{Introduction}
	\label{sec:1}

	Consider a sparse linear regression model $y=  \beta^*_0 + \bx\T\bbeta^* + \varepsilon$, where $y\in \RR$ is the response variable,  $\bx = (x_1,\ldots, x_p)\T$ is the $p$-vector of explanatory variables (covariates), and $\varepsilon\in \RR$ is the observation noise. In high-dimensional settings where the number of covariates considerably exceeds the number of observations,  a common practice is to impose a low-dimensional structure on $\bbeta^*$, the $p$-vector of regression coefficients.  Over the last three decades, various penalized regression methods have been developed for fitting high-dimensional models with low intrinsic dimensions,  typified by the $L_1$-penalized least squares method, also known as the Lasso \citep{Tibs1996,CDS1999}.  We refer to the monographs \cite{buhlmann2011statistics}, \cite{HTW2015}, \cite{W2019} and \cite{FLZZ2020}  for comprehensive expositions of high-dimensional statistical methods and theory.
	
	One of the main challenges in high-dimensional linear regression is that the maximum spurious correlation between the covariates and the realized noise can be large even when the population counterpart is small. 
	Therefore,  the penalized least squares methods are sensitive to the tails of the error distribution, or equivalently, the response distribution. The statistical properties are often derived under exponentially light-tailed error distributions \citep{BRT2009,buhlmann2011statistics}, including but not limited to Gaussian, sub-Gaussian or sub-exponential distributions.   Heavy-tailedness, however, has been frequently observed in empirical data, such as the high-dimensional microarray data as well as financial and economic data. To cope with heavy-tailed error contamination in high dimensions,  many robust penalized regression methods have been proposed; see, for example,  \cite{WLJ2007}, \cite{WJHZ2013}, \cite{LMY2016},  \cite{loh2017statistical}, \cite{FLW2017}, \cite{AMR2018} and \cite{SZF2020}. A common thread in these methods is the use of a robust loss function (that replaces the $L_2$ loss) to achieve either high breakdown point under arbitrary  contamination or near-optimal error bounds under heavy-tailed errors.
	For the latter,    \cite{loh2017statistical} considered the case where $\varepsilon$ has a symmetric distribution, including the standard Cauchy;  \cite{FLW2017} and \cite{SZF2020} provided a concentration study for  penalized Huber regression with a properly tuned cut-off parameter when $\varepsilon$ has a bounded variance but can be skewed/asymmetric.  To achieve robustness against gross outliers, we refer to \cite{SWS2021} the most recent advance and the references therein.

	In this work, we focus on heavy-tailed error contamination in a more general scenario.  When the error distribution is not only heavy-tailed but also asymmetric,   using a classical robust loss function, such as the $L_1$ loss,  the Huber loss and the Tukey loss,  may induce non-negligible bias.  The impact of this bias can be alleviated by letting the cut-off parameter in the Huber/Tukey loss grow with the sample size,  yet we still need $\varepsilon$ to have finite variance in order to achieve (near-)optimal convergence rate, and the parameter tuning is quite delicate in practice.   Although the least absolute deviation (LAD) regression requires no moment condition on $\varepsilon$,  the relative efficiency of the LAD can be arbitrarily small when compared with the least squares \citep{ZY2008}.   To overcome the efficiency loss while being robust against heavy-tailed errors, \cite{ZY2008} introduced the composite quantile regression (CQR), as a robust regression method, by combining quantile information across various quantile levels.  The asymptotic efficiency of the CQR relative to the least squares has a universal lower bound 86.4\% \citep{KLZ2010}. Theoretically,  CQR requires the existence of an everywhere nonvanishing density function of $\varepsilon$ without any moment constraint, thus allowing the infinite variance case.   By complementing a composite loss function with sparsity-inducing penalties,   \cite{BFW2011} and \cite{SparseCQR} further proposed penalized composite quasi-likelihood and quantile regression estimators, respectively.
	
	While  the CQR method inherits the robustness  property of quantile regression \citep{KB1978}, it also inherits the computational hardness especially in high dimensions.   Note that the $L_1$-penalized quantile regression can be recast as a  linear program (LP) \citep{WLJ2007,LZ2008}, solvable by general-purpose optimization toolboxes.  These toolboxes are convenient to use yet are only adapted to small-scale problems \citep{Bach2011}.  \cite{YLW2017} and \cite{Gu2018} proposed more efficient algorithms based on the alternating direction method of multipliers (ADMM).  For penalized CQR, \cite{SparseCQR} proposed an ADMM-based algorithm which we will revisit in Section~\ref{sec:4.1}.  The computational complexity of each ADMM update is of order $O(p n q + (p+q)^2)$, where $q\geq 1$ is the number of quantile levels used in the CQR. This can be computationally intensive when applied to large-scale datasets; see Section~\ref{sec:algo} for more detailed discussions.

	To extend the capability of CQR with large-scale data,  in this paper we propose a convolution-smoothed CQR (SCQR) method, complemented with iteratively reweighted $L_1$-penalization for fitting sparse models.    Convolution smoothing turns the piecewise linear check function  into a twice continuously differentiable, convex and locally strongly convex surrogate.  Its success has recently been witnessed in the context of quantile regression  in both statistical and computational aspects \citep{FGH2021,he2020smoothed}.
	%The smoothed QR estimator is  shown to be  first-order equivalent to the standard QR estimator,  and enjoys desirable statistical properties.
	%To alleviate the computational complexity of
	%\cite{FGH2021} proposed to smooth the check function (quantile loss) via convolution,   which gives rise to a twice continuously differentiable, convex and locally strongly convex surrogate. This admits fast and scalable gradient-based algorithms to perform optimization \citep{he2020smoothed}.   The smoothed QR estimator is  shown to be  first-order equivalent to the standard QR estimator,  and enjoys desirable statistical properties \citep{FGH2021,he2020smoothed}. 
	%Motivated by the success of convolution smoothing for QR,  in this paper we propose a smoothed composite quantile regression (SCQR) method, complemented with iteratively reweighted $L_1$-penalization for fitting sparse models in the presence of heavy-tailed errors.    
	Under a Lipschitz continuity condition on the density of $\varepsilon$ and sub-Gaussian (stochastic) designs,  we show that the $L_1$-penalized SCQR (SCQR-Lasso) estimator achieves the same rate of convergence as the Lasso estimator when $\varepsilon$ is sub-Gaussian.   We do not require the symmetry of the error distribution nor the existence of any moment, including the mean.   Moreover, under a mild minimum signal strength (also known as the beta-min) condition, we show that the iteratively reweighted $L_1$-penalized SCQR estimator converges at a near-oracle rate $O(\sqrt{(s+\log q) /n})$.   This reveals the advantage of folded-concave penalization in terms of its adaptivity to strong signals.  Heuristically, the $L_1$ penalty applies soft-thresholding to all signals ignoring their magnitudes, thus creating a bias that is of order $\lambda$ for all non-zero signals, where $\lambda>0$ is the regularization parameter.  Furthermore, we employ a variant of the local adaptive majorize-minimization (LAMM) algorithm \citep{Fan2018} for solving weighted $L_1$-penalized SCQR estimator.  The main idea is to construct an isotropic quadratic objective function that locally majorizes the smoothed composite quantile loss such that closed-form updates are available at each iteration.  The quadratic coefficient is adaptively chosen so that the objective function is non-increasing along the iteration path.  Compared to ADMM, LAMM is a simpler gradient-based algorithm that is particularly suited for large-scale problems, where the dominant computational effort is a relatively cheap matrix-vector multiplication at each step. The (local) strong convexity of the convolution smoothed loss facilitates the convergence of such a first order method.
	
	Our work complements \cite{SparseCQR} in two aspects.  The theoretical results in \cite{SparseCQR} are derived in the case of fixed designs satisfying conditions (C1) and (C2) therein.  It is unclear whether these conditions hold with high probability for Gaussian or sub-Gaussian covariates.  We provide a random design analysis for sub-Gaussian covariates.  To achieve oracle convergence rate,  our beta-min condition  is weaker than that in \cite{SparseCQR}  by relaxing the $\sqrt{s}$-factor, where $s$ denotes the model sparsity.
	Computationally,  we develop a fast algorithm for penalized CQR without sacrificing statistical efficiency by means of convolution smoothing. We believe that this paper introduces an interesting compromise between
	robustness,  statistical performance and numerical  efficiency for sparse linear regression with heavy-tailed errors.

	This work is also closely related to  \cite{Wang2020}, in which a new robust regression method is proposed  along with a simulation-based procedure for choosing the regularization parameter.   In low dimensions,  we refer to \cite{Wang2020}'s method as pairwise-LAD as it applies LAD regression to the pairwise differences of the observations, namely, $\{ (y_i - y_j, \bx_i - \bx_j )\}_{1\leq i\neq j\leq n}$.  Although CQR and pairwise-LAD are motivated quite differently,  an intriguing connection is that the asymptotic relative efficiency of pairwise-LAD  is equivalent to that of CQR (compared to the least squares) when $q$, the number of quantile levels, goes to infinity.
	Computationally, \cite{Wang2020} reformulates $L_1$-penalized pairwise-LAD as a linear program with $2n^2+2p$ variables and $O(n^2 +p)$ constraints.  Due to the high computational complexity and storage cost,  generic LP solvers can be extremely slow in practice.     To alleviate the computational burden, \cite{Wang2020} suggested using the resampling technique \citep{CBC2016} that is able to reduce the effective sample size $O(n^2)$ (for pairwise differences) to $O(n)$.  
	
	The rest of the paper is organized as follows.  Section \ref{sec2} starts with a brief review of (penalized) composite quantile regression,  followed by the proposed convolution smoothed CQR with iteratively reweighted $L_1$-penalization.  The selection of tuning parameters is discussed  in Section \ref{sec:2.3}.
	Section \ref{sec:theory} provides the statistical guarantees for penalized SCQR, including a bias analysis and rates of convergence of the solution path. In Section \ref{sec:algo},  we first revisit the ADMM-based algorithm proposed in \cite{SparseCQR}, and then introduce a gradient-based LAMM algorithm for convolution smoothed CQR. Numerical comparisons of the three methods, CQR, SCQR and pairwise-LAD, are  conducted in Section \ref{numericalstudy}. All the proofs are placed in the appendix. The Python code for the proposed method and our implementation of the methods in\cite{SparseCQR} and \cite{Wang2020} is available at \href{https://github.com/hsmoonjohn/scqr}{https://github.com/hsmoonjohn/scqr}.

	\section{Sparse composite quantile regression}
	\label{sec2}

	\subsection{Preliminaries}
	\label{subsec:problem}
	
	Suppose we observe $n$ independent samples $\{ (\bx_i, y_i ) \}_{i=1}^n$ of a random variable $(\bx, y) \in \RR^p \times \RR$ satisfying the linear model 
	\#
	y =  \beta^*_0 + \bx\tp \bbeta^* + \varepsilon  =  \beta_0^* + \sum_{j=1}^{p} x_j  \beta_j^* + \varepsilon , \label{linear.model}
	\#
	where $\beta_0^*$ is the intercept, $\bbeta^*= (\beta^*_1,\ldots, \beta^*_p)\tp \in \RR^p$ is the $p$-vector of slope coefficients, and $\varepsilon$ denotes the observation noise. Assume that $\varepsilon$ is independent of $\bx$, and has cumulative distribution function $F(\cdot)$ and probability density function $f(\cdot)$.   Without loss of generality,  we assume $\beta^*_0=0$; otherwise we set $\wt \varepsilon =\beta^*_0 + \varepsilon$ so that the model becomes $y = \bx\T \bbeta^* + \wt \varepsilon$.
	Under these assumptions, the conditional $\tau$-quantile ($0<\tau<1$) of $y|\bx$ is
	$$
	F^{-1}(\tau)+ \sum_{j=1}^{p} x_j  \beta_j^*  ,
	$$
	where $F^{-1}(\tau): = \inf\{ u \in \RR: F(u) \geq \tau \}$ is the $\tau$-quantile of $\varepsilon$.
	
	To robustly estimate $\bbeta^*$ in model \eqref{linear.model}, we consider the composite quantile regression (CQR) approach proposed in \cite{ZY2008}, which delivers consistent estimates even when the error distribution has infinite variance and enjoys high efficiency otherwise \citep{ZY2008,KLZ2010}. Given a positive integer $q$, let $\{ \tau_k \}_{k=1}^q \subseteq (0, 1)$ be an increasing sequence of quantile indexes and write $\alpha^*_k =  F^{-1}(\tau_k)$. When $p<n$, the canonical CQR estimator of $\bbeta^*$, denoted by $\hat \bbeta^{{\rm CQR}}$, is defined as 
	\#
	(\hat \alpha_1, \ldots, \hat \alpha_q, \hat \bbeta^{{\rm CQR}} ) \in \argmin_{  \balpha = (\alpha_1, \ldots, \alpha_q)\T \in \RR^q, \atop  \bbeta \in \RR^p}    \underbrace{  \frac{1}{n q}	\sn\sum_{k=1}^q \rho_{\tau_k}(y_i-\alpha_k-\bx_i^{{\rm T}}\bbeta) }_{=: \, \hat Q(\balpha, \bbeta) } , \label{PlainCQRloss}
	\#
	where $\rho_\tau(u) = \{ \tau - \ind(u<0)\} u $ is the check function. In the special case of $q=1$, this becomes the usual quantile regression \citep{KB1978}. \cite{ZY2008} established the asymptotic normality of $\hat \bbeta^{{\rm CQR}}$ when the density function $f(\cdot)$ of $\varepsilon$ is non-vanishing at the selected quantile levels. Therefore, the root-$n$ consistency of $\hat \bbeta^{{\rm CQR}}$ requires no moment condition on $\varepsilon$, thus allowing very heavy-tailed errors such as the Cauchy error.

	For high-dimensional sparse models in which $\bbeta^*$ is $s$-sparse with $s\ll n$,  \cite{SparseCQR} proposed the penalized CQR estimator, defined as the global optimum to the optimization problem
	\#
	\min_{\alpha_1, \ldots, \alpha_q \in \RR , \atop \bbeta \in \RR^p} \bigg\{  \frac{1}{n q}	\sn\sum_{k=1}^q\rho_{\tau_k}(y_i-\alpha_k-\bx_i^{{\rm T}}\bbeta)  + \sum_{j=1}^p P_\lambda(|\beta_j| ) \bigg\} ,\label{penalizedCQR}
	\#
	where $P_\lambda(\cdot) := \lambda^2 P(\cdot/\lambda)$ for some penalty function $P:[0, \infty) \to [0,\infty)$ and regularization parameter $\lambda>0$. The regularizer $P(\cdot)$ is allowed to be non-convex (concave), which helps reduce the bias and leads to oracle estimators when the signals are sufficiently strong \citep{ZZ2012}. The most commonly used sparsity-inducing penalty functions are 
	\begin{itemize}
		\item[(i)] $L_1$ function \citep{Tibs1996}: $P(t) = t$ for $t\geq 0$. 
		
		\item[(ii)] Smoothly clipped absolute deviation (SCAD) penalty \citep{FL2001}: $P(0)=0$ and $P'(t) = \ind(t\leq 1) + \frac{(a-t)_+}{a-1} \ind(t>1)$ for $t\geq 0$ and some constant $a>2$.
		
		\item[(iii)] Minimax concave (MC) penalty \citep{Z2010}: $P(0)=0$ and $P'(t) = (1- t/a)_+$ for $t\geq 0$ and some constant $a\geq 1$.
	\end{itemize}
	
	Computationally, \cite{SparseCQR} employed the local linear approximation (LLA) algorithm \citep{zou2008one} to obtain an approximate solution to the nonconvex problem \eqref{penalizedCQR}, which enjoys desirable statistical properties. The LLA algorithm for the optimization problem \eqref{penalizedCQR} is iterative, starting at iteration 0 with an initial estimate $\hat \bbeta^0\in \RR^p$. At iteration $t = 1, 2, \ldots$, it combines a weighted $L_1$-penalty with the composite quantile loss to obtain the updated estimates $(\hat \balpha^t, \hat \bbeta^t)$. The procedure involves two steps.
	
	\begin{itemize}
		\item[1)] Using the previous estimate $\hat \bbeta^{t-1} = (\hat \beta^{t-1}_1, \ldots, \hat \beta^{t-1}_p)\tp$, compute the penalty weights
		$$
		w^{t-1}_j = P'_\lambda( | \hat \beta^{t-1}_j | ) = \lambda P'(| \hat \beta^{t-1}_j | / \lambda ) \geq 0  , \ \ j=1,\ldots, p.
		$$
		
		\item[2)]  Solve the convex optimization problem 
		\#
		\min_{\alpha_1, \ldots, \alpha_q \in \RR , \, \bbeta \in \RR^p}  \bigg\{  \frac{1}{n q}	\sn\sum_{k=1}^q\rho_{\tau_k}(y_i-\alpha_k-\bx_i\tp  \bbeta) + \sum_{j=1}^p   w^{t-1}_j | \beta_j |   \bigg\}  \label{weighted.l1.cqr}
		\#
		to obtain $\hat \balpha^t = (\alpha^t_1, \ldots, \alpha^t_q)\tp$ and $\hat \bbeta^t \in \RR^p$.
	\end{itemize}
	Under the following ``beta-min" (minimum signal strength) condition
	\#
	\min_{1\leq j\leq p : \beta^*_j \neq 0} | \beta^*_j | \gtrsim \sqrt{\frac{s \log (p)}{n}}  \label{beta-min.cond1}
	\#
	among other regularity conditions on the non-stochastic design matrix $\mathbf{X}=(\bx_1, \ldots, \bx_n)\tp \in \RR^{n\times p}$, \cite{SparseCQR} showed that initialized with the $L_1$-penalized CQR (CQR-Lasso) estimator, the LLA algorithm converges to the oracle estimator in two iterations with high probability.

	\subsection{Convolution smoothed composite quantile regression}
	\label{subsec:smoothCQR}

	Motivated by the smoothed QR approach \citep{FGH2021} that has attractive statistical properties and computational benefits \citep{he2020smoothed,ncvxQR}, we propose a penalized smoothed CQR estimator by complementing the convolution-smoothed composite quantile loss with a folded concave regularizer.
	We show that the proposed estimator, computed by a combination of the LLA and the iterative local adaptive majorize-minimization (LAMM) \citep{Fan2018} algorithms, achieves oracle statistical properties under a relaxed ``beta-min" condition compared to \eqref{beta-min.cond1}.  
	We refer to Section~\ref{sec:algo} for a computational comparison between ADMM and LAMM.
	%The complexity of each ADMM update is $O(pnq+(p+q)^2)$. Moreover, the ADMM needs to compute and store the inverse of a $(p+q)\times (p+q)$ matrix, which has a complexity $O(n^3)$ with the aid of  the Sherman-Morrison-Woodbury formula. In contrast, LAMM is essentially a proximal gradient method, and hence is computationally cheaper to use.

	%the real computation effort of ADMM \citep{SparseCQR} is to evaluate the inverse of a $p\times p$ or $n\times n$ matrix, which is not only computationally intensive but also requires a large storage cost when both $n$ and $p$ are large. 

	For every $\bbeta \in \RR^p$, let $\hat F(\cdot; \bbeta)$ be the empirical  cumulative distribution function  of the residuals $\{r_i(\bbeta):=y_i-\bx_i\T\bbeta\}_{i=1}^n$. Then, the empirical composite quantile loss in \eqref{PlainCQRloss} can be written as
	\$
	\Hat{Q}(\balpha,\bbeta)=\frac{1}{q}\int_{-\infty}^{\infty}\sum_{k=1}^q \rho_{\tau_k}(u-\alpha_k)\diff\Hat{F}(u;\bbeta) ,  
	\$
	where $\balpha = (\alpha_1, \ldots, \alpha_q)\T \in \RR^q$. 
	Let $K: \RR \to [0, \infty)$ be a symmetric, non-negative kernel function (a function that integrates to 1). For a given sequence of bandwidth parameters $h=h_n >0$, we smooth the loss $\hat Q(\cdot, \cdot)$ by
	\#
	\hat{Q}_h(\balpha,\bbeta) & := \frac{1}{q} \sum_{k=1}^q \int_{-\infty}^\infty\rho_{\tau_k}(u-\alpha_k)   \diff \hat F_h(u, \bbeta)   \label{SECQRloss} \\
	& =\frac{1}{n q}\sn\sum_{k=1}^q \int_{-\infty}^\infty\rho_{\tau_k}(u)K_h  (u+\alpha_k-r_i(\bbeta)  )\diff{u}, \nn
	\#
	where
	$$
	\hat F_h(u, \bbeta)  = \frac{1}{n} \sn  \int_{-\infty}^u   K_h(v- r_i(\bbeta))    \diff{v}
	~~\mbox{ and }~~  K_h(u) = \frac{1}{h} K(u/h). 
	$$
	For each $k=1, \ldots, m$, define the convolution smoothed counterpart of $\rho_{\tau_k}(\cdot)$ as
	\#
	\ell_{h, k} (u) = ( \rho_{\tau_k} * K_h ) (u) = \int_{-\infty}^\infty \rho_{\tau_k }(v)  K_h( u-v ) \diff v , 
	\#
	where $*$ denotes the convolution operator. Consequently, the smoothed composite loss $\hat Q_h(\balpha, \bbeta)$ defined in \eqref{SECQRloss} can be equivalently written as 
	$$
	\hat Q_h(\balpha, \bbeta) = \frac{1}{n q}\sn\sum_{k=1}^q  \ell_{h, k} (y_i - \alpha_k - \bx_i\T \bbeta) . 
	$$

	Starting with an initial estimate $\hat \bbeta_h^0 \in \RR^p$, we define a sequence of iteratively reweighted $L_1$-penalized smoothed CQR estimators 
	$$
	\big\{ \hat \balpha_h^t = (\hat \alpha^t_{h, 1}, \ldots, \hat \alpha^t_{h, q} )\T,  \,  \hat \bbeta_h^t = (\hat \beta^t_{h, 1} , \ldots, \hat \beta^t_{h,p})\T \big\}_{t =1, 2, \ldots }
	$$  
	as follows. At iteration $t = 1, 2, \ldots$, $(\hat \balpha^t_h, \hat \bbeta^t_h)$ is defined as a solution to the convex optimization problem
	\#
	\min_{ \balpha\in \RR^q, \, \bbeta \in \RR^p}  \bigg\{   \hat Q_h(\balpha, \bbeta) + \sum_{j=1}^p P_\lambda'( | \hat \beta^{t-1}_{h, j}  | ) \cdot  | \beta_j | \bigg\} , \label{SPCQR}
	\#
	where $\lambda>0$ is the regularization parameter.

	Section~\ref{sec:theory} establishes the statistical properties of the solution path $\{ (\hat \alpha^t_h, \hat \bbeta^t_h ) \}_{t \geq 1}$ in the stochastic design setting. Specifically, we assume that the random covariate vectors $\bx_i$'s are {\it sub-Gaussian}; see Condition (A3) below.  Recall that the theoretical results in \cite{SparseCQR} are derived in the case of fixed designs satisfying conditions (C1) and (C2) therein. It is natural to question whether or not these conditions hold with high probability for Gaussian or sub-Gaussian covariates.
	In Section~\ref{sec:3.2}, we describe a variant of the LAMM  algorithm for solving \eqref{SPCQR}, which will be compared to the ADMM algorithm for solving \eqref{weighted.l1.cqr} in terms of statistical accuracy and computation time; see Section~\ref{numericalstudy}.

	\subsection{Selection of tuning parameters}
	\label{sec:2.3}

	The penalized smoothed CQR method relies primarily on two key tuning parameters, the regularization parameter $\lambda$ and the bandwith $h$.   As for the quantile indexes,   we follow the suggestion in \cite{ZY2008}   and take $\tau_k = k/(q+1)$, $k=1,\ldots, q$ with $q=19$.  The resulting estimator thus combines the strength across multiple QR estimators at levels $5\%, 10\%, \ldots, 90\%, 95\%$.

	\cite{he2020smoothed} and \cite{ncvxQR} demonstrated numerically that the  convolution-smoothed QR estimator is rather insensitive to the choice of the bandwidth as long as it is in a reasonable range (neither too small nor too large). Motivated by \cite{ncvxQR},  we set the default value of $h$ as $\max\{ 0.01,  \sqrt{\overline{\tau}(1-\overline{\tau})} \{ \log(p)/n \}^{1/4} \}$, where $\overline{\tau}= q^{-1}\sum_{k=1}^q \tau_k$.  
	The penalty level $\lambda$, on the other hand,  has a more visible impact on the performance as  it directly controls the sparsity of the solution. One general approach is to use $K$-fold  cross-validation (e.g. $K=5$ or $10$) when given a set of $\lambda$ values.  If model selection is of more interest than prediction,  information criteria typically produce much smaller models and thus are preferable.  As a variant of the high-dimensional Bayesian information criterion (BIC)  for penalized QR \citep{Lee2014},  \cite{SparseCQR}  considered the following BIC in the context of composite QR:
	\#
	\text{BIC}(\lambda):=\log\bigg(\frac{1}{q}\sn\sum_{k=1}^q\rho_{\tau_k}(y_i-\hat{\alpha}_k(\lambda)-\bx_i\T\hat{\bbeta}(\lambda) )\bigg)+\vert\hat{\cS}_\lambda\vert\frac{C_n\log(p)}{n}, \label{biccriterion}
	\#
	where $(\hat{\balpha}(\lambda) ,\hat{\bbeta}(\lambda))$ is a penalized CQR estimator with regularization parameter $\lambda$, $\hat{\cS}_\lambda$ is the support of $\hat{\bbeta}(\lambda)$, and $C_n$ is a positive number depending on $n$.   Typically $C_n$ is chosen as a slowly growing function of $n$,  e.g., $C_n=\log(\log n)$. 
	
	Motivated by the simulation-based method proposed by \cite{Belloni2011}, we further describe a $\lambda$-tuning procedure that is computationally much cheaper than the cross-validation and BIC methods.  The key is to utilize the the pivotal property of the $L_1$-loss \citep{Belloni2011}.  As we shall see from the theoretical results in Section~\ref{sec:theory},  the magnitude of $\lambda$ depends in theory on $\Vert\bomega^*\Vert_\infty$, where  $\bomega^*=(nq)^{-1}\sn\sum_{k=1}^q\{\bar{K}((\alpha_{h,k}^*-\varepsilon_i)/h)-\tau_k\}\bx_i$, and $\bar{K}((\alpha_{h,k}^*-\varepsilon_i)/h)$ serves as a smoothed proxy of $\ind(\varepsilon_i \leq \alpha^*_{k})$.  Recall that $\PP(\varepsilon_i \leq \alpha^*_{k}) = \tau_k$ for each $k=1,\ldots, q$. Therefore, we consider a pivotal proxy of $\bw^*$, defined as
	\$
	\wt \bw^* = \frac{1}{nq}\sn \sum_{k=1}^q  \{ \ind(u_{i,k} \leq \tau_k ) - \tau_k \} \bx_i , \quad u_{ik} \stackrel{\rm i.i.d.}{\sim} {\rm Unif}(0, 1).
	\$
	For some constant $c>1$ and $\alpha \in (0 , 1)$,  we set
	\#
	\lambda^* = \lambda^*(c, \alpha) =c \cdot F^{-1}_{ \Vert\tilde{\bomega}^*\Vert_\infty | \mathbf X }(1-\alpha) , \label{lambdasimulation}
	\#
	where $F^{-1}_{ \Vert\tilde{\bomega}^*\Vert_\infty | \mathbf X }(1-\alpha)$ denotes the $(1-\alpha)$-quantile of $ \Vert\tilde{\bomega}^*\Vert_\infty$ given  $\mathbf{X}=(\bx_1, \ldots, \bx_n)\T$.  We  can calculate $\lambda^* $ numerically with any specified precision by simulation. {In particular,  we choose $(c, \alpha) = (1.9, 0.05)$ for SCQR-Lasso and  $(c, \alpha) = (3.1, 0.05)$ for SCQR-SCAD,  and simulate the conditional distribution of $\Vert\tilde{\bomega}^*\Vert_\infty$ given $ \mathbf X$ based on 200 replications. }

	%%%%%%%%%%%%%%%%%%%%%%%%%%%%%%%%%
	% Theoretical Properties
	%%%%%%%%%%%%%%%%%%%%%%%%%%%%%%%%%
	%%%%%%%%%%%%%%%%%%%%%%%%%%%%%%%%%
	\section{Statistical analysis}
	\label{sec:theory}
	In this section, we establish the statistical properties of the penalized smoothed CQR estimators $\{ (\hat \balpha^t_h, \hat \bbeta^t_h ) \}_{t\geq 1}$ initialized at $\hat \bbeta^0_h = \mathbf{0}$ and for an ordered sequence of quantile indexes $0< \tau_1   < \cdots < \tau_q <1$ with $q \geq 1$. To begin with, Section~\ref{subsec:Smoothingbias} provides non-asymptotic upper bounds on the smoothing bias. Throughout the section, we assume that $\bbeta^* \in \RR^p$ in model \eqref{linear.model} is $s$-sparse, that is, its support $\cS =\{1\leq j\leq p: \beta^*_j\neq 0\}$ has cardinality $s$.

	\subsection{Smoothing bias}
	\label{subsec:Smoothingbias}
	
	For any $h>0$, define the population composite quantile loss $Q_h(\balpha,\bbeta)=\EE\{\hat{Q}_h(\balpha,\bbeta)\}$ for $\balpha\in \RR^q$ and $\bbeta \in \RR^p$, and its minimizer 
	\#
	(\balpha_h^*, \bbeta_h^*) \in\argmin_{\balpha\in\RR^q , \,  \bbeta\in\RR^p}Q_h(\balpha,\bbeta) . \label{SCQRpopulationloss} 
	\#
	We first show that $Q_h: \RR^{q+p} \to \RR$ is convex  under mild regularity conditions.
	
	\begin{lemma} \label{lem:convexity}
		Assume that the random covariate vector $\bx \in \RR^p$ is non-degenerate with $\bar \Sigma = \EE(\bar \bx \bar \bx\T) \succ \textbf{0}$, where $\bar \bx = (1, \bx\T)\T$. Moreover, let the kernel function $K(\cdot)$ and bandwidth $h>0$ be such that 
		\#
		\min_{k=1,\ldots, q} \int_{-\infty}^\infty K(u) f( F^{-1}(\tau_k) + h u ) {\rm d} u > 0 , \label{reg.cond}
		\#
		where $F$ and $f$ denote the CDF and density function of $\varepsilon$, respectively.
		Then, the population smoothed composite quantile loss $Q_h: \RR^{q +p} \to \RR$ is convex and strictly convex at $(\balpha^*, \bbeta^*)$, where $ \balpha^* = (\alpha^*_1, \ldots, \alpha^*_q)\T$ with $\alpha_k^* =   F^{-1} (\tau_k)$.
	\end{lemma}

	Condition~\ref{reg.cond} can easily be verified if either the kernel function $K(\cdot)$ or the density function $f(\cdot)$ is positive everywhere. Without loss of much generality, we assume the former throughout this section. Intuitively, convolution smoothing induces bias which allows us to think $(\balpha^*_h, \bbeta^*_h) \neq (\balpha^*, \bbeta^*)$ for any given $h>0$. By exploiting the independence of $\varepsilon$ and $\bx$ and a strictly positive kernel (e.g., Gaussian, Laplacian or logistic), we find that $\bbeta^*_h = \bbeta^*$ for any $h>0$, and therefore the proposed smoothing mechanism only generates bias on the intercepts $\alpha^*_1, \ldots, \alpha^*_q$ that are of less interest. 
	
	To obtain explicit upper bounds on the (smoothing) bias, we impose the following regularity conditions of the density function $f(\cdot)$ of $\varepsilon$ as well as the kernel function $K(\cdot)$. 
	\begin{itemize}
		
		\item [(A1)] There exist constants $\underline f, l_0 >0$ such that  $\min_{k=1,\ldots, q} f( F^{-1}(\tau_k))  \geq  \underline f$ and $|f(u) - f(v) |\leq l_0 | u- v|$ for all $u, v \in \RR$.
		
		\item[(A2)] The kernel function $K(\cdot)$ is symmetric around zero and positive everywhere. Moreover, $\kappa_k :=\int_{-\infty}^\infty \vert u\vert^k K(u)\diff u < \infty$ for $k\leq 2$, and $\underline{\kappa}:=\min_{\vert u\vert\leq1}K(u)>0$. 
		
	\end{itemize}
	
	Define the function
	\#
	m_h( \bb ) =  \EE \bigg\{ \frac{1}{q} \sum_{k=1}^q \ell_{h, k} (\varepsilon - b_k) \bigg\}  , \ \  \bb = (b_1, \ldots, b_q)\T \in \RR^q,  \label{def:mh}
	\# 
	where $\ell_{h, k} = \rho_{\tau_k} * K_h$. Under assumptions (A1) and (A2), we will show that $m_h: \RR^q \to \RR$ is strictly convex with a unique minimizer $\bb_h=(b_{h,1} ,\ldots, b_{h, q} )\T$, satisfying $\max_{1\leq k\leq q } | b_{h,k } - F^{-1}(\tau_k) | \lesssim h^2$.  Consequently, the smoothed population composite quantile loss $Q_h:\RR^{q+p}\to \RR$ also has a unique minimizer, which is $(\balpha^*_h , \bbeta^*_h) = (  \bb_h, \bbeta^*)$ and satisfies $\| \balpha^*_h - \balpha^* \|_\infty \lesssim h^2$.

	\begin{proposition} \label{proposition3.1}
		Suppose assumptions (A1) and (A2) hold and that $\bar \Sigma$ is positive definite. Then, the smoothed population composite quantile loss $Q_h: \RR^{q+p} \to \RR$ for any $h>0$ is strictly convex and has a unique minimizer given by $( \bb_h , \bbeta^*)$, where $\bb_h = (b_{h,1}, \ldots, b_{h,q})\T \in \RR^q$ is the unique minimizer of the function $m_h$ defined in \eqref{def:mh}.
		Furthermore, provided $0< h \leq \underline f/(2  \kappa_2^{1/2}l_0 )$, we have
		\# \label{bias.ubd}
		| b_{h, k} - f(F^{-1}(\tau_k) ) | \leq \frac{2  \kappa_2 l_0}{f(F^{-1}(\tau_k) ) } h^2  ~\mbox{ for }~ k = 1, \ldots, q.
		\#
	\end{proposition}

	\subsection{Oracle rate of convergence}
	\label{sec:3.2}
	
	Initialized at $\hat \bbeta_h^0 = \textbf{0}$, let $\{ (\hat \balpha^t_h, \hat \bbeta^t_h ) \}_{t\geq 1}$ be a sequence of penalized smoothed CQR estimators defined in \eqref{SPCQR}. 
	Without loss of generality, we assume $\bmu = \EE(\bx) = \textbf{0}$; otherwise, we can rewrite model \eqref{linear.model} as  $y =  \beta_0^\star   +  (\bx - \bmu)\T \bbeta^* + \varepsilon$ with $  \beta_0^\star := \beta_0^* + \bmu\T \bbeta^*$. Hence, it suffices to work with the centered data $\{(y_i, \bx_i - \bmu) \}_{i=1}^n$. In addition, we assume that the random covariate $\bx$ is sub-Gaussian; see condition (A3) below.
	For the regularizer $P_\lambda : [0, \infty) \to [0, \infty)$, we impose the following conditions that encompass the $L_1$, SCAD and MC penalties.

	\begin{itemize}
		
		\item[(A3)](sub-Gaussian covariates) The covariance matrix $\Sigma= (\sigma_{jk})_{1\leq j, k\leq p}=\EE(\bx \bx\T)$ is positive definite.
		There exist constants $\nu_0,c_0\geq 1$ such that  $\PP(\vert\bar{\bz}\T\bu\vert\geq\nu_0 \|  \bu \|_2 \cdot u)\leq c_0 e^{-u^2/2}$ for all $\bu\in \RR^{p+1}$ and $u \geq 0$, where $\bar \bz = \bar \Sigma^{-1/2} \bar \bx$ and 
		$$
		\bar \Sigma = \EE(\bar \bx \bar \bx\T)  = \begin{bmatrix}
			1 & \textbf{0}_{1 \times p} \\
			\textbf{0}_{p\times 1} & \Sigma 
		\end{bmatrix} .
		$$
		For simplicity, we assume $c_0=1$ and write $\sigma_{\bx}^2=\max_{1\leq j\leq p} \sigma_{jj}$. Moreover, let $\gamma_1 \geq 1 \geq \gamma_p >0$ be the largest and smallest eigenvalues of $\Sigma$.

		\item[(A4)]  $P_\lambda(u)=\lambda^2 P(u/\lambda)$ for $u\geq0$, where the function $P:[0,\infty)\to [0,\infty)$ is non-decreasing, differentiable almost everywhere on $(0, \infty)$, and satisfies $P(0)=0$, $0\leq P'(u)\leq 1$, $\lim_{u \downarrow 0 } P'(u)=1$ and $P'(u_1)\leq P'(u_2)$ for all $u_1\geq u_2\geq 0$.
	\end{itemize}

	Under condition (A4), $(\hat \balpha^1_h, \hat \bbeta^1_h )$ is essentially the $L_1$-penalized smoothed CQR estimator (SCQR-Lasso), that is,
	\#
	(\hat \balpha^1_h, \hat \bbeta^1_h )  \in \argmin_{\balpha\in\RR^q , \bbeta\in\RR^p  }  \big\{ \hat{Q}_h(\balpha,\bbeta)+ \lambda \| \bbeta \|_1 \big\} .   \label{l1SCQR}
	\#
	Without smoothing, \cite{SparseCQR} obtained the convergence rates (under $L_2$-loss) for the $L_1$-penalized CQR (CQR-Lasso) estimator $(\hat \balpha , \hat \bbeta  )  \in \argmin_{\balpha\in\RR^q , \bbeta\in\RR^p  }   \{ \hat{Q} (\balpha,\bbeta)+ \lambda \| \bbeta \|_1 \}$ under fixed designs.
	For sub-Gaussian (stochastic) designs, we first establish estimation error bounds for $(\hat \balpha^1_h, \hat \bbeta^1_h )$, which complement the results in \cite{SparseCQR}.
	
	Recall from Proposition~\ref{proposition3.1} that $(\balpha_h^*, \bbeta^*)$ is the unique minimizer of the population loss $Q_h$. Write $\balpha_h^* = (\alpha^*_{h, 1}, \ldots, \alpha^*_{h, q})\T \in \RR^q$.
	For the smoothed loss $\hat Q_h(\balpha, \bbeta)$, define its partial gradient vectors at $(\balpha^*_h, \bbeta^*)$ as
	\begin{align} 
		\bzeta^*  &:=\nabla_{\balpha} \hat{Q}_h(\balpha_h^*,\bbeta^*) =\frac{1}{n q} \sn \begin{bmatrix}\bar{K}((\alpha_{h,1}^*-\varepsilon_i)/h)-\tau_1\\ \vdots \\ \bar{K}((\alpha_{h,q}^*-\varepsilon_i)/h)-\tau_q \end{bmatrix} \in \RR^q ,\\
		\bomega^* &:=\nabla_{\bbeta} \hat{Q}_h(\balpha_h^*,\bbeta^*) =\frac{1}{n q}	\sn\sum_{k=1}^q   \{\bar{K}((\alpha_{h,k}^*-\varepsilon_i)/h)-\tau_k  \}\bx_i \in \RR^p,
	\end{align}
	where $\bar{K}(u)=\int_{-\infty}^uK(t)\diff t$.
	
	The key elements of our analysis are (i) a cone-like property for $\hat \bbeta^1_h - \bbeta^*$, and (ii) a local restricted strong convexity (RSC) property for the empirical loss $\hat Q_h$, which is based on the function
	\#  
	D(\balpha , \bbeta): = \bigg\langle  \nabla \hat Q_h(\balpha, \bbeta)-\nabla\hat{Q}_h(\balpha_h^*,\bbeta^*) ,  {\scriptsize		 \begin{bmatrix}
			\balpha - \balpha^*_h  \\
			\bbeta - \bbeta^* 
	\end{bmatrix} } \bigg\rangle , \ \ \balpha \in \RR^q, \bbeta \in \RR^p .  \label{BregmanD}
	\#

	For any regularization parameter $\lambda >0$, define the event 
	\#
	\cG(\lambda ):= \big\{  \Vert\bzeta^*\Vert_\infty \leq 3\lambda/(2q), \,  \Vert\bomega^*\Vert_\infty\leq\lambda/2 \big\} \label{eventG}
	\#
	and the restricted (cone-like) set 
	\#
	\cC = \cC(\cS) :=  \bigg\{  {\scriptsize		 \begin{bmatrix}
			\bdelta \\
			\bDelta
	\end{bmatrix} } \in \RR^{q+p} : \| \bDelta_{\cS^{{\rm c}}} \|_1 \leq 3 \| \bDelta_{\cS} \|_1 +  3 q^{-1/2} \| \bdelta \|_2 \bigg\} .
	\#
	It can be shown that $ {\scriptsize		 \begin{bmatrix}
			\hat \balpha^1_h - \balpha^* \\
			\hat \bbeta^1_h - \bbeta^*
	\end{bmatrix} } \in \cC(\cS)$ conditioned on $\cG(\lambda )$.  Proposition \ref{goodeventlemma} below validates that for all sufficiently large $\lambda$,  the event $\cG(\lambda)$ holds with high probability.
	\begin{proposition}\label{goodeventlemma}
		Under  assumption (A3), the event $\cG(\lambda)$ holds with probability at least $1-2q\exp(-9n\lambda^2/2)-2p\exp\{-n\lambda^2/(32\nu_0^2\sigma_x^2)\}$.
	\end{proposition}

	Due to high dimensionality, the empirical loss $\hat Q_h: \RR^q\times \RR^p \to \RR$ does not have a curvature along all directions.  In fact, there exists a subspace with dimension at least $p-n$ of directions in
	which it is completely flat.  Instead, it can be shown that the cone-like subset $\cC = \cC( \cS) $ is well-aligned with the curved directions of the Hessian of $\hat Q_h$ in a local region with high probability, which we refer to as the local RSC property. For $r,l>0$, define the (rescaled) $\ell_2$-ball and $\ell_1$-cone as
	\#
	\BB_{\Omega}(r):=\bigg\{  {\scriptsize		 \begin{bmatrix}
			\bdelta \\
			\bDelta
	\end{bmatrix} } \in \RR^{q+p} :\Vert(\bdelta\T,\bDelta\T)\T\Vert_{\Omega}\leq r  \bigg\}\label{l2ball}\#
	\#
	\text{and}\quad\CC_{\Omega}(l):=\bigg\{  {\scriptsize		 \begin{bmatrix}
			\bdelta \\
			\bDelta
	\end{bmatrix} } \in \RR^{q+p} :\Vert(\bdelta\T,\bDelta\T)\T\Vert_1\leq l\Vert(\bdelta\T,\bDelta\T)\T\Vert_{\Omega}  \bigg\},\label{l1cone}
	\# 
	where $$
	\Omega=  \begin{bmatrix}
		{\rm I}_q & \textbf{0}_{q \times p} \\
		\textbf{0}_{p\times q} & \Sigma 
	\end{bmatrix}  \in \RR^{(q+p) \times (q+p) } .
	$$ 
	For any curvature parameter $c>0$, radius parameter $r>0$, and cone parameter $l>0$,  define the event 
	\#
	\cR(c,r,l):=\Big\{D(\balpha,\bbeta)\geq c\big(\Vert\bDelta\Vert_\Sigma^2+q^{-1}\Vert\bdelta\Vert_2^2\big)\text{ for all }  {\scriptsize		 \begin{bmatrix}
			\bdelta \\
			\bDelta
	\end{bmatrix} }\in\BB_{\Omega}(r)\cap\CC_{\Omega}(l)   \Big\},
	\#
	where $ \bdelta = \balpha-\balpha_h^* \in \RR^q$, $\bDelta = \bbeta-\bbeta^*\in \RR^p$.
	The following proposition shows that under certain conditions on $(s, p, n)$ and $h$,   there exists some curvature parameter $c>0$ such that event $\cR(c,r,l)$ occurs with high probability.

	\begin{proposition}   \label{RSCproperty}
		In addition to assumptions (A1)--(A3), assume $\max_{k=1,\ldots,q}f(F^{-1}(\tau_k))\leq\overline{f}$ for some constant $\overline{f}>0$.
		Then, the event $\cR(c ,r,l)$ with $c = 0.5  \underline{f} \cdot \underline{\kappa}$  holds with probability at least $1-q/(2p)$   as long as 
		\begin{align}
			12 \nu_0^2 r \leq h\leq\underline{f}/\{\max(4\kappa_2^{1/2}l_0 ,2l_0)\}\quad \text{and}\quad n\geq C(\nu_0\sigma_xl/\underline{f}r)^2\overline{f}h\log(2p) ,
		\end{align}
		where  $C>0$ is a constant independent of $(n,s,p,h)$.
	\end{proposition}

	Based on the above preparations, we are now ready to present the first main result  of this subsection regarding the estimation error of the SCQR-Lasso estimator defined in \eqref{l1SCQR}.

	\begin{theorem} \label{Theorem3.1}
		Assume $\max_{k=1,\ldots,q}f(F^{-1}(\tau_k))\leq\overline{f}$ for some constant $\overline{f}>0$. Under   assumptions (A1)-(A3), the SCQR-Lasso estimator $(\hat{\balpha}_h,\hat{\bbeta}_h)$ with $\lambda\asymp\nu_0\sigma_x\sqrt{\log(2p)/n}$ satisfies the following error bound
		\#
		\begin{Vmatrix}
			\frac{\hat{\balpha}_h-\balpha_h^*}{\sqrt{q}}\\\hat{\bbeta}_h-\bbeta^*
		\end{Vmatrix}_{\Omega}\leq C\underline{f}^{-1}s^{1/2}\lambda\quad 
		\#
		with probability at least $1-q/p$, provided that the bandwidth $h$ satisfies
		$$
		\max\Bigg\{\frac{\sigma_x}{\underline{f}}\sqrt{\frac{sq\log(2p)}{n}}, \frac{\sigma_x^2\overline{f}}{\underline{f}^2}\frac{\max(s,q)\cdot\log(2p)}{n}\Bigg\}\lesssim h\lesssim{\underline{f}}/{l_0},
		$$
		where $C>0$ is a constant independent of $(n,s,p,h)$.  
	\end{theorem}
	The above theorem shows that the $L_1$-penalized SCQR estimator achieves the same rate of convergence as the $L_1$-penalized quantile regression estimator \citep{Belloni2011}, with a proper choice of the bandwidth parameter $h$, which is yet flexible. Moreover, Theorem \ref{Theorem3.1} indicates that, the regularization parameter $\lambda\asymp\nu_0\sigma_x\sqrt{\log(2p)/n}$ is independent of the error distibution, which alleviates the difficulty of tuning parameter selection of the LS estimator that typically depends on the standard deviation of the error distribution, and of the high-dimensional CQR estimator by \cite{SparseCQR} that is dependent on the minimum density of the error distribution at each quantile level.
	
	As a corollary, we derive a prediction error bound for $\hat{\bbeta}_h$, which is a direct consequence of Theorem \ref{Theorem3.1}.
	\begin{corollary}\label{predictionerrorbound}
		Under the conditions of Theorem \ref{Theorem3.1},  it holds
		\begin{align}
			\frac{1}{\sqrt{n}}\lVert\bX(\hat{\bbeta}_h-\bbeta^*)\rVert_2\lesssim \underline{f}^{-1}\nu_0\sigma_xs^{1/2}\bigg(\frac{\log p}{n}\bigg)^{1/2}
		\end{align}
		with probability at least $1-(q+2)/p$.
	\end{corollary}
	
	Next, we investigate the statistical properties of the iteratively reweighted $L_1$-penalized  estimators $(\hat{\balpha}_h^{t},\hat{\bbeta}_h^t)$ when $t \geq 2$.  Define the error vectors 
	$$ 
	\btheta^t = 	 \begin{bmatrix}
		\frac{\hat{\balpha}_h^t -\balpha_h^*}{\sqrt{q}}  \\
		\hat{\bbeta}_h^t -\bbeta^*
	\end{bmatrix} \in \RR^{q+p}   ,  \ \  t = 1, 2, \ldots .
	$$ 
	The following result characterizes the dependence of the estimation error $\|	\btheta^t \|_2$ at $t$-th step on $\|	\btheta^{t-1} \|_2$ from the previous step.  It reveals how iteratively reweighted $L_1$-penalization refines the statistical rate when the signals are sufficiently strong.
	We first derive a deterministic bound of the estimation error, conditioned on  some ``good" events.

	\begin{theorem} \label{iterativeboundtheorem}
		Suppose assumptions (A1)--(A4) hold, and let $a_0, c>0$ be such that
		\#
		P'(a_0)>0 ~\mbox{ and }~\sqrt{1+\{P'(a_0)\}^2/2}<a_0c\gamma_p .
		\#
		Furthermore, let $b>0$ satisfy
		\#
		\sqrt{\frac{b^2 + 1}{2}}P'(a_0) +2=a_0c\gamma_p \cdot b,
		\#
		and define $r_{\text{opt}}= a_0 b (\gamma_p s)^{1/2}\lambda$. 
		Then,  conditioned on $\cG(P'(a_0)\lambda)\cap\cR(c,r,l)$ with $r\geq q^{1/2}r_{\text{opt}}$ and $l=4\gamma_p^{-1/2}\sqrt{2\cdot\max(s,q)}$, the sequence of solutions $\{(\hat{\balpha}_h^{t},\hat{\bbeta}_h^t)\}_{t\geq1}$ to programs \eqref{weighted.l1.cqr} satisfies
		\#
		\Vert\btheta^{t}\Vert_{\Omega}\leq  \delta\cdot\Vert\btheta^{t-1}\Vert_{\Omega}+\underbrace{c^{-1}\Big\{\gamma_p^{-1/2}\Vert P_\lambda'((\vert\bbeta_\cS^*\vert-a_0\lambda)_{+})\Vert_2+\Vert\bomega^*_\cS\Vert_2\Big\}}_{=:  \, r_{\text{ora}}} \, +\, c^{-1}\sqrt{q}\Vert\bzeta^*\Vert_2\label{iterativebound},
		\#
		where $\delta :=\sqrt{1+\{P'(a_0)\}^2/2}/(a_0c\gamma_p)\in(0,1)$.
		In addition, it holds for for any $t\geq2$ that
		\#
		\Vert\btheta^{t}\Vert_\Omega\leq\delta^{t-1}r_{\text{opt}}+(1-\delta)^{-1}(r_{\text{ora}}+c^{-1}\sqrt{q}\Vert\bzeta^*\Vert_2).\label{iterativebound2}
		\#
	\end{theorem}

	The above theorem shows that under proper conditions on the curvature parameter $c$ and penalty function,  the estimation error, at least its leading term, can be refined iteratively via reweighted $L_1$-penalization. From \eqref{iterativebound} we see that there are three terms on the right-hand side that cannot be improved, which are
	$$
	\Vert\bomega^*_\cS\Vert_2,\quad \Vert P_\lambda'((\vert\bbeta_\cS^*\vert-a_0\lambda)_{+})\Vert_2 ~~\mbox{ and }~~\sqrt{q}\Vert\bzeta^*\Vert_2.
	$$
	The first term, $\Vert\bomega^*_\cS\Vert_2$,  determines the oracle convergence rate because it corresponds to the estimation error of the oracle SCQR estimator when only the significant covariates (indexed by $\cS$) are used in the fitting. The oracle SCQR estimator is formally defined as 
	\begin{align}
		(\hat{\balpha}^{o},\hat{\bbeta}^{o})=\argmin_{\substack{(\balpha,\bbeta)\in\RR^q\times\RR^p\\\bbeta_{\cS^{{\rm c}}}=\mathbf{0}}}\hat{Q}_h(\balpha,\bbeta),\label{oracleestimator}
	\end{align} where subscript $h$ is ommited for the brevity of notation.
	The second term $\Vert P_\lambda'((\vert\bbeta_\cS^*\vert-a_0\lambda)_{+})\Vert_2$ is the shrinkage bias induced by the penalty function.   For the $L_1$-penalty $P_\lambda(t) = \lambda t$ $(t\geq 0)$,  it is easy to see that this term is of order $s^{1/2} \lambda$ regardless of how large the non-zero coordinates  of $\bbeta^*$ are (in magnitude). For a concave penalty that has a decreasing $P'_\lambda$,  there is a chance that this shrinkage bias might be reduced when the signals are sufficiently strong. A concave penalty function $P_\lambda$ satisfying (A4) is called \textit{folded concave} if it further  satisfies the following property.
	\begin{itemize}
		\item [(A5)]  $a_*:=\inf\{a>0:P'(a)=0\}$ is finite.  
	\end{itemize}
	
	Under assumption (A5) and the minimum signal strength condition (also known as the beta-min condition)  that $\Vert\bbeta_\cS^*\Vert_{\min}\geq(a_0+a_*)\lambda$,  the shrinkage bias becomes zero.  The third term,  $\sqrt{q}\Vert\bzeta^*\Vert_2$,  depends on the partial gradient of the empirical loss with respect to $\balpha$. Therefore,  its order is independent of $p$ and only scales with $q$.

	Recall that Theorem \ref{iterativeboundtheorem} is a deterministic result conditioned on some ``good" event related to the local RSC structure and the magnitude of the gradient of the empirical loss.
	Combining 	Theorem \ref{iterativeboundtheorem} with Propositions \ref{goodeventlemma} and \ref{RSCproperty}, we further provide a complete result characterizing the oracle convergence rate of the iteratively reweighted $L_1$-penalized SCQR estimator under a weaker beta-min condition than needed in \cite{SparseCQR}.

	\begin{theorem}
		Suppose assumptions  (A1)--(A5) hold, and that there exist $a_1 \geq a_* >a_0>0$ such that 
		\#
		P'(a_0)>0 ~~\mbox{ and }~~	\sqrt{4+2\{P'(a_0)\}^2}<a_0 \gamma_p   \underline{\kappa} \underline{f} .
		\#			
		Moreover, let the regularization parameter $\lambda$ and   bandwidth   $h$ satisfy $\lambda\asymp\nu_0\sigma_x\sqrt{\log(2p)/n}$ and
		$$
		\max\bigg\{  C_1\sqrt{\frac{sq\log(2p)}{n}},C_2\frac{\max(s,q)\cdot\log(2p)}{n}\bigg\} \lesssim h\lesssim\underline{f}/l_0,
		$$
		where $C_1=\sigma_x\nu_0^3$ and $C_2=\sigma_x^2\nu_0^6\overline{f}\underline{f}^{-2}$.
		Then, for any $z>0$, under the beta-min condition $\Vert\bbeta_\cS^*\Vert_{\min}\geq(a_0+a_1)\lambda,$ the iteratively reweighted $L_1$-penalized SCQR estimator $(\hat{\balpha}_h^{t},\hat{\bbeta}_h^{t})$ with $t\gtrsim\log \log(2p) /\log(1/\delta)$ satisfies the bounds
		\begin{gather}
			\Vert\hat{\bbeta}_h^{t}-\bbeta^*\Vert_2\lesssim \underline{f}^{-1}\sqrt{\frac{s+\log q +z}{n}},\text{ and } \Vert\hat{\balpha}_h^{t}-\balpha^*\Vert_2\lesssim \underline{f}^{-1}q^{1/2}\bigg(\sqrt{\frac{s+\log q+z}{n}}+h^2\bigg)
		\end{gather}
		with probability at least $1-q/p-2e^{-(s+z)}$ as long as $n\gtrsim s\log p+\log q+t$, where $\delta=\sqrt{4+2\{P'(a_0)\}^2}/(a_0 \gamma_p   \underline{\kappa} \underline{f}  )\in(0,1)$.
		\label{weakoracletheorem}
	\end{theorem}
	
	The above result shows that under a beta-min condition $\| \bbeta^*_{\cS} \|_{\min} \gtrsim \sqrt{\log(p) /n}$,  the proposed estimator (of $\bbeta^*$) achieves a near-oracle rate  $\sqrt{(s + \log q)/n}$ after as many as  $\log(\log p)$ steps, where $q\geq 1$ is a predetermined number of quantile levels.  This complements the strong oracle property established in  \cite{SparseCQR}, which requires the minimum signal strength to be of order $\sqrt{s \log(p)/n}$.
	
	\subsection{Strong oracle property}
	To establish the strong oracle property of our proposed multi-step estimator $(\hat{\balpha}_h^t,\hat{\bbeta}_h^t)$, we need to show that the estimator equals to the oracle estimator defined in \eqref{oracleestimator} for sufficiently large $t$. We define a similar event that resembles the local RSC property. Let 
	\begin{gather}
		D_{rsc}(\balpha_1,\bbeta_1,\balpha_2,\bbeta_2):=\frac{\langle\nabla\hat{Q}_h(\balpha_1,\bbeta_1)-\nabla\hat{Q}_h(\balpha_2,\bbeta_2),(\balpha_1\T-\balpha_2\T,\bbeta_1\T-\bbeta_2\T)\T\rangle}{\Vert\bbeta_1-\bbeta_2\Vert_\Sigma^2+q^{-1}\Vert\balpha_1-\balpha_2\Vert_2^2}.
	\end{gather}
	Given radius parameters $r,l>0$ and a curvature parameter $c>0$, define
	\begin{gather}\label{grsc}
		\cR_{rsc}(r,l,c):=\Big\{D_{rsc}(\balpha_1,\bbeta_1,\balpha_2,\bbeta_2) \geq c\text{ for all } (\balpha_1,\bbeta_1,\balpha_2,\bbeta_2)\in\Lambda(r,l)\Big\},
	\end{gather} where \begin{gather}\Lambda(r,l):=\cap_{k=1}^q\Lambda_{k}(r,l)\cap\{(\balpha_1,\bbeta_2,\balpha_2,\bbeta_2): \begin{bmatrix}\balpha_2-\balpha_h^*\\\bbeta_2-\bbeta^*\end{bmatrix}\in\BB_{\Omega}(r/2), \text{supp}(\bbeta_2)\in\cS\}\nonumber,\\
		\Lambda_{k}(r,l):=\{(\balpha_1,\bbeta_1,\balpha_2,\bbeta_2):\begin{bmatrix}\alpha_{1k}-\alpha_{2k}\\\bbeta_1-\bbeta_2\end{bmatrix}\in \BB_{\bar{\Sigma}}(r)\cap\CC_{\bar{\Sigma}}(l)\}.\end{gather}
	\begin{theorem}\label{deterministicstrongoracle}
		Suppose assumptions (A1)-(A5) hold, and for some predetermined $\delta\in(0,1)$ and $c>0$, there exist constants $a_1>a_0>0$ such that \begin{gather}
			P'(a_1)=0, P'(a_0)>0,\text{ and }a_0\delta c\gamma_p>\sqrt{1+P'(a_0)^2/2}. \label{a0a1}
		\end{gather}
		Moreover, let $r\geq q^{1/2}\gamma_p^{1/2}a_0bs^{1/2}\lambda$ and $l=\sqrt{2}\{2+2/P'(a_0)\}\cdot\big[\max\{q,(1+b^2)s\}/\gamma_p\big]^{1/2}$, where $b>0$ is a constant that satisfies 
		\begin{gather}
			\sqrt{\frac{1+b^2}{2}}P'(a_0)+1=a_0c\gamma_pb.
		\end{gather}
		Assume $\Vert\bbeta_{S}^*\Vert_{\text{min}}\geq(a_0+a_1)\lambda$. Then, conditioned on the event \begin{align}\label{strongoracleevent}\{\Vert\nabla_{\bbeta}\hat{Q}_h(\hat{\balpha}^o,\hat{\bbeta}^o)\Vert_\infty\leq\lambda/2\}&\cap\{\Vert\btheta^o\Vert_\Omega\leq r/2\}\cap\cR_{rsc}(r,l,c)\nonumber\\&\cap\Bigg\{\Vert\hat{\bbeta}^o-\bbeta^*\Vert_\infty\leq\bigg[a_0-\frac{\sqrt{1+\{P'(a_0)/2\}^2}}{\delta c\gamma_p}\bigg]\lambda\Bigg\}\end{align}
		where \begin{gather*}\btheta^o=\begin{bmatrix}
				\frac{\hat{\balpha}^o-\balpha_{h}^*}{\sqrt{q}}\\\hat{\bbeta}^o-\bbeta^*
			\end{bmatrix},
		\end{gather*}the strong oracle property holds: $\hat{\bbeta}^\ell=\hat{\bbeta}^o$ provided $\ell\geq\log(s^{1/2}/\delta)/\log(1/\delta)$.
	\end{theorem}
	Similar to the Theorem \ref{iterativeboundtheorem}, Theorem \ref{deterministicstrongoracle} is a deterministic result that depends on the event described in \eqref{strongoracleevent}. Thus, our next goal is to control the probability of the event \eqref{strongoracleevent}. To control such probability, we require deviation bound and a non-asymptotic Kiefer-Bahadur representation of the oracle estimator that are of independent interest. 
	
	\begin{itemize}
		
		\item[(A1')] In addition to Condition (A1), assume $\sup_{u\in\RR}\vert f_\varepsilon(u)\vert\leq \overline{f}$ for some constant $\overline{f}>0.$

		\item[(A2')]  In addition to Condition (A2), assume $\sup_{u\in\RR}K(u)\leq\kappa_u$ for some $\kappa_u\in(0,1]$.
	\end{itemize}
	Since the oracle estimator is essentially an unpenalized smoothed CQR estimator in the low-dimensional regime where $s<<n$, we need to derive relevant results for the low-dimensional smoothed CQR estimator. The following proposition summarizes those results about the oracle estimator that is essential to deriving necessary conditions for the strong oracle property. Same result has been derived in low dimensional smoothed CQR paper by \cite{YWZ2023}, we refer their paper for detailed proof of the following result.
	\begin{proposition}\label{oracleestimationbahadur}
		Assume Conditions (A1'), (A2'), and (A3) hold. Then, for any $t\geq0$, the oracle estimator $(\hat{\balpha}^o,\hat{\bbeta}^o)$ defined in \eqref{oracleestimator} satisfies
		\begin{gather}
			\begin{Vmatrix}
				\frac{\hat{\balpha}^o-\balpha_h^*}{\sqrt{q}}\\\hat{\bbeta}^o-\bbeta^*
			\end{Vmatrix}_{\Omega}\lesssim\underline{f}^{-1}\sqrt{\frac{s+t}{n}}
		\end{gather} with probability at least $1-2qe^{-t}$. 
		
		Moreover,  let $\bS=\EE(\bx_\cS\bx_\cS\T)$, and  $\bD:=q^{-1}\sum_{k=1}^qf_\varepsilon(F^{-1}(\tau_k))\bS$. Then, 
		\begin{gather}
			\Bigg\Vert\bD(\hat{\bbeta}^o-\bbeta^*)+\frac{1}{nq}\sn\sum_{k=1}^q\{\bar{K}((\alpha_k^*-\varepsilon_i)/h)-\tau_k\}\bx_{i,\cS}\Bigg\Vert_2\lesssim \frac{(s+t)}{h^{1/2}n}+h^{3/2}\sqrt{\frac{q(s+t)}{n}}+h^4
		\end{gather} with probability at least $1-3qe^{-t}$.
		
	\end{proposition} Before presenting our final theorem that characterizes the strong oracle property, there is one more event that we need to make sure that it holds with high probability, which is $\cR_{rsc}(r,l,c)$. The following Proposition characterizes the event and its probability bound.
	\begin{proposition}\label{oraclerscproposition}
		Let $r,l,$ and $h$ satisfy
		\begin{gather}
			24\nu_0^2r\leq h\leq \underline{f}/\{\max(4\kappa_2^{1/2}l_0,2l_0)\}\quad\text{and}\quad nh\geq C\overline{f}\underline{f}^{-2}\max\{s,l^2\log(p)\},
		\end{gather} for some sufficiently large constant $C$ independent of $(n,s,p,h)$. Then, the event $\cR_{rsc}(r,l,c)$ holds with probability at least $1-q/(2p)$ with $c=0.5{\underline{\kappa}\cdot\underline{f}}$.
	\end{proposition}
	With the above preparations, we finally establish the strong oracle property of our iterative estimator. 
	\begin{theorem}\label{Strongoracle}
		Assume (A1'), (A2'), and (A3)-(A5) hold. Assume also that \begin{gather}
			\max_{j\in\cS^c}\Vert\bJ_{j\cS}(\bJ_{\cS\cS})^{-1}\Vert_1\leq A_0
		\end{gather} for some $A_0\geq1$ and $\mu_4:=\sup_{\bu\in\mathbb{S}^p}\EE\vert\bar{\Sigma}^{-1/2}\bar{\bx}\T\bu\vert^4<\infty$ where $\bJ:=q^{-1}\sum_{k=1}^qf_\varepsilon(F^{-1}(\tau_k))\Sigma$.  For a predetermined $\delta\in(0,1)$, suppose there exist $a_1>a_0$ satisfying \eqref{a0a1} with $c=0.5\underline{\kappa}\underline{f}$, and the beta-min condition $\Vert\bbeta_\cS^*\Vert_{\min}\geq(a_0+a_1)\lambda$.  Let the smoothing bandwidth $h\asymp\{\log(p)/n\}^{1/4}$ and $\lambda\asymp\sqrt{\log(p)/n}$. Then, $\hat{\bbeta}^t=\hat{\bbeta}^o$ for all $t\geq \lceil\log(s^{1/2}\delta)/\log(1/\delta)\rceil$ with probability at least $1-2q/p-(5q+1)/n$, provided that the sample size satisfies $n\gtrsim\max\{s^{8/3}/(\log p)^{5/3},s^{4/3}\log(p)\}$.
	\end{theorem} 
	\section{Algorithms}  
	\label{sec:algo}
	
	In this section, we discuss the computational methods for penalized composite quantile regression, with a particular focus on the weighted $L_1$-penalty. We first revisit the ADMM-based algorithm proposed in \cite{SparseCQR}, and then describe a local adaptive
	majorize-minimization (LAMM) for convolution-smoothed CQR.  Complexities of the two algorithms are also discussed.

	\subsection{An alternating direction method of multipliers algorithm}
	\label{sec:4.1}
	
	The computation of either $L_1$-penalized or folded concave penalized composite quantile regression boils down to solving a weighted $L_1$-penalized problem
	\#
	\min_{\balpha,\bbeta} \bigg\{ \frac{1}{n q}\sn\sum_{k=1}^{q}\rho_{\tau_k}(y_i-\alpha_k-\bx_i^{{\rm T}}\bbeta)+ \sum_{j=1}^{p} \lambda_j\vert\beta_j\vert \bigg\} , \label{weightedlassopenalizedCQR}
	\#
	where $\lambda_j \geq 0$ for $j=1,\ldots, d$. A conventional strategy is to formulate \eqref{weightedlassopenalizedCQR} as a linear program, solvable by general-purpose optimization toolboxes. These toolboxes are convenient  to use yet are only adapted to small-scale problems.  To solve \eqref{weightedlassopenalizedCQR} more efficiently under high-dimensional settings, \cite{SparseCQR} proposed an algorithm based on the alternating direction method of multipliers (ADMM).   The idea is to cast \eqref{weightedlassopenalizedCQR} as an equivalent program solvable by ADMM. Specifically, they consider the following reformulation
	\$
	{\rm minimize}~~ &   \frac{1}{n q}\sn\sum_{k=1}^{q}\rho_{\tau_k}(z_{ik})+ \sum_{j=1}^{p} \lambda_j \vert\gamma_j\vert  \\
	\text{subject to}~~  &\bZ=\mathbf{1}_q\T \otimes\by-\mathbf{1}_n\otimes\balpha\T-\mathbf{1}_q\otimes(\bX\bbeta), \bgamma=\bbeta,
	\$
	where  $\bZ=(z_{ik})_{n\times q}\in \RR^{n\times q}$ with $z_{ik}=y_i-\alpha_k-\bx_i\T\bbeta$,   $\bX=(\bx_1,\bx_2,\hdots,\bx_n)\T \in \RR^{n\times p}$ and $\bgamma=(\gamma_1,\hdots,\gamma_p)\T$.
	Here $\otimes$ denotes the Kronecker product.
	Let $\bvarphi = (\balpha\T, \bbeta\T)\T \in \RR^{p+q}$ be the total vector of parameters (of interest).
	The augmented Lagrangian of the above problem is 
	\begin{align}
		L_\sigma(\bvarphi,\bZ,\bgamma,\bU,\bv):=&\frac{1}{n q}\sn\sum_{k=1}^{q}\rho_{\tau_k}(z_{ik})+  \sum_{j=1}^{p} \lambda_j \vert\gamma_j\vert+\langle \vecc(\bU),\vecc(\bZ)+\XX_1\bvarphi\rangle\nonumber\\&+\langle \bv,\bgamma-\XX_2\bvarphi\rangle+\frac{\sigma}{2}\Vert\vecc(\bZ)+\XX_1\bvarphi-\bY\Vert_{{\rm F}}^2+\frac{\sigma}{2}\Vert\bgamma-\XX_2\bvarphi\Vert_2^2,
	\end{align}
	where  $ \bU=(u_{ik}) \in \RR^{n\times q}$ and $\bv=(v_1,\hdots,v_p)\T$ are the
	Lagrangian multipliers,  $\bY=\mathbf{1}_q\otimes\by $ is an $(nq)$-dimensional vector that stacks $q$ copies of $\by \in \RR^n$ one underneath the other,  $\sigma>0$ is a optimization parameter and
	$$
	\XX_1=\begin{pmatrix}\mathbf{1}_n&&\hdots&&\mathbf{0}&&\bX\\\vdots&&\ddots&&\vdots&&\vdots\\\mathbf{0}&&\hdots&&\mathbf{1}_n&&\bX\end{pmatrix} \in \RR^{(nq) \times (p+q)} , \quad  \XX_2=(\mathbf{O}_{p\times q}\quad\mathbf{I}_p) \in \RR^{p \times (p+q)}  .
	$$
	Moreover, define
	$$	
	\text{Prox}_\tau(v,a):=v-\max\{(\tau-1)/a,\min(v,\tau/a)\} ~~\mbox{ and }~~\text{Shrink}(v,a):=\sgn(v)(\vert v\vert-a)_+.
	$$
	The former is  the proximity operator of the check loss $\rho_\tau$ with respect to parameter $a>0$, and the latter is the proximity operator of $|\cdot |$,  also known as the soft-thresholding operator. The ADMM-based algorithm \citep{SparseCQR} for solving problem \eqref{weightedlassopenalizedCQR} is summarized in Algorithm~\ref{ADMMalgotirhm}.
	\begin{algorithm} 
		\caption{The ADMM Algorithm for Solving Weighted $L_1$-penalized CQR}
		\textbf{Input:} Initialize with $(\bvarphi^0,\bZ^0,\bgamma^0,\bU^0,\bv^0)$, where $\bvarphi^0= ( (\balpha^0)\T,(\bbeta^0)\T)\T$
		
		For $t=0,1,\hdots,$ repeat the following steps until convergence.
		\begin{itemize}
			\item [1. ] Update \#\bvarphi^{t+1}=\frac{1}{\sigma}  (\XX_1\T\XX_1+\XX_2\T\XX_2  )^{-1}\cdot\big[\XX_1\T  \{\sigma\bY-\sigma\vecc(\bZ^{t})-\vecc(\bU^{t})  \}+\XX_2\T  (\sigma\bgamma^{t}+\bv^{t}  )\big]\label{admmcomputation}\#
			\item[2. ] Update
			\begin{align}
				z_{ik}^{t+1}&=\text{Prox}_{\tau_k}\bigg( y_i-\alpha_k^{t+1}-\bx_i\T\bbeta^{t+1}-\frac{u_{ik}^{t}}{\sigma},n q\sigma\bigg), 1\leq i\leq n, 1\leq k \leq q,\\
				\gamma_j^{t+1}&=\text{Shrink}\bigg(\beta_j^{t+1}-\frac{v_j^t}{\sigma},\frac{\lambda d_j}{\sigma}\bigg), 1\leq j\leq p.
			\end{align}
			
			\item[3. ] Update
			\begin{align*}
				\vecc(\bU^{t+1})&=\vecc(\bU^t)+\sigma \{\vecc(\bZ^{t+1})+\XX_1\bvarphi^{t+1}-\bY  \} , \\
				\bv^{t+1}&=\bv^t+\sigma (\bgamma^{t+1}-\XX_2\bvarphi^{t+1} ) .
			\end{align*}
		\end{itemize}	\label{ADMMalgotirhm}
	\end{algorithm}

	\subsection{A local adaptive majorize-minimization algorithm for   smoothed CQR}

	In this section, we focus on solving a smoothed version of \eqref{weightedlassopenalizedCQR} with each $\rho_{\tau_k}$ replaced by $\ell_{h, k} = \rho_{\tau_k} \circ K_h$.   To take advantage of the smoothness and the local strong convexity of the smoothed loss, we employ a variant of the  local adaptive majorize-minimization algorithm (LAMM)  proposed by \cite{Fan2018}. The main idea of LAMM  is to construct an isotropic quadratic objective function that locally majorizes the smoothed composite quantile loss such that closed-form updates are available at each iteration. To see this,  recall the smoothed objective function $\hat{Q}_h(\balpha,\bbeta)=(n q)^{-1}\sn\sum_{k=1}^q \ell_{h,k}(y_i-\alpha_k-\bx_i\T\bbeta)$.   For $k=1, 2,\ldots$, let ${\bvarphi}^{k} =( (  {\balpha}^{k})\T,(  {\bbeta}^{k })\T )\T$ be the iterate after the $k$ iteration.  At the $k$-th iteration,  for some sufficiently large quadratic parameter $\phi_k>0$, we define a locally majorizing isotropic quadratic function 
	\begin{align}
		F(\bvarphi;\phi_k,\bvarphi^{k-1}):=\hat{Q}_h(  {\bvarphi}^{k-1})+
		\langle\nabla\hat{Q}_h( {\bvarphi}^{k-1}),\bvarphi-  {\bvarphi}^{k-1}\rangle+
		\frac{\phi_k}{2}\Vert\bvarphi- {\bvarphi}^{k-1}\Vert_2^2,\label{isotropicquadratic}
	\end{align}
	satisfying $F( {\bvarphi}^{k-1} ;\phi_k,\bvarphi^{k-1}) = \hat{Q}_h( {\bvarphi}^{k-1})$.  For a large enough $\phi_k$,  say no less than the largest eigenvalue of $\nabla^2 \hat Q_h(\bvarphi^{k-1})$,  we have $F(\bvarphi;\phi_k,\bvarphi^{k-1}) \geq  \hat{Q}_h(\bvarphi)$ for all $\bvarphi$.
	Then, we define the updated iterate $\bvarphi^{k}=( (  {\balpha}^{k})\T,(  {\bbeta}^{k })\T )\T$ as the solution to
	\#
	\min_{\bvarphi}   \{  F(\bvarphi;\phi_k,\bvarphi^{k-1}) + \| \blambda \circ \bbeta \|_1 \}, \label{lamm.minimization}
	\#
	where $\blambda = (\lambda_1 ,\ldots, \lambda_p)\T$.
	It is easy to see that 
	\begin{align}
		&\hat{Q}_h(  {\bvarphi}^k)+ \Vert \blambda \circ  {\bbeta}^k\Vert_1 \leq F(  {\bvarphi}^k;\phi_k,\bvarphi^{k-1})+ \Vert \blambda \circ {\bbeta}^k\Vert_1  \label{majorize.property} \\
		&\leq F( {\bvarphi}^{k-1};\phi_k,\bvarphi^{k-1})+  \Vert \blambda \circ  {\bbeta}^{k-1}\Vert_1 =\hat{Q}_h( {\bvarphi}^{k-1})+ \Vert \blambda \circ  {\bbeta}^{k-1}\Vert_1.  \nn 
	\end{align}
	This ensures that the objective function (with penalty) decreases after each iteration.  From the first-order optimality condition we obtain the following closed forms for $\balpha^k$ and $\bbeta^k$:
	\$
	{\alpha}_j^k &=  {\alpha}_j^{k-1} - \phi_k^{-1}\partial_{\alpha_j}\hat{Q}_h(  {\bvarphi}^{k-1}),  \quad j=1,2,\ldots, q, \\
	{\beta}_j^k &=\text{Shrink}( {\beta}_j^{k-1}-\phi_k^{-1}\partial_{\beta_j}\hat{Q}_h( {\bvarphi}^{k-1}),\phi_k^{-1}\lambda_j ), \quad j=1, \ldots, p .
	\$ 
	To choose a sufficiently large quadratic coefficient $\phi_k$ that ensures majorization property,  we start from a relatively small number, say $\phi_0=0.01$,  and successively inflate it by a factor $\gamma>1$,  denoted by $\phi_{k,l}=\gamma^{l } \phi_0$ for $l=1,2,\ldots$.  If the solution $\bvarphi^{k, l}$ to \eqref{lamm.minimization} with $\phi_k=\phi_{k, l}$ satisfies \eqref{majorize.property} for some $l\geq 0$,  we stop the search and set $\phi_k = \phi_{k, l}$.  Therefore, the quadratic coefficient $\phi_k$ is automatically determined at each step. By default, we set the optimization parameters to be $(\phi_0, \gamma) =(0.01, 1.25)$.  We summarize the whole procedure in Algorithm~\ref{LAMMalgotirhm}.

	\begin{algorithm}
		\caption{The LAMM Algorithm for Smoothed CQR with Weighted $L_1$-penalization}
		\textbf{Input:} Initialize with  $\balpha^0= \textbf{0}$ and $\bbeta^0 =\mathbf{0}$ 
		
		For $k=0,1,\hdots,$ repeat the following steps until convergence.
		%criterion $\Vert\bvarphi^{k}-\bvarphi^{k-1}\Vert_2^2/\Vert\bvarphi^{k-1}\Vert_2^2\leq\epsilon$ is met.
		\begin{itemize}
			\item[1. ] Set $\phi_k=\max\{\phi_0,\phi_{k-1}/\gamma\}$.
			
			%\item[2. ] \textbf{repeat} Steps 3--5
			
			\item [2. ] Update 
			\$ 
			{\alpha}_j^k & =  {\alpha}_j^{k-1}-\phi_k^{-1}\partial_{\alpha_j}\hat{Q}_h( {\bvarphi}^{k-1}), \quad j=1,2,\ldots, q,   \\
			{\beta}_j^k &=\text{Shrink}( {\beta}_j^{k-1}-\phi_k^{-1}\partial_{\beta_j}\hat{Q}_h( {\bvarphi}^{k-1}),\phi_k^{-1}\lambda_j), \quad j=1.\ldots, p.
			\$
			
			\item[3 ] If $F( {\bvarphi}^k;\phi_k, {\bvarphi}^{k-1})<\hat{Q}_h( {\bvarphi}^k)$, set $\phi_k=\gamma\phi_k$ and repeat Step 2 until $F( {\bvarphi}^k;\phi_k,  {\bvarphi}^{k-1})\geq\hat{Q}_h( {\bvarphi}^k)$.
			
		\end{itemize}
		%\textbf{Ouput:} the updated parameter  $\bvarphi^k:=\{(\balpha^k)\T,(\bbeta^k)\T\}\T$.
		\label{LAMMalgotirhm}
	\end{algorithm}
	
	From Algorithms~\ref{ADMMalgotirhm} and \ref{LAMMalgotirhm} we see that  the dominant computational effort of each LAMM update is the multiplication of a $p\times nq$ matrix and $(nq)$-dimensional vectors, which can be implemented in $O(p n q)$ operations.  In addition to this,  each ADMM update also involves the multiplication of a $(p+q) \times (p+q)$ matrix and $(p+q)$-dimensional vectors with a complexity $O(p n q + (p+q)^2)$.  Moreover, the ADMM needs to compute and store the inverse of $\XX_1\T\XX_1+\XX_2\T\XX_2 \in \RR^{(p+q) \times (p+q)}$,  hence incurring extra computational cost and memory allocation.
	Via the  Sherman-Morrison-Woodbury formula, the real computational effort of this step is to evaluate the inverse of an $n\times n$ matrix (with complexity $O(n^3)$), which is still expensive when $n$ is large.

\begin{figure}[htp!]	
		\centering
		\subfigure[Model error]{\includegraphics[scale=0.47]{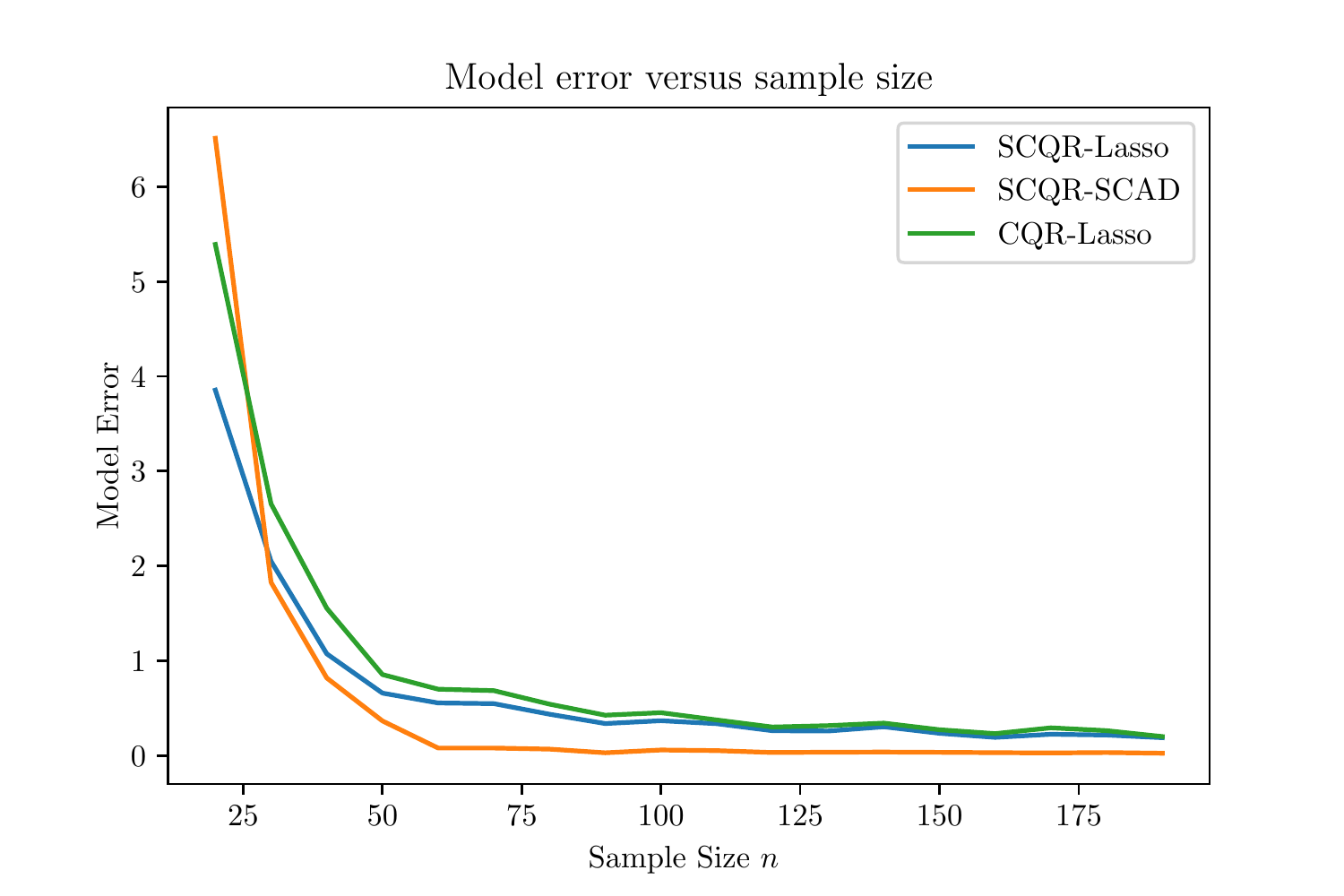}}  \qquad 
		\subfigure[Runtime]{\includegraphics[scale=0.47]{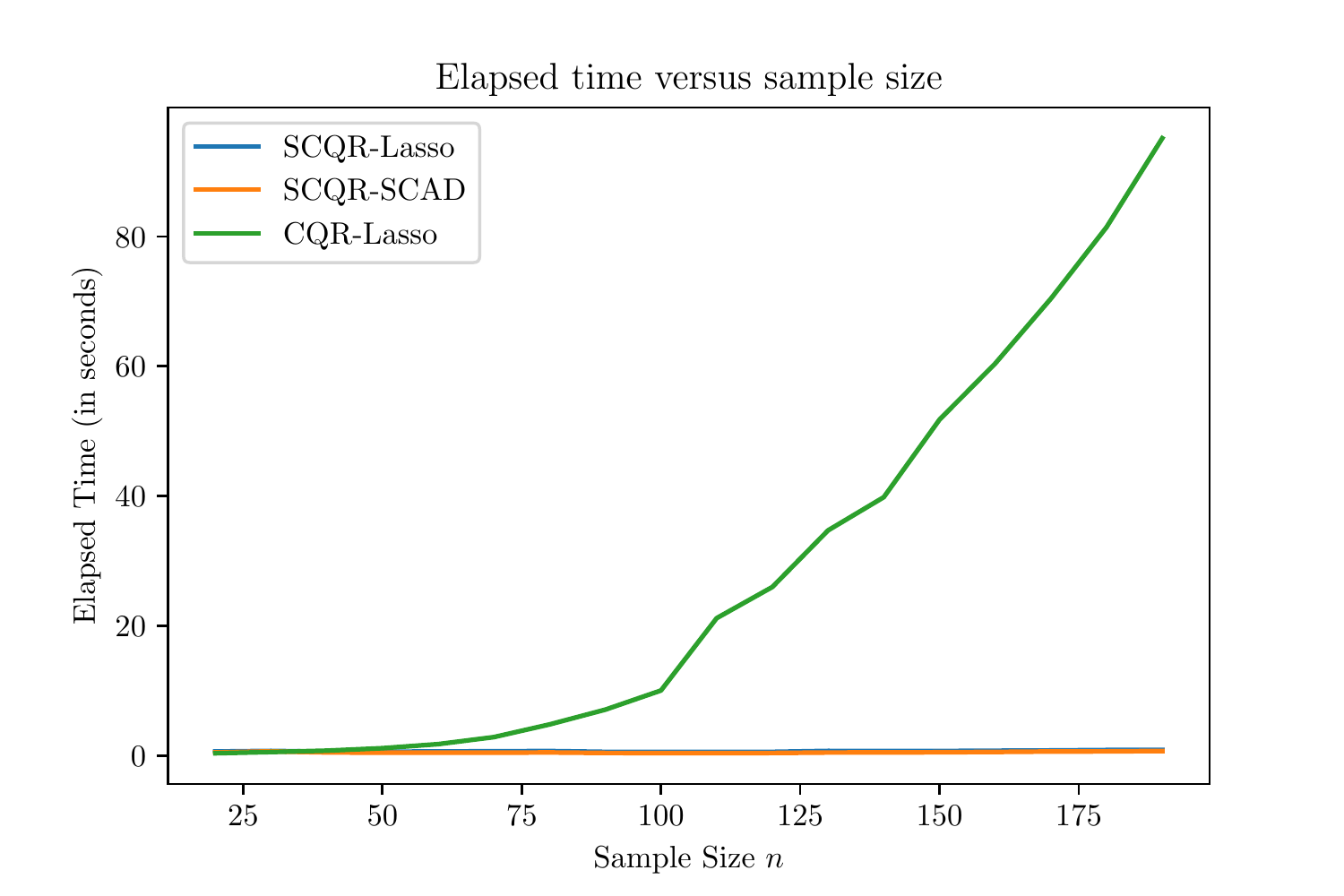}} 
		\caption{A numerical comparison between  CQR via ADMM and SCQR via LAMM.  Panels (a) and (b) display, respectively, the ``model error versus sample size'' curve and the ``runtime versus sample size" curve.  The sample size $n$ increases from {20} to {200}, and the dimension $p$ is set as  $5n$.}
		\label{comparison.fig}
	\end{figure}

	Figure \ref{comparison.fig} shows a preliminary comparison between the ADMM and LAMM algorithms for computing $L_1$-penalized CQR estimators on a simulated dataset with increasing $n, p$ subject to $p=5n$. To make comparisons that are as fair as possible,  each algorithm is implemented in \texttt{Python},  using the \texttt{NumPy} library  for basic linear algebra operations.
	On the statistical aspect, the CQR-Lasso (by ADMM) and SCQR-Lasso (by LAMM) estimators exhibit nearly identical estimation errors (under squared model error); in a speed comparison, the runtime of ADMM grows significantly faster than that of LAMM as the sample size and dimension increase. 
	These preliminary numerical results show evidence that LAMM can also be faster than ADMM by several orders of magnitude.
	%When $(n, p) = (200, 1000)$,  LAMM 
	%Figure \ref{comparison.fig} (b), when the sample size $n=200$ $(p=1000)$, LAMM runs more than a hundred times faster than ADMM.
	More empirical evidence will be given in the next section.

	\section{Numerical studies}
	\label{numericalstudy}
	Recall that composite quantile regression was introduced by \cite{ZY2008} as a robust regression method for linear models with heavy-tailed errors that may have infinite variance.  The relative efficiency of CQR compared to the least squares is at least 70\% regardless the error distribution,  could be arbitrarily close to 95.5\%   in the Gaussian model and arbitrarily large with very heavy-tailed errors. The least absolute deviation (LAD) regression, however, may have an arbitrarily small relative efficiency with respect to the least squares.  Recently, \cite{Wang2020} introduced a new robust method for high-dimensional regression along with a simulation-based  procedure for choosing the regularization parameter.  In the Gaussian model,  their oracle estimator achieves the same asymptotic relative efficiency (with respect to the least squares)  as the CQR.
	
	In the following simulation study, we first compare the penalized SCQR method with its non-smoothed counterpart \citep{SparseCQR},  and then with the robust regression method proposed  by \cite{Wang2020} when the tuning parameters are automatically chosen for both methods. Data are generated independently from the linear model
	\#
	y=\bx\T\bbeta^*+\varepsilon,  ~~~~ \bx \sim N_p( 0 ,\Sigma) , \label{simulationmodel}
	\#
	where $\bbeta^*=(3, 1.5, 0, 0, 2,  0, \ldots, 0)\T \in \RR^p$ and $\Sigma=(0.5^{\vert j-k\vert})_{1\leq j, k\leq p }$. %which is the same model used in the numerical simulation of \cite{SparseCQR}. The covariates of the simulation model is from the multivariate normal distribution, where $\bx$ follows $N_p( 0 ,\Sigma)$ with $\bSigma=(0.5^{\vert i-j\vert})_{ij}$. 
	Independent of $\bx$, the observation noise $\varepsilon$ is generated from one of following four distributions:
	\begin{itemize}
		\item[(a)] The normal distribution with mean 0 and variance 3---$ N(0,3)$.
		\item[(b)] The mixture normal (MN) distribution---$ \sqrt{6}\times\{0.5N(0,1)+0.5N(0,0.5^6)\}$.
		\item[(c)] The $t$-distribution with 3 degrees of freedom---$ t_3$.
		\item[(d)] The standard Cauchy distribution with the density function $ f(t)=1/\{\pi(1+ t^2)\}$.
	\end{itemize}
	We consider two moderate-scale settings with $(n,p)=(100,600)$ and $(n,p)=(200,1200)$.

	The statistical performance of each method is measured via the (average) squared model error (with the standard error in the parenthesis),  which is $\| \hat{\bbeta}-\bbeta^* \|_\Sigma^2$,   the number of false positive results (FP), which is the number of spurious covariates that are selected, and the number of true positive results (TP), which is the number of significant covariates that are selected.   
	For the implementation of CQR,  we set $q=19$ and choose quantile levels $\tau_k = k/20$ for $k=1,\ldots, 19$.

	Table~\ref{table1} summarizes the simulation results for CQR-Lasso, SCQR-Lasso and SCQR-SCAD that uses the SCAD penalty to compute the weights in \eqref{SPCQR}.
	For a fair comparison between the two methods in terms of  statistical and numerical efficiency,  we first compute an ``oracle'' $\lambda$ value based on the true model error $\| \hat \bbeta - \bbeta^* \|_\Sigma$ for each estimator.
	To be specific,   we first compute each estimator along a predetermined sequence of $\lambda$  values, and choose the $\lambda$ that minimizes the true model error averaged over 50 replications. 
	Next,  we run 100 additional simulations for each method using the optimally chosen $\lambda$, and report the results  in Table~\ref{table1}.  When the $L_1$ penalty is used,    the SCQR has slightly lower model errors
	than the CQR yet at the cost of more false positives. From the runtime comparison we see a significant computational advantage of the SCQR via LAMM over the CQR via ADMM.
	As mentioned in Section~\ref{sec:3.2},  both algorithms are implemented in \texttt{Python} using the \texttt{NumPy} library  for basic linear algebra operations.
	Moreover, with the optimally chosen $\lambda$, the SCQR-SCAD estimator  considerably outperforms the Lasso counterparts and achieves oracle-like performance.

	We further implement both methods with $\lambda$ chosen by two data-driven procedures, the  (five-fold) cross-validation and a modified BIC method; see Section~\ref{sec:2.3} for details.
	Under the four error distributions,  Tables~\ref{table2}  and  \ref{bictable} show the simulation results for CQR-Lasso, SCQR-Lasso and SCQR-SCAD with $\lambda$ chosen by five-fold cross-validation and BIC,  respectively. Statistically, the $L_1$-penalized CQR and SCQR methods perform  similarly  in terms of model
	selection accuracy and estimation accuracy (for $\bbeta^*$). This empirically validates  the theoretical results that smoothing only affects the intercept terms and thus does not compromise the estimation of $\bbeta^*$. The runtime comparison, on the other hand, shows that the computational cost of ADMM, combined with either cross-validation or BIC, becomes prohibitive  as soon as the data has
	moderately large scales.

	\begin{table}[!htp]
		\fontsize{7}{8}\selectfont
		\centering
		\caption{Statistical performance comparison between the CQR (via ADMM) and the SCQR (via LAMM) estimators under linear model \eqref{simulationmodel} with four error distributions.  Optimally chosen $\lambda$ values are used for both methods.
		}
		\label{table1}
		\begin{tabular}{ c | c |c c c c |c c c c}
			\hline
			& & \multicolumn{4}{c|}{$n = 100, p = 600$} & \multicolumn{4}{c}{$n = 200, p = 1200$} \\
			Error & Method & ME  & TP & FP & Runtime&ME   & TP & FP &Runtime\\
			\hline
			$N(0,3)$&CQR (Lasso)&0.7680 (0.41)&3&11.62&21.05&0.3581 (0.14)&3&14.66&166.01 \\
			&SCQR (Lasso)&0.6617 (0.26)&3&18.37&0.88&0.3317 (0.15)&3&24.31&1.21\\
			&SCQR (SCAD)&0.1096 (0.99)&3&0.06&0.46&0.0474 (0.04)&3&0.04&0.81\\
			
			\cline{1-10}
			MN&CQR (Lasso)&0.3871 (0.18)&3&10.15&18.95&0.1940 (0.08)&3&12.65&147.39 \\
			&SCQR (Lasso)&0.3298 (0.16)&3&17.87&0.72&0.1808 (0.06)&3&23.89&1.08\\
			&SCQR (SCAD)&0.0484 (0.04)&3&0.06&0.44&0.0290 (0.02)&3&0.03&0.79\\
			\cline{1-10}
			$t_3$&CQR (Lasso)&0.4131 (0.23)&3&10.3&19.78&0.2196 (0.09)&3&12.78&150.94 \\
			&SCQR (Lasso)&0.3661 (0.17)&3&17.57&0.81&0.1839 (0.08)&3&23.58&1.15\\
			&SCQR (SCAD)&0.0561 (0.05)&3&0.04&0.52&0.0246 (0.02)&3&0.03&0.88\\
			\cline{1-10}
			Cauchy&CQR (Lasso)&1.3223 (0.99)&3&13.33&24.64&0.6289 (0.38)&3&17.92&190.73 \\
			&SCQR (Lasso)&1.0474 (0.68)&3&17.57&0.96&0.4655 (0.26)&3&23.36&1.40\\
			&SCQR (SCAD)&0.2350 (0.39)&3&0.41&0.64&0.1174 (0.16)&3&0.69&1.11\\
			\hline
			
		\end{tabular}
	\end{table}

	\begin{table}[!htp]
		\fontsize{7}{8}\selectfont
		\centering
		\caption{Statistical performance comparison between the CQR (via ADMM) and the SCQR (via LAMM) estimators under linear model \eqref{simulationmodel} with four error distributions---$N(0,3)$,  mixture normal, $t_3$ and Cauchy. The average of the squared model $L_2$ error (and standard error), true positives (TP), and false positives (FP), and runtime (in seconds), over 100 replications, are reported.  5-fold CV is used to select $\lambda$. }
		\label{table2}
		\begin{tabular}{ c | c | c c c c }
			\hline
			& & \multicolumn{4}{c}{$n = 100, p = 600$}  \\
			Error& Method & ME  & TP & FP& Runtime  \\
			\hline
			$N(0,3)$&CQR (Lasso)&0.7924 (0.36)&3&3.52&768.38 \\
			&SCQR (Lasso)&0.8066 (0.43)&3&8.77&52.41\\
			&SCQR (SCAD)&0.2074 (0.24)&3&0.31&49.94\\
			\cline{1-6}
			MN&CQR (Lasso)&0.4110 (0.20)&3&4.14&626.67 \\
			&SCQR (Lasso)&0.5066 (0.29)&3&9.69&46.29\\
			&SCQR (SCAD)&0.1383 (0.13)&3&0.29&44.05\\
			\cline{1-6}
			$t_3$&CQR (Lasso)&0.4125 (0.23)&3&4.27&722.36\\
			&SCQR (Lasso)&0.4966 (0.34)&3&8.60&48.12\\
			&SCQR (SCAD)&0.1395 (0.12)&3&0.35&47.68\\
			\cline{1-6}
			Cauchy&CQR (Lasso)&1.2951(1.12)&3&4.74&1158.31\\
			&SCQR (Lasso)&1.2584 (1.06)&3&6.32&63.46\\
			&SCQR (SCAD)&0.3848 (0.48)&2.96&0.22&79.90\\
			\hline
		\end{tabular}
	\end{table}
	
	\begin{table}[!htp]
		\fontsize{7}{8}\selectfont
		\centering
		\caption{Statistical performance comparison between the CQR (via ADMM) and the SCQR (via LAMM) estimators under linear model \eqref{simulationmodel} with four error distributions---$N(0,3)$,  mixture normal, $t_3$ and Cauchy. The average of the squared model $L_2$ error (and standard error), true positives (TP), false positives (FP) and runtime (in seconds), over 200 replications, are reported.  The BIC \eqref{biccriterion} is used to select $\lambda$. }
		\label{bictable}
		\begin{tabular}{ c | c | c c c c }
			\hline
			& & \multicolumn{4}{c}{$n = 100, p = 600$}  \\
			Error& Method & ME  & TP & FP& Runtime  \\
			\hline
			$N(0,3)$&CQR (Lasso)&1.0595 (0.64)&3&0.64&197.86 \\
			&SCQR (Lasso)&0.9438 (0.54)&3&0.62&9.51\\
			&SCQR (SCAD)&0.3394 (0.49)&3&0.95&8.18\\
			\cline{1-6}
			MN&CQR (Lasso)&0.5632 (0.27)&3&0.75&157.95 \\
			&SCQR (Lasso)&0.5159 (0.24)&3&0.67&8.48\\
			&SCQR (SCAD)&0.1208 (0.17)&3&0.49&7.28\\
			\cline{1-6}
			$t_3$&CQR (Lasso)&0.6291 (0.39)&3&0.54&173.23\\
			&SCQR (Lasso)&0.5659 (0.34)&3&0.53&8.81\\
			&SCQR (SCAD)&0.0876 (0.13)&3&0.22&7.81\\
			\cline{1-6}
			Cauchy&CQR (Lasso)&2.7542 (2.64)&2.83&0.30&283.25\\
			&SCQR (Lasso)&2.1228 (1.81)&2.93&0.27&11.51\\
			&SCQR (SCAD)&0.4055 (0.94)&2.88&0.03&12.60\\
			\hline
		\end{tabular}
	\end{table}

	We end this section with a numerical comparison of the (smoothed) composite quantile regression method and the robust regression method proposed by \cite{Wang2020}. 
	The latter is a combination of the pairwise difference technique and LAD regression. For simplicity, we focus on  $L_1$-penalization.  Following the terminology in  \cite{Wang2020},  we refer to their estimator as Rank Lasso,  defined as 
	\#
	\hat{\bbeta}^{{\rm RL}}(\lambda)=\argmin_{\bbeta\in\RR^p}\bigg\{ \frac{1}{n(n-1)}  \sum_{i=1}^n \sum_{j = 1, j\neq i}^n \vert(y_i-\bx_i\T\bbeta)-(y_j-\bx_j\T\bbeta)\vert+\lambda\sum_{k=1}^p\vert\beta_k\vert\bigg\} .\label{ranklassoloss}
	\#
	By utilizing the pivotal property of the $L_1$-loss \citep{Belloni2011},  they further proposed a simulation-based procedure to choose $\lambda$ automatically from the data.  Computationally, 	 \cite{Wang2020} reformulate the optimization problem \eqref{ranklassoloss} as a linear program (LP),  and then use general-purpose optimization toolboxes.  We thus follow this route and implement Rank Lasso using the \texttt{SciPy} library with method ``highs" \citep{highs}.  In the following simulation study, we use equation (7) in \cite{Wang2020} with $c=1.01$ and $\alpha_0=0.1$ to compute the $\lambda$ in \eqref{ranklassoloss}; for SCQR-Lasso,  we  simulate $\lambda$ via  \eqref{lambdasimulation} with $c=1.9$ and $\alpha=0.05$.

	For data-driven Rank Lasso and SCQR-Lasso estimators,	we summarize results on the statistical and computational performance  in Table \ref{ranklassotable} under the four error distributions when $(n, p)=(100, 600)$.  	The data-driven SCQR typically has much smaller estimation errors but more false positives than the data-driven Rank Lasso.  This could just be a consequence of the different tuning  procedures. The runtime comparison confirms SCQR as a practical and computational efficient  approach to robust regression.
	The linear program reformulation of \eqref{ranklassoloss}, on the other hand, involves $2n^2+2p$ variables and $O(n^2 + p)$ constraints. Even the state-of-the-art LP solvers are not adapted to large-scale problems.

	\begin{table}[!htp]
		\fontsize{7}{8}\selectfont
		\centering
		\caption{Statistical and computational performance comparison of the Rank Lasso and the SCQR methods, among four error distributions: $N(0,3)$, MN, $t_3$, and the standard Cauchy, under model \ref{simulationmodel}. The mean (and standard error) of the model estimation error, true positives (TP),  false positives (FP), and runtime (in seconds) are reported. }
		\label{ranklassotable}
		\begin{tabular}{ c | c | c c c c }
			\hline
			& & \multicolumn{4}{c}{$n = 100, p = 600$}  \\
			Error& Method & ME & TP & FP &Runtime  \\
			\hline
			$N(0,3)$&Rank Lasso&1.5222(0.52)&3&0.30&249.28 \\
			&SCQR (Lasso)&0.6970(0.24)&3&17.95&0.62\\
			&SCQR (SCAD)&0.1237(0.13)&3&0.05&0.34\\
			\cline{1-6}
			MN&Rank Lasso&0.8145(0.51)&3&0.55&247.97 \\
			&SCQR (Lasso)&0.3478(0.22)&3&18.30&0.60\\
			&SCQR (SCAD)&0.0726(0.05)&3&0.05&0.38\\
			\cline{1-6}
			$t_3$&Rank Lasso&0.8324(0.48)&3&0.35&241.71\\
			&SCQR (Lasso)&0.3660(0.21)&3&16.20&0.56\\
			&SCQR (SCAD)&0.0551(0.042)&3&0.05&0.40\\
			\cline{1-6}
			Cauchy&Rank Lasso&4.6628(2.89)&3&0.40&241.81 \\
			&SCQR (Lasso)&1.3241(0.84)&3&17.55&0.82\\
			&SCQR (SCAD)&0.4395(1.47)&2.95&0.20&0.43\\
			\hline

		\end{tabular}
	\end{table}

	\newpage
	\appendix
	
	\section{Proof of main results}

	By a change of variable, we can rewrite the smoothed composite quantile loss $\hat{Q}_h: \RR^{q+p} \to \RR$ in \eqref{SECQRloss} as
	$$
	\hat{Q}_h(\balpha,\bbeta)=\frac{1}{q}\sum_{k=1}^q \bigg\{(1-\tau_k)\int_{-\infty}^0\hat{F}_h(u+\alpha_k;\bbeta)\diff u+\tau_k\int_0^\infty(1-\hat{F}_h(u+\alpha_k;\bbeta))\diff u\bigg\} . 
	$$
	Using this expression, we obtain 
	\#
	\partial_{\alpha_k} \hat{Q}_h(\balpha,\bbeta) & =\frac{1}{ q n }\sn  \{ \bar K_h ( \alpha_k-r_i(\bbeta) )-\tau_k  \} , \ \ k= 1, \ldots, q  ,  \label{alphagradient} \\
	\nabla_{\bbeta}\hat{Q}_h(\balpha,\bbeta) & =\frac{1}{q n }\sum_{k=1}^q \sn \{ \bar K_h (  \alpha_k-r_i(\bbeta) )-\tau_k  \}\bx_i , \label{betagradient}
	\#
	where 
	$$
	\bar K(u) = \int_{-\infty}^u K(v) \diff v ~~\mbox{ and }~~ \bar K_h(u) = \bar K(u/h).
	$$ 
	In this notation, we have $\bar K_h'(u) = K_h(u) = (1/h) K(u/h)$.

	\subsection{Proof of Lemma \ref{lem:convexity}}
	
	Combining \eqref{alphagradient} and \eqref{betagradient}, we see that the full gradient of $\hat{Q}_h$ with respect to $(\balpha\T ,\bbeta\T)\T \in \RR^{q+p}$ is
	\#
	\nabla\hat{Q}_h(\balpha,\bbeta)=\frac{1}{q n }\sn\begin{pmatrix} \bar K_h( \alpha_1-r_i(\bbeta) )-\tau_1\\\vdots\\  \bar K_h( \alpha_q-r_i(\bbeta) )-\tau_q\\\sum_{k=1}^q \{ \bar K_h( \alpha_k-r_i(\bbeta) )-\tau_k \}  \bx_i \end{pmatrix}  \in \RR^{q+p} ,   \label{ESCQR_Gradient}
	\#
	where $r_i(\bbeta) = y_i - \bx_i\T \bbeta$.
	For the Hessian, note that for any $1\leq k,l \leq q$ and $1\leq j \leq p$,
	$$
	\frac{\partial^2\hat{Q}_h}{\partial\alpha_k\partial\alpha_l}=\frac{1}{q n }\sn K_h(\alpha_k-r_i(\bbeta))\delta_{kl},\quad \frac{\partial^2\hat{Q}_h}{\partial\beta_j \partial\alpha_k}=\frac{1}{q n}\sn K_h(\alpha_k-r_i(\bbeta))x_{ij} 
	$$
	with $\delta_{kl} = \ind(k=l)$, and
	$$
	\nabla_{\bbeta}^2\hat{Q}_h(\balpha,\bbeta)=\frac{1}{ n q }\sn\sum_{k=1}^q  K_h(\alpha_k-r_i(\bbeta))\bx_i\bx_i\T.
	$$
	For every $\bbeta \in \RR^p$, write
	\$
	\bv_i  &= \bv_i(\bbeta) =(K_h(\alpha_1-r_i(\bbeta)),\cdots,K_h(\alpha_q-r_i(\bbeta)))\T  , \ \ i=1,\ldots, n,  \\
	\bv & = (v_1, \ldots, v_q )\T =   \bv (\bbeta) =(K_h(\alpha_1-r (\bbeta) ),\cdots,K_h(\alpha_q-r (\bbeta)))\T ,
	\$
	where $r (\bbeta) = y - \bx\T \bbeta$.
	It follows that 
	\#
	\nabla^2\hat{Q}_h(\balpha,\bbeta)=\frac{1}{n q}\sn\begin{pmatrix} \diag(\bv_i) & \bv_i\bx_i\T\\\bx_i\bv_i\T &1_q\T\bv_i \cdot \bx_i\bx_i\T\end{pmatrix}   \label{HessianESCQR}
	\#
	and
	\#
	\nabla^2 {Q}_h(\balpha,\bbeta) =  \frac{1}{q}\begin{pmatrix}\EE( \diag(\bv))&&\EE( \bv\bx\T) \\\EE( \bx\bv\T) &&\EE ( 1_q\T\bv \cdot \bx\bx\T )  
	\end{pmatrix} , \label{populationHessianSCQR}
	\#
	where $\mathbf{1}_q=(1, \ldots, 1)\T \in \RR^q$.
	
	For any $\ba=(a_1,\ldots ,a_q)\T \in \RR^q$ and $\bb\in\RR^p$, note that
	\$
	(\ba\T,\bb\T) \nabla^2 {Q}_h(\balpha,\bbeta)(\ba\T,\bb\T)\T  = \frac{1}{q}   \sum_{k=1}^q  \EE \{  v_k ( a_k + \bx\T \bb )^2 \} \geq 0 ,
	\$
	where $v_k = K_h(\alpha_k - r(\bbeta) )$.  This verifies the positive semidefiniteness of $\nabla^2 {Q}_h(\balpha,\bbeta)$ and so is the convexity of $Q_h$.  At $(\balpha,\bbeta)= (\balpha^*, \bbeta^*)$, $v_k= K_h(\alpha^*_k - \beta^*_0 - \varepsilon)= K_h(F^{-1}(\tau_k) - \varepsilon)$. Using the independence of $\varepsilon$ and $\bx$, and condition \eqref{reg.cond}, we obtain that
	\$
	&(\ba\T,\bb\T) \nabla^2 {Q}_h(\balpha,\bbeta)(\ba\T,\bb\T)\T   \big|_{(\balpha,\bbeta)= (\balpha^*, \bbeta^*)}  \\
	& = \frac{1}{q} \sum_{k=1}^q  \EE (a_k + \bx\T \bb)^2 \cdot  \EE K_h( F^{-1}(\tau_k) - \varepsilon)  \\
	& = \frac{1}{q} \sum_{k=1}^q  \EE (a_k + \bx\T \bb)^2 \cdot \int_{-\infty}^\infty K(v) f( F^{-1}(\tau_k)  - hv  ) \diff v >0  
	\$
	for all $\ba \in \RR^q$ and $\bb \in \RR^p$, where the second equality follows from integration by parts and a change of variable.  This proves the strict convexity of  ${Q}_h$ at $(\balpha^*, \bbeta^*)$.

	Turning to the sample Hessian, for any $\ba=(a_1,\ldots ,a_q)\T \in \RR^q, \bb\in\RR^p$ we have 
	\$
	(\ba\T,\bb\T) \nabla^2\hat{Q}_h(\balpha,\bbeta)(\ba\T,\bb\T)\T = \frac{1}{n q} \sn 	\sum_{k=1}^q K_h(\alpha_k  - r_i(\bbeta))  ( a_k + \bx_i\T \bb)^2 \geq 0 .
	\$
	Hence, the empirical composite quantile loss $\hat{Q}_h$ is twice-differentiable and convex. 
	\qed

	\subsection{Proof of Proposition \ref{proposition3.1}}
	
	We first show that the function $m_h: \RR^q \to \RR$ has a unique minimizer, denoted by $\bb_h$. 
	For each $1\leq k\leq q$, define the univariate function $m_{h, k} (b) = \EE \ell_{h, k} (\varepsilon - b)$ whose first and second-order derivatives are
	$$
	m'_{h, k} (b)  =  \int_{-\infty}^\infty K(v) F(b -hv)\diff v-\tau_k , \quad 
	m''_{h, k} (b) = \int_{-\infty}^\infty K(v)  F(b-hv)\diff v   .
	$$
	Since $K$ is positive everywhere, we have $m''_{h, k} (b) >0$ for all $b$. Therefore, $m_{h, k}(\cdot)$ is strictly convex and has a unique minimizer, denoted by $b_{h,k}$.
	Noting further that $\nabla^2 m_h(\bb ) = q^{-1} \diag(\{ m''_{h, 1} (b_1) , \ldots ,  m''_{h, q} (b_q) \} )$,  the function $m_h: \RR^q \to \RR$ is also strictly convex with a unique minimizer $\bb_h =(b_{h, 1}, \ldots, b_{h, q})\T$.

	For any $\balpha\in\RR^q, \bbeta\in\RR^p$, we write $ \Delta_{\bx} =\bx\T(\bbeta - \bbeta^*)$ and obtain that
	\$
	Q_h(\balpha,\bbeta) & = \frac{1}{q} \sum_{k=1}^q \EE  \ell_{h, k} (\varepsilon + \beta^*_0 - \alpha_k - \Delta_{\bx} )\\
	& = \EE \bigg[  \frac{1}{q} \sum_{k=1}^q  \EE \{ \ell_{h, k} (\varepsilon + \beta^*_0 - \alpha_k  -\Delta_{\bx}  )  | \bx \} \bigg] \\
	& = \EE \{  m_h ( \alpha_1 + \Delta_{\bx} - \beta^*_0,  \ldots, \alpha_q + \Delta_{\bx} - \beta^*_0)  \} \\
	& \geq  m_h(\bb_h)= Q_h(  \beta^*_0 + \bb_h ,  \bbeta^*) .
	\$
	In other words, the function $Q_h: \RR^{q + p} \to \RR$ is minimized at $(\beta^*_0 + \bb_h ,  \bbeta^*)$.
	With a everywhere positive kernel, $Q_h$ is strictly convex so that $(\beta^*_0 + \bb_h ,  \bbeta^*)$ is the unique minimizer, implying $\balpha_h^* = \beta^*_0 + \bb_h$ and $\bbeta^*_h = \bbeta^*$ as claimed. 
	
	Finally, it remains to bound $| b_{h, k} - F^{-1}(\tau_k) |$.
	For each $k=1,\ldots, q$, define $\wt b_k = F^{-1}(\tau_k) + t_k \{ b_{h, k} - F^{-1}(\tau_k) \}$ with 
	$$
	t_k = \sup\{t\in [0, 1] : t |b_{h, k} - F^{-1}(\tau_k)  | \leq \kappa_2^{1/2} h \} 
	\begin{cases} 
		= 1 & \mbox{ if }~ |b_{h, k} - F^{-1}(\tau_k)| \leq \kappa_2^{1/2} h   , \\
		\in (0, 1) & \mbox{ otherwise}. 
	\end{cases}
	$$
	Set $\delta_k = \wt b_k - F^{-1} (\tau_k)$, satisfying $|\delta_k| \leq \kappa_2^{1/2} h$ and in particular, $|\delta_k| = \kappa_2^{1/2} h$ if $|b_{h, k} - F^{-1}(\tau_k)| > \kappa_2^{1/2} h$.
	By Lemma~\ref{lem:convexity} and the fact that $m'_{h,k}(b_{h,k}) = 0$, we have
	\$
	\{ m'_{h,k}(\wt b_k) - m'_{h,k}( F^{-1}(\tau_k) ) \} \delta_k  \leq \{ m'_{h,k}( b_{h,k}) - m'_{h,k}( F^{-1}(\tau_k) ) \} \delta_k  \leq | m'_{h,k}( F^{-1}(\tau_k) )|    \cdot | \delta_k | .
	\$
	For the left-hand side,
	\#
	& m'_{h,k}(\wt b_k) - m'_{h,k}( F^{-1}(\tau_k) )   = \int_{F^{-1}(\tau_k) }^{ \wt b_k } m_{h,k}''(t) \, {\rm d}t  = \int_{F^{-1}(\tau_k) }^{ \wt b_k }  \int_{-\infty}^\infty K(u) f(t-hu) \, {\rm d} u  \, {\rm d}t \nn \\
	& =  f(F^{-1}(\tau_k)) \delta_k   + \int_{F^{-1}(\tau_k) }^{ \wt b_k }  \int_{-\infty}^\infty K(u)    \{ f (t-hu)  -  f(F^{-1}(\tau_k)) \} \, {\rm d} u  \, {\rm d}t . \nn
	\#
	This, together with the Lipschitz continuity of $f$, implies
	\#
	\{ m'_{h,k}(\wt b_k) - m'_{h,k}( F^{-1}(\tau_k) ) \} \delta_k \geq  f(F^{-1}(\tau_k)) \delta_k^2 - \frac{l_0}{2} |\delta_k |^3 - l_0 \kappa_1 h \cdot  \delta_k^2 .\nn
	\#
	On the other hand, $|m_{h,k}'(F^{-1}(\tau_k))| = | \int_{-\infty}^\infty  K(u)  \{ F (F^{-1}(\tau_k) -hu) -F (F^{-1}(\tau_k)) \} \, {\rm d} u  | \leq  l_0  \kappa_2 h^2 /2$. Putting together the pieces, we conclude that
	\#
	f(F^{-1}(\tau_k))    \delta_k^2 \leq  \frac{l_0}{2} \kappa_2 h^2 | \delta_k | + \frac{l_0}{2} | \delta_k |^3 + l_0 \kappa_1  h    \delta_k^2 <   \kappa_2 l_0 h^2 |\delta_k | + \frac{1}{2}  f(F^{-1}(\tau_k))    \delta_k^2   \nn
	\#
	provided $0< h \leq   f(F^{-1}(\tau_k))/(2  \kappa_2^{1/2}l_0 )$, where the second inequality follows from the fact that $\kappa_1 < \kappa_2^{1/2}$.
	Canceling out $|\delta_k |$ from both sides yields
	\$
	| \delta_k | <   \frac{ 2 \kappa_2^{1/2} l_0  h }{ f (F^{-1}(\tau_k) )  }  \cdot    \kappa_2^{1/2} h \leq  \kappa_2^{1/2} h. 
	\$
	By the definition of $\wt b_k$, we must have $|b_{h, k} - F^{-1}(\tau_k)  | \leq \kappa_2^{1/2} h$; otherwise $| \delta_k | = \kappa_2^{1/2}h$ which contradicts the above inequality. Consequently, $t_k = 1$ and $\wt b_k = b_{h,k}$, thus implying the claimed bound \eqref{bias.ubd}. \qed

	\subsection{Proof of Proposition \ref{goodeventlemma}}
	
	We first consider $\Vert\bzeta^*\Vert_\infty = \max_{1\leq k\leq q} | \zeta^*_k|$, where 
	$$
	\zeta^*_k= \frac{1}{nq }\sn\{\bar{K}((\alpha_{h,k}^*-\varepsilon_i)/h)-\tau_k\}
	$$ 
	is the $k$-th coordinate of $\bzeta^*$. Note that  $\bar{K}((\alpha_{h,k}^*-\varepsilon_i)/h)-\tau_k \in [-\tau_k , 1-\tau_k  ]$.  Hence, applying Hoeffding's inequality yields
	$$
	\PP\{\vert\zeta^*_k\vert>3\lambda/(2q)\}\leq  2 e^{ -2 n (3\lambda /2)^2 } =2 e^{ -9n\lambda^2/2 }.
	$$
	Taking the union bound over $k=1,\ldots, q$, 	 it follows that $\PP\{ \Vert\bzeta^*\Vert_\infty > 3\lambda/(2q) \} \leq 2 q e^{-9n \lambda^2/2}$.
	
	For $\bw^* = (\omega_1^*, \ldots, \omega^*_p)\T \in\RR^p$,  let
	$$z_{ij } =\frac{x_{ij}}{q}\sum_{k=1}^q\{\bar{K}((\alpha_{h,k}^*-\varepsilon_i)/h)-\tau_k\},$$
	so that  $\omega_j^*=(1/n) \sn z_{ij}$.
	Note that $\EE(z_{ij})=0$, $\vert z_{ij}\vert\leq\vert x_{ij}\vert$, and  by assumption (A3),  $	\PP(\vert x_{ij}\vert\geq\nu_0\sigma_{jj}^{1/2} t)\leq e^{-t^2/2}$ for all $t\geq 0$.  It thus follows from Proposition 2.5 of \cite{W2019} that 
	\#
	\PP\bigg(\bigg\vert\frac{1}{n}\sn z_{ij}\bigg\vert>t\bigg)\leq 2\exp\{-nt^2/(8\nu_0^2\sigma_{jj})\}.
	\#
	Finally, taking $t=\lambda/2$ and applying the union bound over $j=1,\ldots, p$ prove the claimed bound. \qed

	\subsection{Proof of Proposition \ref{RSCproperty}}
	
	We have
	\#
	D({\balpha},{\bbeta})=\frac{1}{n q}\sn\sum_{k=1}^q\Bigg\{\bar{K}\bigg(\frac{{\alpha}_{k}-r_i({\bbeta})}{h}\bigg)-\bar{K}\bigg(\frac{\alpha_{h,k}^*-r_i(\bbeta^*)}{h}\bigg)\Bigg\}\langle\bar{\bx_i},({\delta_k},{\bDelta})\rangle,  \label{Bregman Divergence}
	\#
	by letting $\bar{\bx_i}=(1,\bx_i\T)\T$, where ${\bdelta}={\balpha}-\balpha_h^*, {\bDelta}={\bbeta}-\bbeta^*.$

	We restrict our focus on the symmetrized Bregmann divergence with each quantile index $k=1,\ldots,q$, by letting
	$$D_k({\balpha},{\bbeta}):=\frac{1}{n}\sn\left\{\bar{K}\left(\frac{{\alpha}_k-r_i({\bbeta})}{h}\right)-\bar{K}\left(\frac{\alpha_{h,k}^*-r_i(\bbeta^*)}{h}\right)\right\}\langle\bar{\bx_i},\bDelta_k\rangle,$$
	where $\bDelta_k:=({\delta_k},{\bDelta})$.
	
	We first claim that, when $\Vert\bDelta_k\Vert_{\bar{\Sigma}}\leq r,$ we have the following lower bound $$D_k(\balpha,\bbeta)\geq c\Vert\bDelta_k\Vert_{\bar{\Sigma}}^2$$ with high probability for some positive $c>0$, so that we can combine all the bounds for each $k$ to derive the desired result of the Proposition \ref{RSCproperty}.
	Let us define an event in the neighborhood of the true parameter, $\cE_{i,k}:=\{\vert\varepsilon_i-\alpha_{h,k}^*\vert\leq h/2\}\cap\{\vert\langle \bar{\bx_i},\bDelta_k\rangle\vert/\Vert\bDelta_k\Vert_{\bar{\Sigma}}\leq h/(2r)\},$ with $r>0$. We can lower bound $D_k({\balpha},{\bbeta})$ by 
	\#\frac{\underline{\kappa}}{nh}\sn\langle\bar{\bx_i},\bDelta_k\rangle^2\ind_{\cE_{i,k}},\label{dklower}\# where $\underline{\kappa}:=\min_{\vert x\vert\leq1}K(x)$. 
	Also, by using similar smoothing technique from \cite{loh2017statistical}, we define a Lipshitz continuous function for $R>0$, as
	\#
	\varphi_R(u):=u^2\ind(\vert u\vert\leq R/2)+(u-R)^2\ind(R/2<u\leq R)+(u+R)^2\ind(-R\leq u<R/2).\label{Lipshitzconti}
	\#
	Then, we can further lower bound \eqref{dklower} by
	\#
	\frac{\underline{\kappa}}{nh}\sn\langle\bar{\bx_i},\bDelta_k\rangle^2\ind_{\cE_{i,k}}\geq\underline{\kappa}\cdot\underbrace{\frac{1}{nh}\sn\varphi_{h/2}(\langle\bDelta_k,\bar{\bx_i}\rangle)\cdot\chi_{i,k}}_{=:D_{0,k}(\balpha,\bbeta)},
	\#
	where $\chi_{i,k}=\ind(\vert\varepsilon_i-\alpha_{h,k}^*\vert\leq h/2)$. 
	
	To prove our claim, now it suffices to show that when $\Vert\bDelta_k\Vert_{\bar{\Sigma}}=\delta r,$ for each $(\delta\in(0,1]),$ we have
	\#
	\frac{\underline{\kappa}}{nh}\sn\varphi_{\delta (h/2)}(\langle\bDelta_k,\bar{\bx_i}\rangle)\cdot\chi_{i,k}\geq c(\delta r)^2.\label{fixedradius}
	\#
	
	If it holds for $\delta=1$, then 
	$$\frac{\underline{\kappa}}{nh}\sn\varphi_{h/2}(\langle\bDelta_k/\delta,\bar{\bx_i}\rangle)\cdot\chi_{i,k}\geq cr^2,$$ which gives 
	$$
	\frac{\underline{\kappa}}{nh}\sn\varphi_{\delta (h/2)}(\langle\bDelta_k,\bar{\bx_i}\rangle)\cdot\chi_{i,k}\geq c(\delta r)^2.
	$$
	Hence, we only need to prove when $\Vert\bDelta_k\Vert_{\bar{\Sigma}}=r$. Suppose $\Vert\bDelta_k\Vert_{\bar{\Sigma}}=r$, we have
	$$\vert\EE\chi_{i,k}-hf(\alpha_{h,k}^*)\vert\leq\int_{\alpha_{h,k}^*-h/2}^{\alpha_{h,k}^*+h/2}\vert f_\varepsilon(t)-f_\varepsilon(\alpha_{h,k}^*)\vert\diff{t}\leq\frac{l_0}{4}h^2$$
	with Proposition \ref{proposition3.1}. Moreover, we have the following lower bound
	\#
	\EE\chi_{i,k}\geq hf_\varepsilon(\alpha_{h,k}^*)-\frac{l_0}{4}h^2\geq h(\underbar{$f$}-l_0C_bh^2-\frac{l_0}{4}h)\geq\frac{3}{4}\underbar{$f$}h\label{chiik}
	\#
	when $h\leq\min\{\underbar{$f$}/(4\kappa_2^{1/2}l_0),\underbar{$f$}/(2l_0)\}$, where $C_b:={2\kappa_2l_0}/{\underbar{$f$}}$ is an upper bound of the bias $\vert b_{h,k}-f(F^{-1}(\tau_k))\vert$ that derived in the Proposition \ref{proposition3.1}. Then, using above results, we get $$\EE\{h^{-1}\varphi_{h/2}(\langle\bDelta_k,\bar{\bx_i}\rangle)\chi_{i,k}\}\geq\frac{3}{4}\underline{f}\EE\varphi_{h/2}(\langle\bDelta_k,\bar{\bx_i}\rangle)\geq\frac{3}{4}\underline{f}[r^2-\EE\{\langle\bDelta_k,\bar{\bx_i}\rangle^2\ind\{\vert\langle\bDelta_k,\bar{\bx_i}\rangle\vert\geq h/4)\}].$$
	The last term of the above inequality is equal to
	\#\label{expectationlowerbound}
	\frac{3}{4}\underline{f}r^2\big(1-\EE\{\bxi_{\bDelta_k}^2\ind_{\vert\bxi_{\bDelta_k}\vert\geq h/(4r)}\}\big),
	\#
	where $\bxi_{\bDelta_k}=\langle\bDelta_k,\bar{\bx_i}\rangle/\Vert\bDelta_k\Vert_{\bar{\Sigma}}$.
	
	Since we have sub-Gaussian covariates, for any $u>0$, we get
	\begin{align*}
		\EE\{\bxi_{\bDelta_k}^2\ind(\vert\bxi_{\bDelta_k}\vert>u)\}&=2\EE\Bigg\{\int_0^\infty t\cdot\ind(\vert\bxi_{\bDelta_k}\vert>t)\ind(\vert\bxi_{\bDelta_k}\vert>u)\diff{t}\Bigg\}\\&=2\EE\int_0^ut\cdot\ind(\vert\bxi_{\bDelta_k}\vert>t)\ind(\vert\bxi_{\bDelta_k}\vert>u)\diff{t}+2\EE\int_u^\infty t\cdot\ind(\vert\bxi_{\bDelta_k}\vert>t)\diff{t}\\&=u^2\PP(\vert\bxi_{\bDelta_k}\vert>u)+2\int_u^\infty t\cdot\PP(\vert\bxi_{\bDelta_k}\vert>t)\diff{t}\\&\leq u^2e^{-(u/\sqrt{2}\nu_0)^2}+2\nu_0^2\int_{u/\nu_0}^\infty t\cdot\PP(\vert\bxi_{\bDelta_k}\vert>\nu_0t)\diff{t}\leq(u^2+2\nu_0^2)e^{-(u/\sqrt{2}\nu_0)^2},
	\end{align*}
	where we use the condition (A3) to establish the last two inequalities.
	
	To simplify, define $L_\delta:=\min\big\{L: \EE(\bv\T\bar{\bx})^2\cdot\ind(\vert\bv\T\bar{\bx}\vert>L)\leq\delta \text{ for all } \bv\in\RR^{p+1}, \Vert\bv\Vert_{\bar{\Sigma}}=1\big\}$.
	%Then, by letting $u=h/(4r)\geq3\nu_0^2$ gives $\EE\big(\bxi_{\bDelta_k}^2\ind_{\vert\bxi_{\bDelta_k}\vert\geq h/(4r)}\big)<1/8.$
	Then, we get $L_{1/8}\leq h/(4r)$ when $h/(4r)\geq3\nu_0^2$., which leads to
	\#
	\inf_{\Vert\bDelta_k\Vert_{\bar{\Sigma}}=r}\EE\{D_{0,k}(\balpha,\bbeta)\}>\frac{21}{32}\underline{f}r^2.\label{D0k}
	\#
	
	Now, we need to bound $\vert D_{0,k}(\balpha,\bbeta)-\EE\{ D_{0,k}(\balpha,\bbeta)\}\vert$. The domain of interest is $\BB_{\Omega}(r)\cap\CC_{\Omega}(l)$, in particular $\Vert(\bdelta\T,\bDelta\T)\T\Vert_{\Omega}\leq r$.  Thus, when $\Vert\bDelta_k\Vert_{\bar{\Sigma}}=r$, it also satisfies that $\Vert\bDelta_k\Vert_1\leq l\Vert\bDelta_k\Vert_{\bar{\Sigma}}$.
	Let us define
	$$Z_n(l):=\sup_{\Vert\bDelta_k\Vert_1\leq lr}\vert D_{0,k}(\balpha,\bbeta)-\EE\{ D_{0,k}(\balpha,\bbeta)\}\vert.$$
	Then, we have $0\leq h^{-1}\varphi_{h/(2r)}(\bxi_{\bDelta_k})\leq (4r)^{-2}h,$ and
	\begin{align}
		\EE\{h^{-2}\varphi_{h/(2r)}^2(\bxi_{\bDelta_k})\chi_{i,k}\}&\leq(h/4r)^2h^{-2}\EE\big(\bxi_{\bDelta_k}^2\cdot\chi_{i,k}\big)\nonumber\\
		&\leq(4r)^{-2}\cdot\{hf_\varepsilon(\alpha_{h,k}^*)+l_0h^2/4\}\nonumber\\
		&\leq(4r)^{-2}({5\overline{f}h}/{4})\label{secondmomment},
	\end{align}
	where the last inequality follows from  Proposition \ref{proposition3.1}, with $h\leq\min\{\underbar{$f$}/(4\kappa_2^{1/2}l_0),\underbar{$f$}/(2l_0)\}$.
	\begin{comment}
		Then, since $(nh)^{-1}\varphi_{h/2}(\cdot)\leq h/(16n)$, using the bounded difference inequality, we get
		\#
		\PP[Z_n(l)-\EE\{Z_n(l)\}\geq s]\leq\exp(-{16ns^2}/{h^2}).\label{BDI}
		\#
	\end{comment}
	
	To control $Z_n(l)$ defined above, let us divide it by $r^2$ for convenience, which gives
	\#
	\frac{1}{r^2}\cdot Z_n(l)=\sup_{\Vert\bDelta_k\Vert_1\leq l\Vert\bDelta_k\Vert_{\bar{\Sigma}}}\Bigg\vert\frac{1}{nh}\sn\{\varphi_{h/(2r)}(\bar{\bx_i}\T\bDelta_k/\Vert\bDelta_k\Vert_{\bar{\Sigma}})-\EE\varphi_{h/(2r)}(\bar{\bx_i}\T\bDelta_k/\Vert\bDelta_k\Vert_{\bar{\Sigma}})\}\Bigg\vert.\label{Znl}
	\#
	With \eqref{secondmomment} and above preparations on $Z':=Z_n(l)/r^2$, we can apply Theorem 7.3 of \cite{bousquet2003concentration}, a refined Talagrand's inequality, which gives
	\#
	\PP\Big\{Z'\leq\EE(Z')+h/(16r^2n)\sqrt{{40n\overline{f}r^2x}/{h}+2\EE (Z')x}+{hx}/(48r^2n)\Big\}\leq e^{-x}.\label{refTalagrand}
	\#
	We further simplify the inequality above as
	\begin{align}
		Z'&\leq\EE(Z')+\{\EE(Z')\}^{1/2}\sqrt{\frac{hx}{4r^2n}}+(4r)^{-1}\cdot2\overline{f}^{1/2}\sqrt{\frac{hx}{n}}+\frac{hx}{48r^2n}\nonumber\\&\leq\frac{5}{4}\EE(Z')+\sqrt{\frac{\overline{f}hx}{4r^2n}}+\frac{hx}{3r^2n}\label{TalagrandBound}
	\end{align}
	with probability at least 1$-e^{-x}$, where the last inequality follows from $ab\leq a^2/4+b^2$. 
	
	By taking expectation on the right-hand side of \eqref{Znl}, we apply Talagrand's contraction principle in Theorem 4.12 of \cite{ledoux2013probability}, which leads to
	\begin{align}
		\EE(Z')&\leq\frac{l}{r}\cdot\EE\Bigg\Vert\frac{1}{n}\sn e_i\chi_{i,k}\bar{\bx_i}\Bigg\Vert_\infty,\label{Contraction}
	\end{align}
	where $e_i$'s are independent Rademacher random variables.
	
	We have $\EE(e_i\chi_{i,k}x_{ij})=0$ and $\EE(e_i\chi_{i,k}x_{ij})^2\leq\sigma_{jj}c_h$, where $c_h=(9/8)\overline{f}h+l_0h^2/4$. Also, for $k=3,4,\dots,$
	\begin{align}
		\EE\vert e_i\chi_{i,k}x_{ij}\vert^k&\leq c_h\cdot k\int_0^\infty u^{k-1}\PP(\vert x_{ij}\vert\geq u)\diff{u} \nonumber\\
		&\leq c_h\nu_0^k\sigma_{jj}^{k/2}\cdot k\int_0^\infty\PP(\vert x_{ij}\vert\geq\nu_0\sigma_{jj}^{1/2}t)\diff{t}\nonumber\\
		&\leq c_h\nu_0^k\sigma_{jj}^{k/2}\cdot k\int_0^\infty e^{-t^2/2}t^{k-1}\diff{t}\nonumber\\&\leq\frac{k!}{2}\cdot \nu_0^2\sigma_{jj}c_h\cdot(2\nu_0\sigma_{jj}^{1/2})^{k-2}.\label{Bernsteintype}
	\end{align}
	Then, following the proof of Theorem 2.10 in \cite{boucheron2013concentration}, letting $v=\nu_0^2\sigma_x^2c_hn, c=2\nu_0\sigma_x$ and using Bernstein's inequality, we get
	\#
	\EE(Z')\leq2\nu_0\sigma_x\frac{l}{r}\Bigg(\sqrt{\frac{\overline{f}h\log(2p)}{n}}+\frac{\log(2p)}{n}\Bigg).\label{expectedZnl}
	\#
	
	Hence, combining bounds \eqref{TalagrandBound} and \eqref{expectedZnl} with $x=\log(2p)$, we get
	\#
	Z'\leq \frac{1}{2}(1+5\nu_0\sigma_xl)\sqrt{\frac{\overline{f}h\log(2p)}{r^2n}}+2.5\nu_0\sigma_xl\frac{\log(2p)}{rn}+\frac{h\log(2p)}{3r^2n}\label{combinedZnllbound}
	\#
	with probability at least $1-(2p)^{-1}$. Then, provided that $n\geq C(\nu_0\sigma_xl/\underline{f}r)^2\overline{f}h\log(2p)$ for some large constant $C$, we get 
	\#
	D_k(\balpha,\bbeta)\geq c\Vert\bDelta_k\Vert_{\bar{\Sigma}}^2
	\#
	with probability at least $1-(2p)^{-1}$ where $c=0.5\underline{f}\cdot\underline{\kappa}$.
	Summing up these results for $k=1\hdots,q$, we get the desired RSC property with probability at least $1-q/(2p).$   \qed

	\subsection{Proof of Theorem \ref{Theorem3.1}}
	
	Throughout the proof, we write $\hat \balpha = \hat \balpha_h$ and $\hat \bbeta = \hat \bbeta_h$ for simplicity.
	To prove Theorem \ref{Theorem3.1}, we first derive an upper bound on the symmetrized Bregman divergence given in \eqref{BregmanD}, along with a cone property for the estimator.
	Next, we prove a local RSC property based on Proposition \ref{RSCproperty}, which in turns implies a lower bound on the Bregman divergence. Combining these upper and lower bounds yields the claimed estimation error bound.
	
	Set $\hat\bdelta=\hat\balpha-\balpha_h^* \in \RR^q$ and $\hat\bDelta=\hat\bbeta-\bbeta^* \in \RR^p$. 
	Conditioned on the event $\cG(\lambda)$ defined in \eqref{eventG}, we have 
	\#
	D(\hat{\balpha},\hat{\bbeta})\leq\lambda(\Vert\hat{\bDelta}_S\Vert_1-\Vert\hat{\bDelta}_{S^{{\rm c}}}\Vert_1)+\frac{3\lambda}{2{q}}\Vert\hat{\bdelta}\Vert_1+\frac{\lambda}{2}\Vert\hat{\bDelta}\Vert_1\leq\frac{3\lambda}{2}\big(\Vert\hat{\bDelta}_S\Vert_1+q^{-1}\Vert\hat{\bdelta}\Vert_1\big) . \label{hatupper}
	\# 
	Recall from Proposition \ref{goodeventlemma} that a lower bound for $D( \balpha , \bbeta )$ holds when $(\bdelta\T, \bDelta\T)\T$ is in a cone-like set, where $\bdelta = \balpha - \balpha^*_h$ and $\bDelta = \bbeta - \bbeta^*$. We thus need to show that the estimator satisfies a cone-like property (with high probability).
	%			We first establish the cone-like property of the estimator that is derived from Proposition \ref{goodeventlemma}, so that 
	%$ {\scriptsize		 \begin{bmatrix}
			%					\hat \bdelta \\
			%					\hat \bDelta
			%			\end{bmatrix} } \in \cC(\cS)$.
	Using the optimality of $(\hat \balpha, \hat \bbeta)$ and the convexity of $\hat Q_h$, we have
	\#	0
	& \geq\hat{Q}_h(\hat{\balpha},\hat{\bbeta})-\hat{Q}_h(\balpha_h^*,\bbeta^*)+\lambda(\Vert\hat{\bbeta}\Vert_1-\Vert\bbeta^*\Vert_1)\label{goodeventcondition} \\
	&  \geq{\bzeta^*}\T(\hat{\balpha}-\balpha_h^*)+{\bomega^*}\T(\hat{\bbeta}-\bbeta^*)+\lambda(\Vert\hat{\bbeta}\Vert_1-\Vert\bbeta^*\Vert_1) \nn  \\
	& \geq  -\Vert\bzeta^*\Vert_\infty\Vert\hat{\balpha}-\balpha_h^*\Vert_1-\Vert\bomega^*\Vert_\infty\Vert\hat{\bbeta}-\bbeta^*\Vert_1+\lambda(\Vert\hat{\bbeta}_{S^{{\rm c}}}-\bbeta_{S^{{\rm c}}}^*\Vert_1-\Vert\hat{\bbeta}_S-\bbeta_S^*\Vert_1). \nn 
	\#
	It follows that
	\#
	(\lambda-\Vert\bomega^*\Vert_\infty)\Vert\hat{\bbeta}_{S^{{\rm c}}}-\bbeta_{S^{{\rm c}}}^*\Vert_1\leq(\lambda+\Vert\bomega^*\Vert_\infty)\Vert\hat{\bbeta}_{S}-\bbeta_{S}^*\Vert_1+\Vert\bzeta^*\Vert_\infty\Vert\hat{\balpha}-\balpha_h^*\Vert_1,\label{conelike}
	\#
	which further implies
	\begin{gather}\label{conelikeproperty}
		\Vert\hat{\bbeta}_{S^{{\rm c}}}-\bbeta_{S^{{\rm c}}}^*\Vert_1\leq3\Vert\hat{\bbeta}_{S}-\bbeta_{S}^*\Vert_1+3q^{-1/2}\Vert\hat{\balpha}-\balpha_h^*\Vert_2
	\end{gather}
	conditioned on $\cG(\lambda)$.
	Using the above bound and Cauchy-Schwarz inequality, we get 
	\$
	\Vert(\hat{\bdelta}\T,\hat{\bDelta}\T)\T\Vert_1 & \leq  4s^{1/2}\Vert\hat{\bDelta}_\cS\Vert_2+\big(q^{1/2}+3q^{-1/2}\big)\Vert\hat{\bdelta}\Vert_2\\
	&\leq   4 \max(s,q)^{1/2} (\Vert\hat{\bDelta}_\cS\Vert_2+\Vert\hat{\bdelta}\Vert_2)\\
	& \leq  4\sqrt{2} \cdot \max(s,q)^{1/2}\gamma_p^{-1/2}\Vert(\hat{\bdelta}\T,\hat{\bDelta}\T)\T\Vert_\Omega,
	\$
	so that $(\hat{\bdelta}\T,\hat{\bDelta}\T)\T\in\CC_\Omega(l)$ with $l=4\gamma_p^{-1/2}\sqrt{2\cdot\max(s,q)}$, where $\CC_\Omega(l)$ is defined in \eqref{l1cone}.
	
	Note further that the RSC property only holds in a local neighborhood of $\balpha^*_h$ and $\bbeta^*$, for which $(\hat{\balpha},\hat{\bbeta})$ does not necessarily satisfy. We thus employ a localized argument complemented with proof by contradiction.
	%		From Proposition \ref{RSCproperty}, we only have the RSC property with high probability within a local neighborhood of $(\balpha_h^*,\bbeta^*)$, which does not necessarily contain $(\hat{\balpha},\hat{\bbeta})$. 
	%To ensure the estimator lie on the local neighborhood of the true parameter, we define following quantity,
	For some $r>0$ to be determined, define $\eta:=\sup\big\{u\in[0,1]: u(\hat{\bdelta},\hat{\bDelta})\in\BB_{  {\Omega}}(r)\big\}$, where $\BB_{\Omega}(r)$ is defined in \eqref{l2ball}. By definition, $\eta=1$ when $(\hat{\bdelta},\hat{\bDelta}) \in \BB_{\Omega}(r)$, and $\eta\in(0,1)$ otherwise. 
	Then define an intermediate ``estimate" $(\tilde{\balpha},\tilde{\bbeta})=(\balpha_h^*,\bbeta^*)+\eta(\hat{\bdelta},\hat{\bDelta})$, which satisfies (i) $(\tilde{\balpha},\tilde{\bbeta})= (\hat{\balpha},\hat{\bbeta})$ if $(\hat{\bdelta},\hat{\bDelta}) \in \BB_{\Omega}(r)$, and (ii) $(\tilde{\balpha},\tilde{\bbeta})$ lies on the boundary of $(\balpha_h^*,\bbeta^*)+\BB_{\Omega}(r)$ if $(\hat{\bdelta},\hat{\bDelta}) \notin \BB_{\Omega}(r)$.
	Moreover, $(\tilde{\balpha},\tilde{\bbeta})$ inherits the cone property of $(\hat{\balpha},\hat{\bbeta})$ conditioned on $\cG(\lambda)$. Applying Proposition \ref{RSCproperty}, we obtain that
	\#
	D(\tilde{\balpha},\tilde{\bbeta})\geq c\bigg(\Vert\tilde{\bDelta}\Vert_\Sigma^2+\frac{1}{q}\Vert\tilde{\bdelta}\Vert_2^2\bigg) \label{tildelower}
	\# with probability at least $1-q/(2p)$ conditioned on $\cG(\lambda)$, where $c=0.5\underline{f}\cdot\underline{\kappa}$. On the other hand, Lemma F.2 in the supplementary material of \cite{Fan2018} states that $D(\tilde{\balpha},\tilde{\bbeta})\leq\eta D(\hat{\balpha},\hat{\bbeta})$. Combining the upper and lower bounds in \eqref{hatupper} and \eqref{tildelower}, we obtain
	\#
	c\bigg(\Vert\tilde{\bDelta}\Vert_\Sigma^2+\frac{1}{q}\Vert\tilde{\bdelta}\Vert_2^2\bigg)& \leq\frac{3\lambda}{2}\bigg(\Vert\tilde{\bDelta}_S\Vert_1+\frac{1}{\sqrt{q}}\Vert\tilde{\bdelta}\Vert_2\bigg)\nn  \\ 	
	& \leq  \frac{3\lambda}{2} \bigg(  s^{1/2}\Vert\tilde{\bDelta}\Vert_2 +\frac{1}{\sqrt{q}} \Vert  \wt \bdelta  \Vert_2  \bigg)			 \nn \\
	& \leq \frac{3}{\sqrt{2}} s^{1/2} \lambda \cdot   \Vert(\tilde{\bdelta}\T/\sqrt{q},\tilde{\bDelta}\T)\Vert_2 \leq \frac{3}{\sqrt{2}} \gamma_p^{-1/2} s^{1/2} \lambda \cdot   \Vert(\tilde{\bdelta}\T/\sqrt{q},\tilde{\bDelta}\T)\Vert_{\Omega}  . \nn
	\#
	Canceling out the common factor $\Vert(\tilde{\bdelta}\T/\sqrt{q},\tilde{\bDelta}\T)\Vert_{\Omega}$  from both sides yields
	\#
	\Vert(\tilde{\bdelta}\T/\sqrt{q},\tilde{\bDelta}\T)\Vert_{\Omega} \leq \frac{3}{\sqrt{2} c}\gamma_p^{-1/2}s^{1/2} = \frac{3\sqrt{2}}{\underline{f}\underline{\kappa} \gamma_p^{1/2}} s^{1/2} \lambda .
	\#

	In view of Proposition~\ref{RSCproperty}, we choose $r= h/(12 \nu_0^2)$ so that $\Vert(\tilde{\bdelta}\T,\tilde{\bDelta}\T)\Vert_{\Omega} < r$ provided 	
	$$ 
	\frac{3\sqrt{2}}{\underline{f}\underline{\kappa}}\gamma_p^{-1/2} (s q)^{1/2}\lambda< \frac{h}{12\nu_0^2}.
	$$  
	In this case, $(\tilde\balpha,\tilde\bbeta)$ falls into the interior of $\BB_{\Omega}(r)$ and we claim  that  $(\hat{\bdelta},\hat{\bDelta}) \in \BB_{\Omega}(r)$ and thus $\eta=1$. Otherwise if $(\hat{\bdelta},\hat{\bDelta}) \notin \BB_{\Omega}(r)$, $(\tilde\balpha,\tilde\bbeta)$ is constructed to be on the boundary of $(\balpha^*_h, \bbeta^*) + \BB_\Omega(r)$ so that $\Vert(\tilde{\bdelta}\T,\tilde{\bDelta}\T)\Vert_{\Omega} = r$. This contradicts the above, and therefore proves the claim. The desired estimation error bound then holds on the event $\cR(c,r,l)\cap\cG(\lambda)$. Finally, from Propositions \ref{goodeventlemma} and \ref{RSCproperty} we see that event $\cR(c,r,l)\cap\cG(\lambda)$  with $\lambda\asymp\nu_0\sigma_x\sqrt{\log(2p)/n}$ occurs with probability at least $1-q/p$ as long as {$n\gtrsim sq\log(p)$}.\qed

	\subsection{Proof of Corollary \ref{predictionerrorbound}}
	From \eqref{conelike}, we know that $\Vert\hat{\bbeta}_{\cS^c}-\bbeta_{\cS^c}^*\Vert_1\leq 3\Vert\hat{\bbeta}_{\cS}-\bbeta_{\cS}^*\Vert_1+3q^{-1}\Vert\hat{\balpha}-\balpha_{h}^*\Vert_1$. Also, we have $3q^{-1}\Vert\hat{\balpha}-\balpha_{h}^*\Vert_1\leq 3\vert\hat{\alpha}_j-\alpha_{h,1}^*\vert$ for some $j\in\{1,\hdots,q\},$ and assume $j=1$ satisfy the condition without loss of generality. Let $\bar{\bx}_i\T:=(q^{-1/2},\bx_i\T)$, $\bar{\bX}=(\bar{\bx}_1,\hdots,\bar{\bx}_n)\T\in\RR^{n\times (p+1)}$, $\bar{\bSigma}:=\EE\bar{\bx}_i\bar{\bx}_i\T$, which notation will be only used in this proof. Moreover, let $\bPsi=\bar{\bX}\bar{\bSigma}^{-1/2},$ $\bA=\bar{\bSigma}^{1/2}$, and $\hat{\bDelta}_1\T:=
	(\hat{\alpha}_1-\alpha_{h,1}^*,\hat{\bbeta}_h\T-{\bbeta^*}\T)$.  Then, Definition 1 in \cite{RZ2013} holds with $s_0=s, k_0=3, A=\bA$, and $K(s_0,k_0,A)=\{\min(\gamma_p,1/q)\}^{-1/2}$. Then, by using Theorem 16 of \cite{RZ2013}, we obtain that, with probability at least $1-2p^{-1}$, $$\frac{1}{\sqrt{n}}\Vert\bX(\hat{\bbeta}_h-\bbeta^*)\Vert_2\leq\frac{1}{\sqrt{n}}\Vert\bar{\bX}\hat{\bDelta}_1\Vert_2\leq2\Vert\bA\hat{\bDelta}_1\Vert\leq2\begin{Vmatrix}
		\frac{\hat{\balpha}_h-\balpha_h^*}{\sqrt{q}}\\\hat{\bbeta}_h-\bbeta^*
	\end{Vmatrix}_{\Omega}.$$ Then, the result follows from Theorem \ref{Theorem3.1}.

	\subsection{Proof of Theorem \ref{iterativeboundtheorem}}
	
	The idea behind this proof is that we need to control the magnitude of false discoveries at each step to refine the estimation error. The larger value of $\lambda_j$ with  $j\in\cS^{{\rm c}}$ tends to penalize the false discoveries harder. Throughout the proof, we write $\hat \balpha^t = \hat \balpha_h^t$ and $\hat \bbeta^t = \hat \bbeta^t_h$ for simplicity.
	Let us define a sequence of sets for $t\geq1$ as follows
	\#
	\cS_{t}:=\cS\cup\{1\leq j\leq p: \lambda_j^{t-1}=P_\lambda'(\vert\hat{\beta}_j^{t-1}\vert)<P'(a_0)\lambda\}.\label{indexset}
	\#
	Each set depends on the estimator of the previous iterative step. Using above definition of the index set $\cS_{t}$, we claim that 
	\#
	\vert\cS_{t}\vert<(b^2+1)s, \quad\text{ and }\quad\Vert\blambda_{\cS_{t}^{{\rm c}}}^{t-1}\Vert_{\min}\geq P'(a_0)\lambda.\label{indexsetclaim}
	\#
	We first assume that the above claime holds. On $\cG(P'(a_0)\lambda)$, using  \eqref{conelike}, we can derive that $(\hat{\bdelta}^t,\hat{\bDelta}^t)\in\CC_{\Omega}(l)$ with $l=4\gamma_p^{-1/2}\sqrt{2\cdot\max(s,q)}$, where $(\hat{\bdelta}^t,\hat \bDelta^t)=(\hat \balpha^t-\balpha_h^*,\hat\bbeta^t-\bbeta^*)$, and $\CC_{\Omega}(l) $ is defined in \eqref{l1cone}.  Consider the symmetrized Bregmann divergence 
	\begin{align*}
		D(\hat{\balpha}^{t},\hat{\bbeta}^{t})&=\langle-\blambda\circ\hat{\bg},\hat{\bDelta}^{t}\rangle+\langle-\bzeta^*,\hat{\bdelta}^{t}\rangle+\langle-\bomega^*,\hat{\bDelta}^{t}\rangle,
	\end{align*}
	where  $\hat{\bg}\in\partial\Vert\hat{\bbeta}^{t}\Vert_1$.
	For the first term of the right-hand side of the equality above, we split it into three parts, which leads to
	\begin{align*}
		\langle\blambda\circ\bg,\hat{\bDelta}^{t}\rangle&=\langle(\blambda\circ\bg)_\cS,\hat{\bDelta}_\cS^{t}\rangle+\langle(\blambda\circ\bg)_{\cS_t\setminus\cS},\hat{\bDelta}_{\cS_t\setminus\cS}^{t}\rangle+\langle(\blambda\circ\bg)_{\cS_t^{{\rm c}}},\hat{\bDelta}_{\cS_t^{{\rm c}}}^{t}\rangle\\
		&\geq-\Vert\blambda_\cS\Vert_2\Vert\hat{\bDelta}_\cS\Vert_2+0+\Vert\blambda_{\cS_t^{{\rm c}}}\Vert_{\min}\Vert\hat{\bDelta}_{\cS^{{\rm c}}}^{t}\Vert_1.
	\end{align*}
	The inequality above is derived using $\bbeta_{S^{{\rm c}}}^*=0$ and the property of subdifferential.
	Then, combining above result with H\"older's inequality, we get the following upper bound of the symmetrized Bregmann divergence
	\begin{align}
		D(\hat{\balpha}^{t},\hat{\bbeta}^{t})&\leq\Vert\bzeta^*\Vert_2\Vert\hat{\bdelta}^{t}\Vert_2+\Vert\bomega_{\cS_t}^*\Vert_2\Vert\hat{\bDelta}_{\cS_t}^{t}\Vert_2+\Vert\blambda_\cS\Vert_2\Vert\hat{\bDelta}_\cS\Vert_2+(\Vert\bomega_{\cS_t^{{\rm c}}}^*\Vert_\infty-\Vert\blambda_{\cS_t^{{\rm c}}}\Vert_{\min})\Vert\hat{\bDelta}_{\cS^{{\rm c}}}^{t}\Vert_1\nonumber\\
		&\leq\Vert\bzeta^*\Vert_2\Vert\hat{\bdelta}^{t}\Vert_2+\Vert\bomega_{\cS_t}^*\Vert_2\Vert\hat{\bDelta}_{\cS_t}^{t}\Vert_2+\Vert\blambda_\cS\Vert_2\Vert\hat{\bDelta}_\cS\Vert_2\label{ellupperbound}\\
		&\leq q\Vert\bzeta^*\Vert_\infty\Vert\hat{\bdelta}^{t}/\sqrt{q}\Vert_2+(0.5P'(a_0)s^{1/2}\sqrt{b^2+1}+s^{1/2})\lambda\Vert\hat{\bDelta}^{t}\Vert_{2}\nonumber\\
		&<\gamma_p^{-1/2}s^{1/2}\lambda\big(P'(a_0)\sqrt{(b^2+1)/2}+2\big)\Vert\btheta^{t}\Vert_{\Omega}=c\cdot r_{\text{opt}}\Vert\btheta^{t}\Vert_{\Omega}.
	\end{align}
	As in the proof of Theorem \ref{Theorem3.1}, define an intermediate vector  $(\tilde{\balpha}^t,\tilde{\bbeta}^t):=(\balpha_{h}^*,\bbeta^*)+\eta(\hat{\bdelta}^t,\hat{\bDelta}^t)$ with $\eta:=\sup\big\{u\in[0,1]: u(\hat{\bdelta}^t,\hat{\bDelta}^t)\in\BB_{{\Omega}}(r)\big\}$, where $\BB_{{\Omega}}(r)$ is defined in \eqref{l2ball}. On event $\cR(c,r,l)\cap\cG(P'(a_0)\lambda)$, we can ensure that $\eta=1$, since $D(\tilde{\balpha}^t,\tilde{\bbeta}^t)\leq\eta D(\hat{\balpha}^{t},\hat{\bbeta}^{t})$ from Lemma F.2 in the supplementary material of \cite{Fan2018} combined with the RSC property gives $\Vert\tilde{\btheta}^t\Vert_\Omega< r_{\text{opt}}$, which implies $(\tilde\bdelta^t,\tilde\bDelta^t)\in\BB_{\Omega}(r)$, thus ensuring $\eta=1$ via proof by contradiction. 
	
	Now, we need to verify the claim \eqref{indexsetclaim}. For the second part of the claim, it holds trivially for $t=1$. Assume that it holds for $1,\hdots,t$. Using the definition of the index set, for $j\in\cS_{t+1}^{{\rm c}}$, we have $\lambda_j^{t}\geq P'(a_0)\lambda$, which verifies the second part of \eqref{indexsetclaim}. For the first part of \eqref{indexsetclaim}, since $\cS_{1}=\cS$, it holds for $t=1$ trivially. Suppose it holds for some $t\geq1$. Then, we get $P_\lambda'(\vert\hat{\beta}_j^{t}\vert)<P'(a_0)\lambda=P_\lambda'(a_0\lambda)$ for $j\in\cS_{t+1}\setminus\cS$, which implies that $\vert\hat{\beta}_j^{t}\vert>a_0\lambda$ due to the monotonicity of $P'(\cdot).$ Thus, we get an upper bound for the size of the set as follows.
	\#
	\vert\cS_{t+1}\setminus\cS\vert^{1/2}<(a_0\lambda)^{-1}\Vert(\hat{\bbeta}^{t}-\bbeta^*)_{\cS_{t+1}\setminus\cS}\Vert_2\leq(a_0\lambda)^{-1}\gamma_p^{-1/2}r_{\text{opt}}=bs^{1/2}.
	\#
	Hence, we get $\vert\cS_{t+1}\vert=\vert\cS\vert+\vert\cS_{t+1}\setminus\cS\vert< s+b^2s,$ which verifies the first part of the claim \eqref{indexsetclaim}. 
	
	To refine the rate at each step, we need to control the terms in \eqref{ellupperbound}. For each $j$, we consider two cases, where the first case is when $\vert\hat{\beta}_j^{t-1}-\beta_j^*\vert\geq a_0\lambda$ which gives $a_0^{-1}\vert\hat{\beta}_j^{t-1}-\beta_j^*\vert\geq\lambda\geq\lambda_j^{t-1}$, and the other case is when $\vert\hat{\beta}_j^{t-1}-\beta_j^*\vert< a_0\lambda$ which gives $\lambda_j^{t-1}\leq P_\lambda'((\vert\beta_j^*\vert-a_0\lambda)_+)$ due to the monotonicity of $P'(\cdot)$ and the fact that $\vert\beta_j^*\vert-a_0\lambda<\vert\hat{\beta}_j^{t-1}\vert$ using the triangle inequality. Then, we get following bounds
	\#
	\Vert\blambda_\cS\Vert_2\leq\Vert P_\lambda'((\vert\beta_\cS^*\vert-a_0\lambda)_+)\Vert_2+a_0^{-1}\Vert\hat{\bDelta}_\cS^{t-1}\Vert_2,
	\#
	and
	\begin{align}
		\Vert\bomega_{\cS_t}^*\Vert_2&\leq\Vert\bomega_{\cS}^*\Vert_2+\vert\cS_{t}\setminus\cS\vert^{1/2}\Vert\bomega^*_{\cS^{{\rm c}}}\Vert_\infty\nonumber\\&\leq\Vert\bomega_{\cS}^*\Vert_2+(a_0\lambda)^{-1}\Vert\bomega^*_{\cS^{{\rm c}}}\Vert_\infty\Vert\hat{\bDelta}_{\cS_{t}\setminus\cS}^{t-1}\Vert_2\nonumber\\&\leq\Vert\bomega_{\cS}^*\Vert_2+\frac{P'(a_0)}{2a_0}\Vert\hat{\bDelta}_{\cS_{t}\setminus\cS}^{t-1}\Vert_2,
	\end{align}
	which gives 
	\#
	\Vert\blambda_\cS\Vert_2+\Vert\bomega_{\cS_t}^*\Vert_2\leq\Vert P_\lambda'((\vert\beta_\cS^*\vert-a_0\lambda)_+)\Vert_2+ \Vert\bomega_{\cS}^*\Vert_2+a_0^{-1}\sqrt{1+\{P'(a_0)/2\}^2}\Vert\hat{\bDelta}_{\cS_{t}}^{t-1}\Vert_2.
	\#
	Substituting this results into \eqref{ellupperbound}, we get
	$$
	D(\hat{\balpha}_h^{t},\hat{\bbeta}_h^{t})\leq\bigg\{\sqrt{q}\Vert\bzeta^*\Vert_2+\gamma_p^{-1/2}\bigg(\Vert P_\lambda'((\vert\beta_\cS^*\vert-a_0\lambda)_+)\Vert_2+ \Vert\bomega_{\cS}^*\Vert_2+a_0^{-1}\sqrt{1+\{P'(a_0)/2\}^2}\Vert\hat{\bDelta}_{\cS_{t}}^{t-1}\Vert_2\bigg)\bigg\}\Vert\btheta^{t}\Vert_{\Omega}.
	$$
	Then, combining with the RSC property, we get
	\begin{align}
		\Vert\btheta^{t}\Vert_{\Omega}&\leq \delta\cdot\Vert\btheta^{t-1}\Vert_{\Omega}+c^{-1}\Big(\gamma_p^{-1/2}\Vert P_\lambda'((\vert\bbeta_\cS^*\vert-a_0\lambda)_{+})\Vert_2+\Vert\bomega^*_\cS\Vert_2\Big)+c^{-1}\sqrt{q}\Vert\bzeta^*\Vert_2,
	\end{align}
	where $\delta=\sqrt{1+\{P'(a_0)/2\}^2}/(ca_0\gamma_p)\in(0,1)$.\qed
	
	\subsection{Proof of Theorem \ref{weakoracletheorem}}
	
	Let $\lambda=8P'(a_0)^{-1}\nu_0\sigma_x\sqrt{\log(2p)/n}$ , and let  $\lambda_1=P'(a_0)^{-1}\sqrt{\{s+\log(q)+z\}/n}\leq\lambda$. If the inequality does not hold, then  let $\lambda=P'(a_0)^{-1}\sqrt{\{s+\log(q)+z\}/n}$. Then, with slight modification of the proof of Proposition \ref{goodeventlemma}, the event $\{\Vert\bzeta^*\Vert_\infty\leq3P'(a_0)\lambda_1/(2q), \Vert\bomega^*\Vert_\infty\leq P'(a_0)\lambda/2\}\subset\cG(P'(a_0)\lambda)$ holds with probability at least $1-q/(2p)-e^{-2(s+z)}.$
	
	On $\cG(P'(a_0)\lambda)$, we can use the results from Proposition \ref{RSCproperty} and Theorem \ref{iterativeboundtheorem}. 
	Consider $c=0.5\underline{f}\cdot\underline{\kappa}$, $l=4\gamma_p^{-1/2}\sqrt{2\cdot\max(s,q)}$ and $b$ such that satisfies
	$$\sqrt{2}P'(a_0)(b^2+1)^{1/2}+4=a_0\underline{\kappa}\underline{f}\gamma_pb.$$
	Also, let $r\geq q^{1/2}r_{\text{opt}}=a_0b(\gamma_psq)^{1/2}\lambda$.
	Then, using Proposition \ref{RSCproperty}  with proper conditions of sample size and the smoothing parameter decribed therein, the event $\cR(c,r,l)$ holds with probability at least $1-q/(2p)$. With above preparations, we get the following estimation error bound using Theorem \ref{iterativeboundtheorem} on event $\{\Vert\bzeta^*\Vert_\infty\leq3P'(a_0)\lambda_1/(2q), \Vert\bomega^*\Vert_\infty\leq P'(a_0)\lambda/2\}\cap\cR(c,r,l)\subset\cG(P'(a_0))\cap\cR(c,r,l)$,
	\#
	\Vert\btheta^{t}\Vert_{\Omega}\leq\delta^{t-1}r_{\text{opt}}+(1-\delta)^{-1}(r_{\text{ora}}+c^{-1}\sqrt{q}\Vert\bzeta^*\Vert_2)\label{iterativebound3}
	\#  where $r_{\text{ora}}=c^{-1}\big\{\gamma_p^{-1/2}\Vert P_\lambda'((\vert\bbeta_\cS^*\vert-a_0\lambda)_{+})\Vert_2+\Vert\bomega^*_\cS\Vert_2\big\}$ is defined in Theorem \ref{iterativeboundtheorem}.
	Since we have $t\gtrsim\log\{\log(2p)\}/\log(1/\delta)$, we get $\delta^{t-1}r_{\text{opt}}\lesssim\sqrt{s/n}$. 
	
	Remaining quantity to bound is $r_{\text{ora}}+c^{-1}\sqrt{q}\Vert\bzeta^*\Vert_2$ from \eqref{iterativebound3}. 
	The second term $c^{-1}\sqrt{q}\Vert\bzeta^*\Vert_2$ is bounded by $(3/2)P'(a_0)^{-1}c^{-1}\sqrt{\{s+\log(q)+z\}/n}$ on the event $\{\Vert\bzeta^*\Vert_\infty\leq3P'(a_0)\lambda_1/(2q), \Vert\bomega^*\Vert_\infty\leq P'(a_0)\lambda/2\}\subset\cG(P'(a_0)\lambda)$. By using the beta-min condition, the shrinkage bias term $\Vert P_\lambda'((\vert\bbeta_\cS^*\vert-a_0\lambda)_{+})\Vert_2$ vanishes. The only term remaining to bound is $\Vert\bomega_{\cS}^*\Vert_2$. 
	
	Consider 
	\#
	\Vert\bS^{-1/2}\bomega_{\cS}^*\Vert_2=\bigg\Vert\frac{1}{n q}\sn\sum_{k=1}^q\big\{\bar{K}_h(\alpha_{h,k}^*-\varepsilon_i)-\tau_k\big\}\bS^{-1}\bx_{i,\cS}\bigg\Vert_2,
	\#
	where $\bS:=\EE(\bx_{\cS}\bx_{\cS}\T)$. Since $\vert(\bS^{-1/2}\bomega_{\cS}^*)_j\vert=\vert q^{-1}\sum_{k=1}^q\big\{\bar{K}_h(\alpha_{h,k}^*-\varepsilon_i)-\tau_k\big\}(\bS^{-1/2}\bx_{i,\cS})_j\vert\leq\vert(\bS^{-1/2}\bx_{i,\cS})_j\vert$, we have $\Vert\bS^{-1/2}\bomega_{\cS}^*\Vert_2\leq\Vert\frac{1}{n}\sn\bS^{-1/2}\bx_{i,\cS}\Vert_2$.
	Then, since each $\bS^{-1/2}\bx_{i,\cS}$ is an $s$-dimensional sub-Gaussian random vector with parameter $2\sqrt{2}\nu_0\sigma_x$, we get 
	\begin{align}
		\Vert\bS^{-1/2}\bomega_{\cS}^*\Vert_2\leq8{\nu_0\sigma_x}\sqrt{\frac{2s}{n}}+4{\nu_0\sigma_x}\sqrt{\frac{2\log(1/\epsilon)}{n}}
	\end{align}
	with probability at least $1-\epsilon.$ By letting $\epsilon=e^{-(s+z)}$ and combining all the bounds we have for \eqref{iterativebound2}, we get the desired bound with probability at least $1-q/p-2e^{-(s+z)}$. Seperate bounds comes from $\Vert\hat{\bbeta}_h^{t}-\bbeta^*\Vert_{\Sigma},\Vert(\hat{\balpha}_h^{t}-\balpha_h^*)/\sqrt{q}\Vert_2\leq\Vert\btheta^{t}\Vert_{\Omega}$. Finally, the $h^2$ bias term comes from $\Vert\balpha_h^*-\balpha^*\Vert_2$, which was derived in the Proposition~\ref{proposition3.1}. \qed
	\subsection{Proof of Theorem \ref{deterministicstrongoracle}}
	For $t=1,2,\hdots,$ let $\cT_{t}=\cS\cup\{1\leq j\leq p: \lambda_j^{t-1}=P_\lambda'(\vert\hat{\beta}_j^{t-1}\vert)<P'(a_0)\lambda\}$, and $k=|\cT_{t}|$ . By using the optimality we get
	\begin{gather*}
		0=\Bigg\langle\nabla\hat{Q}_h(\hat{\balpha}^t,\hat{\bbeta}^t)+\lambda\circ\hat{\bg},\begin{bmatrix}
			\hat{\bdelta}^o\\\hat{\bDelta}^o
		\end{bmatrix}\Bigg\rangle\\=\Bigg\langle\nabla\hat{Q}_h(\hat{\balpha}^t,\hat{\bbeta}^t)-\nabla\hat{Q}_h(\hat{\balpha}^o,\hat{\bbeta}^o),\begin{bmatrix}
			\hat{\bdelta}^o\\\hat{\bDelta}^o
		\end{bmatrix}\Bigg\rangle+\Bigg\langle\nabla\hat{Q}_h(\hat{\balpha}^o,\hat{\bbeta}^o),\begin{bmatrix}
			\hat{\bdelta}^o\\\hat{\bDelta}^o
		\end{bmatrix}\Bigg\rangle+\Bigg\langle\lambda\circ\hat{\bg},\begin{bmatrix}
			\hat{\bdelta}^o\\\hat{\bDelta}^o
		\end{bmatrix}\Bigg\rangle\\\geq-\Vert\bzeta^o\Vert_\infty\Vert\hat{\bdelta}^o\Vert_1-\Vert\bomega^o\Vert_\infty\Vert\hat{\bDelta}^o\Vert_1+\Vert\lambda_{\cT_t^c}\Vert_{\min}\Vert\hat{\bDelta}^o_{\cT^c}\Vert_1-\Vert\lambda_{\cT_t}\Vert_{\infty}\Vert\hat{\bDelta}^o_{\cT}\Vert_1,
	\end{gather*}
	where $\hat{\bdelta}^o=\hat{\balpha}^t-\hat{\balpha}^o, \hat{\bDelta}^o=\hat{\bbeta}^t-\hat{\bbeta}^o, \bzeta^o=\nabla_{\balpha}\hat{Q}_h(\hat{\balpha}^o,\hat{\bbeta}^o)$, and $\bomega^o=\nabla_{\bbeta}\hat{Q}_h(\hat{\balpha}^o,\hat{\bbeta}^o)$.
	By rearranging terms and using the optimality ($\Vert\bzeta^o\Vert_\infty=0$), we get
	\begin{gather}
		(\Vert\lambda_{\cT_t^c}\Vert_{\min}-\Vert\bomega^o\Vert_\infty)\Vert\hat{\bDelta}^o_{\cT_t^c}\Vert_1\leq(\Vert\bomega^o\Vert_\infty+\Vert\lambda_{\cT_t}\Vert_\infty)\Vert\hat{\bDelta}_{\cT_t}^o\Vert_1,
	\end{gather}
	which leads to $$\Vert\hat{\bDelta}^o_{\cT_t^c}\Vert_1\leq\{1+2/P'(a_0)\}\Vert\hat{\bDelta}^o_{\cT_t}\Vert_1,$$
	thus \begin{gather}
		\begin{Vmatrix}
			\hat{\bdelta}^o\\\hat{\bDelta}^o
		\end{Vmatrix}_{1}\leq\{2+2/P'(a_0)\}(\Vert\hat{\bdelta}^o\Vert_1+\Vert\hat{\bDelta}_{\cT_t}^o\Vert_1)\leq\{2+2/P'(a_0)\}\bigg\{\frac{2\max(q,k)}{\gamma_p}\bigg\}^{1/2}\begin{Vmatrix}
			\hat{\bdelta}^o\\\hat{\bDelta}^o
		\end{Vmatrix}_{\Omega},
	\end{gather} which explains the choice of $l$ in the theorem.
	
	Now, using the optimality and properties of subdifferential, we get \begin{gather}\Bigg\langle\nabla\hat{Q}_h(\hat{\balpha}^t,\hat{\bbeta}^t)-\nabla\hat{Q}_h(\hat{\balpha}^o,\hat{\bbeta}^o),\begin{bmatrix}
			\hat{\bdelta}^o\\\hat{\bDelta}^o
		\end{bmatrix}\Bigg\rangle=\Bigg\langle-\lambda\circ\hat{\bg}-\nabla\hat{Q}_h(\hat{\balpha}^o,\hat{\bbeta}^o),\begin{bmatrix}
			\hat{\bdelta}^o\nonumber\\\hat{\bDelta}^o
		\end{bmatrix}\Bigg\rangle\\\leq\Vert\bomega_{\cT_t}^o\Vert_2\Vert\hat{\bDelta}_{\cT_{t}}^o\Vert_2-(\Vert\blambda_{\cT_t^c}\Vert_{\min}-\Vert\bomega_{\cT_t^c}\Vert_\infty)\Vert\hat{\bDelta}_{\cT_{t}^c}^o\Vert_1+\Vert\blambda_{\cS}\Vert_2\Vert\hat{\bDelta}_{\cS}^o\Vert_2\label{strongoraclebound}\\\leq(s^{1/2}+0.5P'(a_0)k^{1/2})\lambda\Vert\hat{\bDelta}^o\Vert_2\leq\sqrt{2}(s^{1/2}+0.5P'(a_0)k^{1/2})\lambda\gamma_p^{-1/2}\Vert\hat{\btheta}^{ot}\Vert_\Omega\end{gather}
	Using the similar argument given in the proof of Theorem \ref{iterativeboundtheorem}, we can get $|\cT_{t}|\leq(1+b^2)s$ and $q^{1/2}\Vert\hat{\btheta}^{ot}\Vert_\Omega<r,$ which makes $$\begin{Vmatrix}
		\hat{\balpha}^t-\hat{\balpha}^o\\\hat{\bbeta}^t-\hat{\bbeta}^o
	\end{Vmatrix}_\Omega\leq r.$$
	Now, define $\cS_t:=\{1\leq j\leq p: |\hat{\beta}_j^t-\beta_j^*|>a_0\lambda\}$, which makes $\cS_0=\cS$.
	We have $$\lambda_j^{t-1}=P_\lambda'\big(|\hat{\beta}_j^{t-1}|\big)\leq P_\lambda'(\vert\beta_j^*|-a_0\lambda)$$ if $j\in\cS\cap\cS_{t-1}^c.$, and $\lambda_j^{t-1}\leq\lambda$ for remaining $j$. Thus, we get
	\begin{gather}
		\Vert\blambda_\cS^{t-1}\Vert_2\leq\Vert P_\lambda'(|\bbeta_\cS^*|-a_0\lambda)\Vert_2+\lambda|\cS\cap\cS_{t-1}|^{1/2}=\lambda|\cS\cap\cS_{t-1}|^{1/2}.
	\end{gather}
	For each $j\in\cT_t\setminus\cS$, $\beta_j^*=0$ and $\lambda_j^{t-1}=P_\lambda'\big(|\hat{\beta}_j^{t-1}|\big)<P_\lambda'(a_0\lambda)$, which leads to $|\hat{\beta}_j^{t-1}-\beta_j^*|>a_0\lambda$, thus we get $\cT_t\setminus\cS\subseteq\cS_{t-1}\setminus\cS$. Then, we get $\Vert\bomega_{\cT_t}^o\Vert_2\leq\Vert\bomega^o\Vert_\infty|\cT_{t}\setminus\cS|^{1/2}\leq\Vert\bomega^o\Vert_\infty|\cS_{t-1}\setminus\cS|^{1/2}$ since $\bomega_\cS^o=\mathbf{0}$. Then, by combining above results with \eqref{strongoraclebound}, we get\begin{gather}
		c\Vert\btheta^{ot}\Vert_{\Omega}^2\leq\{\Vert\bomega^o\Vert_\infty|\cS_{t-1}\setminus\cS|^{1/2}+\lambda|\cS\cap\cS_{t-1}|^{1/2}\}\gamma_p^{-1/2}\Vert\btheta^{ot}\Vert_{\Omega},
	\end{gather}
	which leads to
	\begin{gather}
		\Vert\hat{\bbeta}^t-\hat{\bbeta}^o\Vert_2\leq\gamma_p^{-1/2}\Vert\btheta^{ot}\Vert_{\Omega}\leq \frac{\sqrt{1+\{P'(a_0)/2\}^2}}{c\gamma_p}|\cS_{t-1}|^{1/2}\lambda.
	\end{gather}
	By the definition of $\cS_t$, we have $\min_{j\in\cS_{t}}|\hat{\beta}_j^t-\hat{\beta}_j^o|>a_0\lambda-\Vert\hat{\bbeta}^o-\bbeta^*\Vert_\infty$. Thus, provided that 
	$$\Bigg\{\Vert\hat{\bbeta}^o-\bbeta^*\Vert_\infty\leq\bigg[a_0-\frac{\sqrt{1+\{P'(a_0)/2\}^2}}{\delta c\gamma_p}\bigg]\lambda\Bigg\},$$
	we have $$|\cS_t|^{1/2}<\frac{\Vert(\hat{\bbeta}^t-\hat{\bbeta}^o)_{\cS_t}\Vert_2}{a_0\lambda-\Vert\hat{\bbeta}^o-\bbeta^*\Vert_\infty}\leq\delta|\cS_{t-1}|^{1/2},$$
	which completes the proof.
	\subsection{Proof of Proposition \ref{oracleestimationbahadur}}
	We need to first establish the estimation error bound of the oracle estimator, which is essentially the estimation error bound of low dimensional smoothed CQR estimator. In this proof, with abuse of notations, use same notaion we used for the high-dmensional estiamtion error bound, except for the dimension which is now $s<<n$. Proof strategy is similar to the proof of Theorem \ref{Theorem3.1}, but now it is low dimensional, so that some steps can be omitted. We establish upper and lower bounds of $D(\balpha,\bbeta)$ in low dimension. Here, let $$\cR(c,r):=\Big\{D(\balpha,\bbeta)\geq c\big(\Vert\bDelta\Vert_\Sigma^2+q^{-1}\Vert\bdelta\Vert_2^2\big)\text{ for all }  {\scriptsize		 \begin{bmatrix}
			\bdelta \\
			\bDelta
	\end{bmatrix} }\in\BB_{\Omega}(r)   \Big\}.$$ We first prove that $\cR(c,r)$ holds with high probability. We follow the proof of Proposition \ref{RSCproperty} until \eqref{TalagrandBound}, with slight modification that we no longer require $\CC_{\Omega}(l)$, thus taking supremum only on $\Vert\bDelta_k\Vert_{\bar{\Sigma}}=r$. Thus, just denote $Z_n(l)$ as $Z_n$ in this proof. Then, we need to bound $\EE(Z')$ to establish the RSC property. Using Rademacher symmetrization and Talagrand's contraction principle on \eqref{Znl}, we get \begin{gather}
		\EE(Z')\leq\frac{1}{r}\cdot\EE\Bigg\Vert\frac{1}{n}\sn e_i\chi_{i,k}\bar{\bx_i}\Bigg\Vert_2\leq\overline{f}^{1/2}\sqrt{\frac{hs}{r^2n}}.
	\end{gather} Thus, as long as $n\geq C\overline{f}hs/r^2$ for sufficiently large $C>0$, we get the desired RSC property with probability at least $1-qe^{-(s+t)}.$ Next, we need to get an upper bound of $D(\balpha,\bbeta)$. Using the optimality, we get $D(\hat{\balpha}^o,\hat{\bbeta}^o)\leq\Vert\bzeta^*\Vert_\infty\Vert\hat{\balpha}^o-\balpha_h^*\Vert_1+\Vert\Sigma^{-1/2}\bomega^*\Vert_2\Vert\hat{\bbeta}^o-\bbeta^*\Vert_\Sigma$. Then, using the result of the Proposition \ref{goodeventlemma}, we have $\Vert\bzeta^*\Vert_\infty\Vert\hat{\balpha}^o-\balpha_h^*\Vert_1\leq2\lambda\Vert(\hat{\balpha}^o-\balpha_h^*)/q^{1/2}\Vert_2$ with probability at least $1-2qe^{-8n\lambda^2}$. Setting $\lambda=\sqrt{(s+t)/n}$ gives the desired probability bound. Now it remains to bound $\Vert\Sigma^{-1/2}\bomega^*\Vert_2$. Let $\xi_i:=q^{-1}\sum_{k=1}^q\{\bar{K}((\alpha_{h,k}^*-\varepsilon_i)/h)-\tau_k\}$. We have $\vert\xi_i\vert\leq1.$ Then, $\Vert\Sigma^{-1/2}\bomega^*\Vert_2=\Vert(1/n)\sn\xi_i\bw_i\Vert_2$, where $\bw_i=\Sigma^{-1/2}\bx_i$. Using a covering argument, for any $\epsilon\in(0,1),$ there exist an $\epsilon-$net $\cN_\epsilon$ of the unit sphere with cardinality $|\cN_\epsilon|\leq(1+2/\epsilon)^s$ such that \begin{gather}
		\Vert\Sigma^{-1/2}\bomega^*\Vert_2\leq(1-\epsilon)^{-1}\max_{\bu\in\cN_\epsilon}\bigg\langle\bu,\frac{1}{n}\sn\xi_i\bw_i\bigg\rangle.
	\end{gather}
	Then, we have, for $k=2,3,\hdots$
	\begin{align*}
		\EE\big(\vert\langle\bu,\xi_i\bw_i\rangle\vert^k\big) &\leq\EE\vert\langle\bu,\bw_i\rangle\vert^k\\
		&\leq\nu_0^k\int_{0}^{\infty}\PP(\vert\langle\bu,\bw_i\rangle\vert\geq\nu_0t)kt^{k-1}\diff t\\
		&\leq\nu_0k\int_{0}^\infty t^{k-1}e^{-t}dt\\
		&\leq\frac{k!}{2}\nu_0^2(2\nu_0)^{k-2}.
	\end{align*}
	Now, using Bernstein's inequality and applying union bound over $\cN_\epsilon$, we get \begin{gather}
		\Vert\Sigma^{-1/2}\bomega^*\Vert_2\leq\frac{\nu_0}{1-\epsilon}\bigg(\sqrt{\frac{2u}{n}}+\frac{2u}{n}\bigg)
	\end{gather} with probability at least $1-e^{\log(1+2/\epsilon)s-u}.$ Choosing $\epsilon=2/(e^2-1)$ ahaend $u=2s+t$ gives the desired upper bound with probability at least $1-e^{-t}$. Thus, following the similar argument used in the Theorem \ref{Theorem3.1} to combine the lower and upper bounds of the Bregmann divergence, we get the desired estimation error bound for the oracle estimator. For the Bahadur representation part, we refer to the Theorem 2 of \cite{YWZ2023}.

	\subsection{Proof of Proposition \ref{oraclerscproposition}}
	We restrict our focus on the symmetrized Bregmann divergence with each quantile index $k=1,\ldots,q$, by letting
	$$D_{rsc}^k({\balpha_1},{\bbeta_1},\balpha_2,\bbeta_2):=\frac{1}{n}\sn\left\{\bar{K}\left(\frac{{\alpha}_{1k}-r_i({\bbeta_1})}{h}\right)-\bar{K}\left(\frac{\alpha_{2k}-r_i(\bbeta_2)}{h}\right)\right\}\langle\bar{\bx_i},\bDelta_k\rangle,$$
	where $\bDelta_k:=(\alpha_{1k}-\alpha_{2k},\bbeta_1\T-\bbeta_2\T)\T$. Then, we have
	\#D_{rsc}^k({\balpha_1},{\bbeta_1},\balpha_2,\bbeta_2)\geq\frac{\underline{\kappa}}{nh}\sn\langle\bar{\bx_i},\bDelta_k\rangle^2\ind_{E_{i,k}}\label{dkrsc}\# where \begin{gather}E_{i,k}=\{\vert\varepsilon_i-\alpha_{h,k}^*\vert\leq h/2\}\cap\{\vert\bar{\bx_i}\T\bDelta_k^*\vert\leq h/4\}\cap\{\vert\langle\bx_i,\bDelta_k\rangle\vert/\Vert\bDelta_k\Vert_{\bar{\Sigma}}\leq h/(4r)\},\nonumber\\\bDelta_k^*:=\begin{pmatrix}
			\alpha_{2k}-\alpha_{h,k}^*\\\bbeta_2-\bbeta^*
		\end{pmatrix}.\end{gather}In addition to $\varphi_R$, let 
	$$\phi_R(u):=\ind(|u|< R/2)+2\{1-|u|/R\}\ind(R/2\leq|u|\leq R).$$ Then, we can further lower bound \eqref{dkrsc} by 
	\begin{gather}
		D_{rsc}^k(\balpha_1,\bbeta_1,\balpha_2,\bbeta_2)\geq \underline{\kappa}\Vert\bDelta_k\Vert_{\bar{\Sigma}}^2\cdot\underbrace{\frac{1}{nh}\sn\chi_{i,k}\cdot\varphi_{h/(4r)}\big(\bar{\bx_i}\T\bDelta_k/\Vert\bDelta_k\Vert_{\bar{\Sigma}}\big)\phi_{h/4}\big(\bar{\bx_i}\T\bDelta_k^*\big)}_{=:D_{rsc}^{0,k}(\balpha_1,\bbeta_1,\balpha_2,\bbeta_2)}
	\end{gather} Similar to the proof of the Proposition \ref{RSCproperty}, we have $3\underline{f}h/4\leq\EE\chi_{i,k}\leq5\overline{f}h/4 \text{ almost surely.}$ Using the sub-Gaussianity and the similar analyses following \eqref{expectationlowerbound}, we have
	\begin{align}
		&\EE\big\{\chi_{i,k}\varphi_{h/(4r)}\big(\bar{\bx_i}\T\bDelta_k/\Vert\bDelta_k\Vert_{\bar{\Sigma}}\big)\phi_{h/4}\big(\bar{\bx_i}\T\bDelta_k^*\big)\big\}\nonumber\\&\geq\frac{3}{4}\underline{f}h\EE\big\{\varphi_{h/(4r)}\big(\bar{\bx_i}\T\bDelta_k/\Vert\bDelta_k\Vert_{\bar{\Sigma}}\big)\phi_{h/4}\big(\bar{\bx_i}\T\bDelta_k^*\big)\big\}\nonumber\\&\geq\frac{3}{4}\underline{f}h\big(1-\EE\{\bxi_{\bDelta_k}^2\ind_{\vert\bxi_{\bDelta_k}\vert\geq h/(8r)}\}-\EE\{\bxi_{\bDelta_k}^2\ind_{\vert\bx_i\T\bDelta_k^*\vert\geq h/8}\}\big)>\frac{9}{16}\underline{f}h
	\end{align} when $h/(8r)>3\nu_0^2$.  Now, we need to bound $\vert -D_{rsc}^{0,k}(\balpha_1,\bbeta_1,\balpha_2,\bbeta_2)+\EE\{D_{rsc}^{0,k}(\balpha_1,\bbeta_1,\balpha_2,\bbeta_2)\}\vert$ uniformly over $\Lambda(r,l)$.
	Let $$Z_k(r,l):=\sup_{\Lambda(r,l)}\vert -D_{rsc}^{0,k}(\balpha_1,\bbeta_1,\balpha_2,\bbeta_2)+\EE\{D_{rsc}^{0,k}(\balpha_1,\bbeta_1,\balpha_2,\bbeta_2)\}\vert.$$ If we denote $D_{rsc}^{0,k}(\balpha_1,\bbeta_1,\balpha_2,\bbeta_2)=(1/n)\sn w_k(\bx_i,\varepsilon_{i})$, where $$w_k(\bx_i,\varepsilon_{i}):=(\chi_{i,k}/h)\cdot\varphi_{h/(4r)}\big(\bar{\bx_i}\T\bDelta_k/\Vert\bDelta_k\Vert_{\bar{\Sigma}}\big)\phi_{h/4}\big(\bar{\bx_i}\T\bDelta_k^*\big),$$ we have $$0\leq w_k(\bx_i,\varepsilon_{i})\leq h/(8r)^2, \quad \text{and}\quad\EE w_k^2(\bx_i\varepsilon_i)\leq(8r)^{-2}\cdot5\overline{f}h/4.$$ Again using Talagrand's inequality as in the proof of Proposition \ref{RSCproperty}, for any $t>0$, we get \begin{gather}
		Z_k(r,l)\leq \frac{5}{4}\EE Z_k(r,l)+\sqrt{\frac{\overline{f}ht}{16r^2n}}+\frac{ht}{12r^2n}
	\end{gather} with probability at least $1-e^{-t}$.  From here, we closely follow the proof of Lemma E.2 of \cite{ncvxQR} to bound $\EE Z_k(r,l).$ With Rademacher symmetrization and using the connection between Gaussian and Rademacher complexities that in Lemma 4.5 of \cite{ledoux2013probability}, we get \begin{gather}
		\EE Z_k(r,l)\leq 2\sqrt{\frac{\pi}{2}}\cdot\EE\Bigg\{\sup_{\Lambda(r,l)}\GG_k(\balpha_1,\bbeta_1,\balpha_2,\bbeta_2)\Bigg\}\label{gaussrademacher}
	\end{gather} where $\GG_k(\balpha_1,\bbeta_1,\balpha_2,\bbeta_2):=(nh)^{-1}\sn g_{i,k}\chi_{i,k}\cdot\varphi_{h/(4r)}\big(\bar{\bx_i}\T\bDelta_k/\Vert\bDelta_k\Vert_{\bar{\Sigma}}\big)\phi_{h/4}\big(\bar{\bx_i}\T\bDelta_k^*\big)$, and $g_{i,k}$ are independent standard normal random variables. Denote $\EE^*$ be the conditional expectation given data $\{(y_i,\bx_i)\}_{i=1}^n$. Then, $\{\GG_k(\balpha_1,\bbeta_1,\balpha_2,\bbeta_2)\}_{\Lambda(r.l)}$ is a Gaussian process, zero at the true parameter. Now, apply the Gaussian comparison theorem to bound $\EE^*\{\sup_{\Lambda(r,l)}\GG_k(\balpha_1,\bbeta_1,\balpha_2,\bbeta_2)\}$. 
	Let $\bgamma_1\T=(\balpha_1\T,\bbeta_1\T), \bgamma_2\T=(\balpha_2\T,\bbeta_2\T)$
	Since only $k$-th element of $\balpha$ contributes to $\GG_k$,   for $(\bgamma_1,\bgamma_2),(\bgamma_1',\bgamma_2')\in\Lambda(r,l)$, denote $\bgamma_{1,k}\T=(\alpha_{1k},\bbeta_1\T), \bgamma_{2,k}\T=(\alpha_{2k},\bbeta_2\T), {\bgamma_{1,k}'}\T=(\alpha_{1k}',{\bbeta_1'}\T), {\bgamma_{2,k}'}\T=(\alpha_{2k}',{\bbeta_2'}\T)$, and abbreviate the notation to $\GG_k(\bgamma_{1,k},\bgamma_{2,k})$. 
	
	Let ${\bDelta_k'}\T=(\alpha_{1k}'-\alpha_{2k}',{\bbeta_1'}\T-{\bbeta_2'}\T), {{\bDelta_k'}^*}\T=(\alpha_{2k}'-\alpha_{h,k}^*,{\bbeta_2'}\T-{\bbeta^*}\T) $. Then, we have
	\begin{gather*}
		\GG_k(\bgamma_{1,k},\bgamma_{2,k})-\GG_k(\bgamma_{1,k}',\bgamma_{2,k}')\\=\GG_k(\bgamma_{1,k},\bgamma_{2,k})-\GG_k(\bgamma_{1,k}'+\bDelta_k',\bgamma_{2,k}')+\GG_k(\bgamma_{1,k}'+\bDelta_k',\bgamma_{2,k}')-\GG_k(\bgamma_{1,k}',\bgamma_{2,k}')\\=\frac{1}{nh}\sn g_{i,k}\chi_{i,k}\cdot\varphi_{h/(4r)}(\bar{\bx_i}\T\bDelta_k/\Vert\bDelta_k\Vert_{\bar{\Sigma}})\{\phi_{h/4}(\bar{\bx_i}\T\bDelta_k^*)-\phi_{h/4}(\bar{\bx_i}\T{\bDelta_k'}^*)\}\\+\frac{1}{nh}\sn g_{i,k}\chi_{i,k}\cdot\phi_{h/4}(\bar{\bx_i}\T{\bDelta_k'}^*)\{\varphi_{h/(4r)}(\bar{\bx_i}\T\bDelta_k/\Vert\bDelta_k\Vert_{\bar{\Sigma}})-\varphi_{h/(4r)}(\bar{\bx_i}\T\bDelta_k'/\Vert\bDelta_k'\Vert_{\bar{\Sigma}})\}.
	\end{gather*} Now, using the Lipshitz continuity of $\phi_R, \varphi_R$ and $\varphi_R\leq(R/2)^2$, we get \begin{align}
		&\EE^*\big\{\GG_k(\bgamma_{1,k},\bgamma_{2,k})-\GG_k(\bgamma_{1,k}'+\bDelta_k',\bgamma_{2,k}')\big\}^2\nonumber\\&\leq\frac{1}{n^2}\sn\frac{h^2}{(8r)^4}\Big(\frac{8}{h}\Big)^2\langle\bar{\bx_i},\bgamma_{2,k}-\bgamma_{2,k}'\rangle^2\chi_{i,k}=\bigg(\frac{1}{8r^2n}\bigg)^2\sn\langle\bar{\bx_i},\bgamma_{2,k}-\bgamma_{2,k}'\rangle^2\chi_{i,k}\label{Gfirstsquare}
	\end{align} and 
	\begin{align}
		&\EE^*\big\{\GG_k(\bgamma_{1,k}'+\bDelta_k',\bgamma_{2,k}')-\GG_k(\bgamma_{1,k}',\bgamma_{2,k}')\big\}^2\nonumber\\&\leq\frac{1}{(nh)^2}\sn\big\{\varphi_{h/(4r)}(\bar{\bx_i}\T\bDelta_k/\Vert\bDelta_k\Vert_{\bar{\Sigma}})-\varphi_{h/(4r)}(\bar{\bx_i}\T\bDelta_k'/\Vert\bDelta_k'\Vert_{\bar{\Sigma}})\big\}^2\chi_{i,k}\nonumber\\&\leq\bigg(\frac{1}{4rn}\bigg)^2\sn\big(\bar{\bx_i}\T\bDelta_k/\Vert\bDelta_k\Vert_{\bar{\Sigma}}-\bar{\bx_i}\T{\bDelta_k'}\Vert\bDelta_k'\Vert_{\bar{\Sigma}}\big)\chi_{i,k}\label{Gsecondsquare}
	\end{align} Now, we have an inequality 
	\begin{align*}
		\EE^*\{\GG_k(\bgamma_{1,k},\bgamma_{2,k})-\GG_k(\bgamma_{1,k}',\bgamma_{2,k}')\}^2&\leq2\EE^*\big\{\GG_k(\bgamma_{1,k},\bgamma_{2,k})-\GG_k(\bgamma_{1,k}'+\bDelta_k',\bgamma_{2,k}')\big\}^2\\&+ 2\EE^*\big\{\GG_k(\bgamma_{1,k}'+\bDelta_k',\bgamma_{2,k}')-\GG_k(\bgamma_{1,k}',\bgamma_{2,k}')\big\}^2
	\end{align*} which can be bounded by using \eqref{Gfirstsquare} and \eqref{Gsecondsquare}. Define another Gaussian process $\{\ZZ_k(\bgamma_{1},\bgamma_{2})\}_{\Lambda(r,l)}$ as 
	\begin{align*}
		\ZZ_k(\bgamma_{1},\bgamma_{2})&=\frac{\sqrt{2}}{8r^2n}\sn g_{i,k}'\langle\bar{\bx_i},\bDelta_k^*\rangle\chi_{i,k}+\frac{\sqrt{2}}{4rn}\sn g_{i,k}''\frac{\langle\bar{\bx_i},\bDelta_k\rangle}{\Vert\bDelta_k\Vert_{\bar{\Sigma}}}\chi_{i,k}\\&=\frac{\sqrt{2}}{8r^2n}\sn \langle g_{i,k}'\bar{\bx}_{i,\cS},\bDelta_{k,\cS}^*\rangle\chi_{i,k}+\frac{\sqrt{2}}{4rn}\sn g_{i,k}''\frac{\langle\bar{\bx_i},\bDelta_k\rangle}{\Vert\bDelta_k\Vert_{\bar{\Sigma}}}\chi_{i,k},
	\end{align*}
	where $\bar{\bx}_{i,\cS}=(1,\bx_{i,\cS}\T)\T,$ $\bDelta_{k,\cS}^*=(\alpha_{2k}-\alpha_{h,k}^*,\bbeta_{2,\cS}\T-{\bbeta^*}\T)\T$, and $g_{1,k}', g_{1,k}'',\hdots, g_{n,k}', g_{n,k}''$ are i.i.d. standard normal random variable that are independent of other variables.. We can also abbreviate the notation as $\ZZ_k(\bgamma_{1,k},\bgamma_{2,k})$. Then, we have an inequality $\EE^*\{\GG_k(\bgamma_{1,k},\bgamma_{2,k})-\GG_k(\bgamma_{1,k}',\bgamma_{2,k}')\}^2\leq\EE^*\{\ZZ_k(\bgamma_{1,k},\bgamma_{2,k})-\ZZ_k(\bgamma_{1,k}',\bgamma_{2,k}')\}^2$. Applying Sudakov-Fernique's Gaussian comparison inequality (see, e.g. Theorem 7.2.11 in \cite{V2018}), we get 
	\begin{gather}
		\EE^*\bigg\{\sup_{\Lambda(r,l)}\GG_k(\bgamma_{1},\bgamma_2)\bigg\}\leq\EE^*\bigg\{\sup_{\Lambda(r,l)}\ZZ_k(\bgamma_{1},\bgamma_2)\bigg\}.\label{GZbound}
	\end{gather}The above remains valid if we replace $\EE^*$ by $\EE$. We use the cone-like constraint $\Vert\bDelta_k\Vert_1\leq l\Vert\bDelta_k\Vert_{\bar{\Sigma}}$, and $\Vert\bDelta_k^*\Vert_{\bar{\Sigma}}\leq r/2,$ which leads to 
	\begin{align}
		\EE\bigg\{\sup_{\Lambda(r,l)}\ZZ_k(\bgamma_{1},\bgamma_{2})\bigg\}&\leq\frac{\sqrt{2}}{16r}\EE\bigg\Vert\frac{1}{n}\sn g_{i,k}'\chi_{i,k}\bar{\bS}^{-1/2}\bar{\bx}_{i,\cS}\bigg\Vert_2+\frac{\sqrt{2}l}{4r}\EE\bigg\Vert\frac{1}{n}\sn g_{i,k}''\chi_{i,k}\bar{\bx_i}\bigg\Vert_\infty\nonumber\\&\leq\frac{\sqrt{2}}{16r}\sqrt{\frac{5\overline{f}h}{4}\frac{s}{n}}+\frac{\sqrt{2}l}{4r}\EE\bigg\Vert\frac{1}{n}\sn g_{i,k}''\chi_{i,k}\bar{\bx_i}\bigg\Vert_\infty \label{Zsup}.
	\end{align}  where $\bar{\bS}:=\EE\bar{\bx}_\cS\bar{\bx}_\cS\T.$ Then, from \eqref{gaussrademacher}, \eqref{GZbound}, and \eqref{Zsup}, we get
	\begin{gather}
		\EE Z_k(r,l)\leq\sqrt{\pi}\bigg\{\frac{\sqrt{5}}{16}\sqrt{\frac{hs}{r^2n}}+\frac{l}{2r}\EE\bigg\Vert\frac{1}{n}\sn g_{i,k}''\chi_{i,k}\bar{\bx_i}\bigg\Vert_\infty\bigg\} \label{EZkbound}
	\end{gather}. To bound the second term of the right-hand side of \eqref{EZkbound}, Let $G_{j}=\sn g_{i,k}\chi_{i,k}\bar{x}_{ij}$ for $j=1,\hdots,p+1.$ Using the sub-Gaussianity (A3), for $k\geq3$, we have
	$$\EE\vert\bar{x}_j\vert^k\leq2\nu_0^k\sigma_{jj}^{k/2}k\int_{0}^\infty t^{k-1}e^{-t^2/2}\diff t=2^{k/2}\nu_0^k\sigma_{jj}^{k/2}k\Gamma(k/2).$$ By using the identity $\Gamma(k)\Gamma(k+1/2)=2^{1-2k}\sqrt{\pi}\Gamma(2k)$, we get 
	$$\EE\vert g_{i,k}\bar{x}_{ij}\vert\leq2^{k/2}\frac{\Gamma(\frac{k+1}{2})}{\sqrt{\pi}}\cdot2^{k/2}\nu_0^k\sigma_{jj}^{k/2}k\Gamma(k/2)=2\nu_0^k\sigma_{jj}^{k/2}k!.$$ Thus, for any $0\leq\lambda<(2\nu_0\sigma_{jj}^{1/2})^{-1}$,\begin{align*}
		\EE e^{\lambda g_{i,k}\chi_{i,k}\bar{x}_{ij}}&\leq1+\frac{1}{2}\cdot\frac{5\overline{f}h}{4}\sigma_{jj}\lambda^2+2\cdot\frac{5\overline{f}h}{4}\sum_{k=3}^\infty\frac{\lambda^{2k}}{(2k)!}\nu_0^{2k}\sigma_{jj}^{2k}(2k)!\\&\leq1+\frac{1}{2}\cdot\frac{5\overline{f}h}{4}\sigma_{jj}\nu_0^2\sum_{k=2}^\infty\lambda^k(2\nu_0\sigma_{jj}^{1/2})^{k-2}\\&\leq1+\frac{1}{2}\cdot\frac{5\overline{f}h}{4}\cdot\frac{\nu_0^2\sigma_{jj}\lambda^2}{1-2\nu_0\sigma_{jj}^{1/2}\lambda}
	\end{align*} which leads to $\log\EE e^{\lambda G_j}\leq\frac{1}{2}\cdot\frac{5\overline{f}h}{4}\cdot\frac{\nu_0^2\sigma_{jj}\lambda^2n}{1-2\nu_0\sigma_{jj}^{1/2}\lambda}$. We can apply same to $-G_j$ using symmetry. By Corollary 2.6 in \cite{boucheron2013concentration}, we have \begin{gather}
		\EE\bigg\Vert\frac{1}{n}\sn g_{i,k}''\chi_{i,k}\bar{\bx_i}\bigg\Vert_\infty\leq\nu_0\sigma_{\bx}\Bigg\{\frac{5}{2}\sqrt{\frac{\overline{f}h\log(2p)}{n}}+\frac{2\log(2p)}{n}\Bigg\}
	\end{gather}. Then, taking $r=h/(24\nu_0^2)$ and combining above result, we get $Z_{k}(r,l)\leq\underline{f}/16$ with probability at least $1-q/(2p)$, which leads to the conclusion by combining those for all $k=1,\hdots, q$.
	\subsection{Proof of Theorem \ref{Strongoracle}}
	To prove Theorem \ref{Strongoracle}, we need to verify the event in Theorem \ref{deterministicstrongoracle} holds with high probability.  We closely follow the proof of Lemma E.3 in \cite{ncvxQR}. First, we bound $\Vert\nabla_{\bbeta}\hat{Q}_h(\hat{\balpha}^o,\hat{\bbeta}^o)\Vert_\infty.$ Let $\bomega_h(\balpha,\bbeta)=\nabla_{\bbeta}\hat{Q}_h(\balpha,\bbeta)-\nabla_{\bbeta}{Q}_h(\balpha,\bbeta)$ We have 
	$$\Vert\nabla_{\bbeta}\hat{Q}_h(\hat{\balpha}^o,\hat{\bbeta}^o)\Vert_\infty\leq\Vert\bomega_h(\hat{\balpha}^o,\hat{\bbeta}^o)-\bomega_h(\balpha_{h}^*,{\bbeta}^*)\Vert_\infty+\Vert\nabla_{\bbeta}{Q}_h(\hat{\balpha}^o,\hat{\bbeta}^o)\Vert_\infty+\Vert\bomega^*\Vert_\infty.$$
	Let $\bgamma\T=(\balpha\T,\bbeta\T), {\bgamma_h^*}\T=(\balpha_h^*,{\bbeta^*}\T)$. Define the oracle neighborhood $\bTheta_{\cS}^*(r)=\{\bgamma\in\bgamma_h^*+\BB_{\Omega}(r), \bbeta_{\cS^{{\rm c}}}=\mathbf{0}\}.$ Conditioned on the event $\{\hat{\bgamma}^o\in\bgamma_h^*+\BB_{\Omega}(r)\}$, we have \begin{gather}
		\Vert\bomega_h(\hat{\balpha}^o,\hat{\bbeta}^o)-\bomega_h(\balpha_{h}^*,{\bbeta}^*)\Vert_\infty\leq\sup_{\bTheta_{\cS}^*(r)}\Vert\bomega_h(\balpha,\bbeta)-\bomega_h(\balpha_{h}^*,{\bbeta}^*)\Vert_\infty
	\end{gather} Then, we can bound the above term by bounding it for each quantile index $k=1.\hdots, q$. Let $\bDelta_{k,\cS}^*=\begin{pmatrix}
		\alpha_k-\alpha_{h,k}^*\\\bbeta_\cS-\bbeta_\cS^*
	\end{pmatrix}$. Also, let $W_{kj}(\bDelta_{k,\cS}^*)=(1/n)\sn(w_{ikj}-\EE w_{ikj})$ where\begin{gather}
		w_{ikj}:=\bigg\{\bar{K}\bigg(\frac{\alpha_{h,k}^*-\varepsilon_i+\langle\bar{\bx}_{i,\cS},\bDelta_{k,\cS}^*\rangle}{h}\bigg)-\bar{K}\bigg(\frac{\alpha_{h,k}^*-\varepsilon_i}{h}\bigg)\bigg\}\frac{x_{ij}}{\sigma_{jj}^{1/2}}.
	\end{gather} Then, we have  \begin{gather}
		\sup_{\bTheta_{\cS}^*(r)}\Vert\bomega_h(\balpha,\bbeta)-\bomega_h(\balpha_{h}^*,{\bbeta}^*)\Vert_\infty\leq\sigma_{\bx}\frac{1}{q}\sum_{k=1}^q\max_{1\leq j\leq p}\sup_{\Vert\bDelta_{k,\cS}^*\Vert\leq r}\vert W_{kj}(\bDelta_{k,\cS}^*)\vert
	\end{gather} Using  (A1') and (A2'), we get $\vert w_{ikj}\vert\leq h^{-1}\vert x_{ij}\langle\bar{\bx}_{i,\cS},\bDelta_{k,\cS}^*\rangle/\sigma_{jj}^{1/2}\vert, \vert\EE(w_{ikj})\vert\leq\overline{f}\Vert\bDelta_{k,\cS}^*\Vert_{\bar{\bS}}$. It can be also shown that $$\EE(w_{ijk}^2|\bx_{i})\leq\overline{f}h^{-1}(x_{ij}^2/\sigma_{jj})\langle\bar{\bx}_{i,\cS},\bDelta_{k,\cS}^*\rangle^2$$ using Minkowski's integral inequality. Above inequalities lead to $$\EE\{(w_{ijk}-\EE w_{ijk})^2|\bx_{i}\}\leq 2\overline{f}^2\Vert\bDelta_{k,\cS}^*\Vert_{\bar{\bS}}^2+2\overline{f}h^{-1}(x_{ij}^2/\sigma_{jj})\langle\bar{\bx}_{i,\cS},\bDelta_{k,\cS}^*\rangle^2.$$ For $\lambda\in\RR, $let $\lambda_*=\lambda/\Vert\bDelta_{k,\cS}^*\Vert_{\bar{\bS}}$, and let $\bDelta_{k,\cS}^{**}=\bDelta_{k,\cS}^*/\Vert\bDelta_{k,\cS}^*\Vert_{\bar{\bS}}.$ Then, using $\vert e^u-1-u\vert\leq(u^2/2)e^{u\vee0}$ we obtain \begin{align}
		&\EE e^{\lambda_* W_{kj}(\bDelta_{k,\cS}^*)}=\prod_{i=1}^n\EE e^{\frac{\lambda_*}{n}(w_{ikj}-\EE w_{ikj})}\nonumber\\
		&\leq \prod_{i=1}^n\EE\bigg\{1+\frac{\lambda_*^2}{2n^2}(w_{ikj}-\EE w_{ikj})^2e^{\frac{\vert\lambda_*\vert}{n}\vert w_{ikj}-\EE w_{ikj}\vert}\bigg\}\nonumber\\&\leq\prod_{i=1}^n\bigg\{1+\frac{\lambda^2\overline{f}^2}{n^2}e^{\frac{\vert\lambda\vert\overline{f}}{n}}\EE e^{\frac{\vert\lambda\vert}{nh}\vert \hat{x}_{ij}\langle\bar{\bx}_{i,\cS},\bDelta_{k,\cS}^{**}\rangle\vert}+\frac{\lambda^2\overline{f}}{n^2h}e^{\frac{\vert\lambda\vert\overline{f}}{n}}\EE\hat{x}_{ij}^2\langle\bar{\bx}_{i,\cS},\bDelta_{k,\cS}^{**}\rangle^2e^{\frac{\vert\lambda\vert}{nh}\vert \hat{x}_{ij}\langle\bar{\bx}_{i,\cS},\bDelta_{k,\cS}^{**}\rangle\vert}\bigg\},\nonumber
	\end{align} where $\hat{x}_{ij}=x_{ij}/\sigma_{jj}^{1/2}.$
	Applying H\"{o}lder's inequality, we get, for any $t>0$,
	$$\EE\hat{x}_{ij}^2\langle\bar{\bx}_{i,\cS},\bDelta_{k,\cS}^{**}\rangle^2e^{t\vert \hat{x}_{ij}\langle\bar{\bx}_{i,\cS},\bDelta_{k,\cS}^{**}\rangle\vert}\leq\big\{\EE\hat{x}_{ij}^2e^{t\hat{x}_{ij}^2}\big\}^{1/2}\cdot \big(\EE\langle\bar{\bx}_{i,\cS},\bDelta_{k,\cS}^{**}\rangle^4e^{t\langle\bar{\bx}_{i,\cS},\bDelta_{k,\cS}^{**}\rangle^2}\big)^{1/2}$$ and $$\EE e^{t\vert\hat{x}_{ij}\langle\bar{\bx}_{i,\cS},\bDelta_{k,\cS}^{**}\rangle\vert}\leq\big(\EE e^{t\hat{x}_{ij}^2}\big)^{1/2}\cdot\big(\EE e^{t\langle\bar{\bx}_{i,\cS},\bDelta_{k,\cS}^{**}\rangle^2}\big)^{1/2}.$$ For a unit vector $\bu\in\mathbb{S}^{p}, $let $Z_{\bu}=(\bz\T\bu)^2/(4\nu_0^2)$, where $\bz=\bar{\Sigma}^{-1/2}\bar{\bx}.$ Then, using sub-Gaussianity, we can show that
	$$\EE e^{Z_{\bu}}=1+\int_{0}^{\infty}e^u\PP(Z_{\bu}\geq u)\diff u\leq3,\quad \EE Z_{\bu}^2e^{Z_{\bu}}=\int_{0}^{\infty}(u^2+2u)e^u\PP(Z_{\bu}\geq u)\diff u\leq 8.$$ Then, we obtain 
	$$\EE e^{\lambda_* W_{kj}(\bDelta_{k,\cS}^*)}\leq\prod_{i=1}^n\{1+C\nu_0^4\overline{f}/(n^2h)\}\leq e^{C\nu_0^4\overline{f}/(nh)},$$
	for $\vert\lambda\vert\leq\min\{nh/(4\nu_0^2),n/\overline{f}\},$ where $C>0$ is an absolute constant. Similarily, for each pair $(\bDelta_{k,\cS}^*,\bDelta_{k,\cS}^{*'})$, we have a bound
	$$\EE e^{\lambda\{W_{kj}(\bDelta_{k,\cS}^*)-W_{kj}(\bDelta_{k,\cS}^{*'})\}/\Vert\bDelta_{k,\cS}^{*}-\bDelta_{k,\cS}^{*'}\Vert_{\bar{\Sigma}}}\leq e^{C\nu_0^4\overline{f}/(nh)}.$$ Then, we can use Corollary 2.2 in \cite{S2012} since above satisfies condition ($\cE d$) of \cite{S2012}. Thus, with probability at least $1-e^{-u}$, $$\sup_{\Vert\bDelta_{k,\cS}^*\Vert\leq r}\vert W_{kj}(\bDelta_{k,\cS}^*)\vert\lesssim\nu_0^2\overline{f}^{1/2}\sigma_{\bx}r\sqrt{\frac{s+u}{nh}}$$ provided $nh\gtrsim(s+u)^{1/2}$ Taking $u=\log(2p)$ and combining bounds for $k=1,\hdots,q$, we get \begin{gather}
		\sup_{\bTheta_{\cS}^*(r)}\Vert\bomega_h(\balpha,\bbeta)-\bomega_h(\balpha_{h}^*,{\bbeta}^*)\Vert_\infty\lesssim\sigma_{\bx}r\sqrt{\frac{s+\log p}{nh}}
	\end{gather} with probability at least $1-q/(2p)$ as long as $nh\gtrsim(s+\log p)^{1/2}$. 
	
	Now, we need to bound $\Vert\nabla_{\bbeta}{Q}_h(\hat{\balpha}^o,\hat{\bbeta}^o)\Vert_\infty.$ Since $\nabla_{\bbeta}{Q}_h(\hat{\balpha}^o,\hat{\bbeta}^o)_\cS=\mathbf{0}$, only need to bound $\cS^c$ part. For $\bgamma\in\bTheta_{\cS}^*(r)$, we have\begin{gather*}
		\nabla_{\bbeta}{Q}_h({\balpha},{\bbeta})_{\cS^c}-\nabla_{\bbeta}{Q}_h({\balpha_h^*},{\bbeta^*})_{\cS^c}\\=\frac{1}{q}\sum_{k=1}^q\EE\int_{-\infty}^\infty K(u)\big\{F_\varepsilon\big(\bar{\bx}_\cS\T\bDelta_{k,\cS}^*+F_\varepsilon^{-1}(\tau_k)-hu\big)-F_\varepsilon\big(F_{\varepsilon}^{-1}(\tau_k)-hu\big)\big\}\diff u\cdot\bx_{\cS^c}.
	\end{gather*} Using Taylor expansion, we get 
	\begin{gather*}F_\varepsilon\big(\bar{\bx}_\cS\T\bDelta_{k,\cS}^*+F_\varepsilon^{-1}(\tau_k)-hu\big)-F_\varepsilon\big(F_{\varepsilon}^{-1}(\tau_k)-hu\big)\\=f_\varepsilon\big(F_{\varepsilon}^{-1}(\tau_k)\big)\cdot\bar{\bx}_\cS\T\bDelta_{k,\cS}^*+\int_{0}^{\bar{\bx}_\cS\T\bDelta_{k,\cS}^*}\{f_\varepsilon(t-hu+F_\varepsilon^{-1}(\tau_k))-f_\varepsilon(F_\varepsilon^{-1}(\tau_k))\}\diff t.\end{gather*} Let $\bJ_{\cS^c{\cS}}=q^{-1}\sum_{k=1}^qf_\varepsilon(F^{-1}(\tau_k))\EE(\bx_{\cS^c}{\bx}_{\cS}\T)$, note that $\EE\bx_{\cS^c}=0,$ then above displays imply\begin{gather*}
		\Vert\nabla_{\bbeta}{Q}_h({\balpha},{\bbeta})_{\cS^c}-\nabla_{\bbeta}{Q}_h({\balpha_h^*},{\bbeta^*})_{\cS^c}-\bJ_{\cS^c{\cS}}(\bbeta-\bbeta^*)\Vert_\infty\\\leq0.5l_0\max_{j\in\cS^c}\EE \bigg\{\frac{1}{q}\sum_{k=1}^q(\bar{\bx}_\cS\T\bDelta_{k,\cS}^2*)^2\vert x_j\vert\bigg\}+\max_{j\in\cS^c}\bigg(\frac{1}{q}\sum_{k=1}^q\kappa_1h\EE\vert x_j\bar{\bx}_\cS\T\bDelta_{k,\cS}^*\vert\bigg)\\\leq0.5l_0\sigma_{\bx}\mu_4^{1/2}\begin{Vmatrix}
			(\balpha-\balpha_h^*)/\sqrt{q}\\\bbeta_\cS-\bbeta_\cS^*
		\end{Vmatrix}_{\Omega_\cS}^2+l_0\kappa_1h\sigma_{\bx}\cdot\frac{1}{q}\sum_{k=1}^q\Vert\bDelta_{k,\cS}^*\Vert_{\bar{\bS}}.
	\end{gather*} Thus, it gives \begin{gather}
		\Vert\nabla_{\bbeta}{Q}_h({\hat{\balpha}^o},{\hat{\bbeta}^o})_{\cS^c}-\bJ_{\cS^c{\cS}}(\hat{\bbeta}^o-\bbeta^*)_\cS\Vert_\infty\leq0.5l_0\sigma_{\bx}\mu_4^{1/2}r^2+l_0\kappa_1h\sigma_{\bx}r.
	\end{gather} Now. it remains to bound. $\Vert\bJ_{\cS^c{\cS}}(\hat{\bbeta}^o-\bbeta^*)_\cS\Vert_\infty.$ Using the condition given in the statement of the theorem, we obtain \begin{gather}
		\Vert\bJ_{\cS^c{\cS}}(\hat{\bbeta}^o-\bbeta^*)_\cS\Vert_\infty=\Vert\bJ_{\cS^c{\cS}}(\bJ_{\cS\cS})^{-1}\bJ_{\cS\cS}(\hat{\bbeta}^o-\bbeta^*)_\cS\Vert_\infty\nonumber\\\leq \max_{j\in\cS^c}\Vert\bJ_{j\cS}(\bJ_{\cS^c{\cS}})^{-1}\Vert_1\cdot\Vert\bJ_{\cS{\cS}}(\hat{\bbeta}^o-\bbeta^*)_\cS\Vert_\infty\leq A_0\cdot\Vert\bJ_{\cS{\cS}}(\hat{\bbeta}^o-\bbeta^*)_\cS\Vert_\infty.
	\end{gather} Instead of using the trivial $\ell_2 $ bound for $\ell_\infty-$norm, we have Proposition \ref{oracleestimationbahadur}, which gives a Bahadur representation of the oracle estimator\begin{gather}
		\Bigg\Vert\bD(\hat{\bbeta}^o-\bbeta^*)_\cS+\frac{1}{nq}\sn\sum_{k=1}^q\{\bar{K}((\alpha_k^*-\varepsilon_i)/h)-\tau_k\}\bx_{i,\cS}\Bigg\Vert_2\lesssim \frac{(s+t)}{h^{1/2}n}+h^{3/2}\sqrt{\frac{q(s+t)}{n}}+h^4 \label{bahadurbound}
	\end{gather} with probability at least $1-3qe^{-t}$, where $\bD=\bJ_{\cS{\cS}}$. This gives \begin{align}
		&\Vert\bD(\hat{\bbeta}^o-\bbeta^*)_\cS\Vert_\infty\nonumber\\&\leq\Vert\bD(\hat{\bbeta}^o-\bbeta^*)_\cS+\nabla_{\bbeta}\hat{Q}_h(\balpha_h^*,\bbeta^*)_\cS\Vert_\infty+\Vert\nabla_{\bbeta}\hat{Q}_h(\balpha_h^*,\bbeta^*)_\cS\Vert_\infty\nonumber\\&\lesssim\frac{(s+t)}{h^{1/2}n}+h^{3/2}\sqrt{\frac{q(s+t)}{n}}+h^4 +\sqrt{\frac{\log(s)+t}{n}},
	\end{align} where $$\Vert\nabla_{\bbeta}\hat{Q}_h(\balpha_h^*,\bbeta^*)_\cS\Vert_\infty\lesssim\sqrt{\frac{\log(s)+t}{n}}$$ with probability at least $1-e^{-t}$ by using the proof of Proposition \ref{goodeventlemma}. Then, combining all results above, we get
	\begin{align}
		\Vert\nabla_{\bbeta}\hat{Q}_h(\hat{\balpha}^o,\hat{\bbeta}^o)\Vert_\infty&\lesssim \sqrt{\frac{\log(2p)}{n}}+\sqrt{\frac{s+\log p}{nh}}\sqrt{\frac{q(s+t)}{n}}\nonumber\\&+A_0\bigg\{\sqrt{\frac{\log(s)+t}{n}}+\frac{(s+t)}{h^{1/2}n}+h^{3/2}\sqrt{\frac{q(s+t)}{n}}+h^4\bigg\}
	\end{align} with probability at least $1-q/p-(5q+1)e^{-t},$ provided that $\sqrt{(s\vee\log p+t)/n}\lesssim h\lesssim1.$ 
	
	Now, take $t=\log(n)$. Using the conditions from Theorem \ref{deterministicstrongoracle} and Proposition \ref{oraclerscproposition}, set $r=h/(24\nu_0^2)$, $l=\sqrt{2}\{2+2/P'(a_0)\}\cdot\big[\max\{q,(1+b^2)s/\gamma_p\}\big]^{1/2}$,$c=0.5\underline{\kappa}\underline{f}$, and choose the bandwidth parameter $h\asymp\{\log(p)/n\}^{1/4}$, we get, with probability at least $1-2q/p-(5q+1)/n$, \begin{gather}
		\Vert\nabla_{\bbeta}\hat{Q}_h(\hat{\balpha}^o,\hat{\bbeta}^o)\Vert_\infty\lesssim\sqrt{\frac{\log(p)}{n}}, \Vert\btheta^o\Vert_{\Omega}\lesssim\sqrt{\frac{s+\log(n)}{n}},\text{ and } \Vert\hat{\bbeta}^o-\bbeta^*\Vert_\infty\lesssim\sqrt{\frac{\log(p)}{n}}\nonumber
	\end{gather} provided that $n\gtrsim\max\{s^{8/3}/(\log p)^{5/3},\log(p)\}$. Then, as in \eqref{strongoracleevent} required in Theorem \ref{deterministicstrongoracle}, choosing $\lambda=C\sqrt{\log(p)/n}$ with sufficiently large $C>0,$ we have \eqref{strongoracleevent} with probability at least $1-2q/p-(5q+1)/n$, provided that $n\gtrsim\max\{s^{8/3}/(\log p)^{5/3},s^{4/3}\log(p)\}$, thus provind the strong oracle property.
\end{document}